\documentclass[
reprint,
amsmath,amssymb,
aps, prx,
]{revtex4-2}

\usepackage[title]{appendix}
\usepackage{graphicx}
\usepackage{dcolumn}
\usepackage{dsfont}
\usepackage{bm}
\usepackage[hidelinks]{hyperref}
\usepackage[capitalise]{cleveref}
\usepackage{lipsum}
\usepackage[dvipsnames]{xcolor}
\usepackage{amsfonts}
\usepackage{yfonts}
\usepackage{booktabs}

\usepackage{amsthm}
\usepackage[scr=boondoxo,scrscaled=1.05]{mathalfa}
\theoremstyle{definition}

\newcommand{\rmd}{\mathrm{d}}

\DeclareFontFamily{U}{futm}{}
\DeclareFontShape{U}{futm}{m}{n}{
  <-> s * [.97534] fourier-bb 
  }{}
\DeclareMathAlphabet{\mathbbs}{U}{futm}{m}{n}

\begin{document}

\preprint{APS/123-QED}

\title{The Memory Hidden in Response Fluctuations: Trajectory-Level Fluctuation-Response Theory and Inequalities for Non-Markovian Jump Dynamics}
\author{Jiming Zheng} \email{jiming@unc.edu}
\author{Zhiyue Lu}
\affiliation{Department of Chemistry, University of North Carolina-Chapel Hill, NC}

\date{\today}

\begin{abstract}
Modern experiments often record nonequilibrium dynamics as sequences of discrete events whose rates depend on the realized past. We develop a fluctuation-response theory for such non-Markovian jump processes directly on the observed event record. Memory can destroy a closed master equation for the state probabilities. Each transition count nevertheless obeys an exact stochastic equation. After the history-dependent mean event tendency is subtracted, the remaining random increment is a martingale increment—the part of the event that cannot be predicted from the past. Martingale increments associated with different transitions and different times are orthogonal. These increments form a complete orthogonal basis for the random deviation of any observable measured from the record, such as a current, occupation time, or event count. The coefficient of a given increment is the event-consequence kernel. It measures how that event changes the predicted final observable, relative to continuing at the instant without the event, for the particular history already realized. Multiplying this kernel by the event intensity gives the history-conditioned response to perturbing the corresponding transition rate. Thus, the intensity-normalized response is exactly the expansion coefficient of that event in the observable fluctuation. This identification yields exact finite-time and finite-frequency fluctuation-response relations. Averaging over histories leaves a nonnegative response-heterogeneity gap. The gap measures how strongly the consequence of the same event varies across histories and tests whether a proposed memory coordinate is response-sufficient. Finite-history versions can be estimated from spontaneous trajectories. Bounding the logarithmic rate sensitivity further gives response-kinetic uncertainty relations controlled by dynamical activity. Applied to BK-channel experimental data, the theory reveals opposite orderings of statistical dwell-time memory and response-relevant memory in cell-membrane and mitochondrial channels. Correlations alone therefore do not reveal which memories matter dynamically.
\end{abstract}

\maketitle

\section{Introduction}
\label{sec:introduction}

Over the past three decades, stochastic thermodynamics has turned stochastic trajectories---and, in the discrete-state setting, Markov jump dynamics---into one of the most successful general frameworks for nonequilibrium physics. It has established trajectory-level definitions of work, heat, and entropy production \cite{sekimoto2010stochastic,seifert2005entropy,seifert2012stochastic,peliti2021stochastic}; fluctuation theorems \cite{jarzynski1997nonequilibrium,crooks1999entropy,lebowitz1999gallavotti,evans2002fluctuation}; information thermodynamics \cite{sagawa2010generalized,parrondo2015thermodynamics,horowitz2014thermodynamics,mandal2012work}; thermodynamic and kinetic uncertainty relations \cite{gingrich2016dissipation,horowitz2020thermodynamic,hasegawa2019fluctuation,dieball2023direct}; and stochastic speed limits \cite{shiraishi2018speed,ito2020stochastic}. Together, these developments connect thermodynamics, fluctuations, information, and kinetics within a single trajectory-level theory. This coherence, however, has been developed almost exclusively for Markov jump dynamics, in which the present state and time suffice to determine the statistics of the next transition.

Modern experiments increasingly resolve nonequilibrium dynamics as sequences of discrete events along individual trajectories, and the resolved event sequences are often not memoryless. Patch-clamp recordings of single ion channels \cite{neher1976single} reveal cumulative inactivation and recovery that depend on previous stimulation protocols \cite{mickus1999properties}; single-molecule enzymology records turnover-to-turnover correlations and dynamic disorder \cite{lu1998single,english2006ever}; neuronal spike trains display serial correlations, refractoriness, and adaptation \cite{farkhooi2009serial}, and synaptic transmission facilitates, depresses, and recovers depending on the presynaptic spike history \cite{markram1996redistribution}; transcription proceeds through refractory promoter dynamics with stimulus-dependent bursting \cite{suter2011mammalian,molina2013stimulus}; and quantum-dot blinking exhibits statistical aging and nonergodicity \cite{brokmann2003statistical} yet admits optical control \cite{shi2021all}.

Operationally, these observations share a common structure: the conditional rate of the next observed event is a functional of the realized history rather than a function only of the present observed state and time. This does not imply that the underlying microscopic dynamics is fundamentally non-Markovian, since non-Markovianity depends on the chosen state space and temporal resolution \cite{lapolla2019manifestations}. At the resolved level, however, the consequences are direct \cite{zwanzig1961memory,mori1965transport}. The present state no longer suffices to predict the next event. Low-order trajectory statistics may fail to identify the dynamical consequences of unresolved conformations \cite{zwanzig1992dynamical,lu1998single,english2006ever}, adaptation \cite{benda2003universal,chacron2001negative}, depletion \cite{dobrunz1997heterogeneity,tsodyks1997neural,abbott1997synaptic}, refractory states \cite{berry1997refractoriness,suter2011mammalian}, or age-dependent kinetics \cite{montroll1965random,brokmann2003statistical}.

Response to controlled perturbations provides the discrimination that such low-order statistics lack: it reveals which variables govern the evolution, on what timescales they act, and how an intervention propagates through the dynamics \cite{kubo1957statistical}. This role becomes decisive when the observed dynamics carries memory. Two trajectories can occupy the same observed state and receive the same perturbation, yet respond differently. Their distinct prehistories modify both the instantaneous susceptibility of the event rate and the subsequent propagation of the perturbation through future events. Dynamical mechanisms that are indistinguishable through unperturbed waiting-time distributions or low-order correlation functions can therefore exhibit sharply different responses to the same controlled change. Sensing, adaptation, reliability, and control are all expressed through input-output behavior. A response theory for non-Markovian event dynamics must therefore be formulated at the level of the trajectory preceding the perturbation, where the same intervention may have different consequences across realizations.

Following the equilibrium fluctuation-dissipation paradigm \cite{kubo1957statistical,marconi2008fluctuation}, response theory for Markov jump dynamics has become comparably structured. Away from equilibrium, it now encompasses generalized trajectory-action formulations \cite{maes2020response,baiesi2009fluctuations,seifert2010fluctuation,speck2006restoring}; dissipation- and steady-state-based generalizations of the equilibrium FDT \cite{harada2005equality,prost2009generalized}; fluctuation-response relations (FRRs) \cite{aslyamov2025nonequilibrium,ptaszynski2026nonequilibrium,aslyamov2026dynamical,bao2024nonlinear}; fluctuation-response inequalities (FRIs) \cite{dechant2020fluctuation,zheng2025universal,zheng2025unified,zheng2026nonlinear,dechant2026finite}; thermodynamic and topological response bounds \cite{aslyamov2025nonequilibrium,zheng2026thermodynamic,owen2020universal,fernandes2023topologically,owen2023size}; and mutual linearity \cite{floyd2025local,harunari2024mutual,bebon2026mutual,zheng2026mutualpre,zheng2026mutual}. Most recently, a stochastic calculus has unified pathwise observables, covariations, response, thermodynamic inequalities, and diffusion limits for Markov jump processes \cite{stutzer2026stochastic}. Together, these results reveal remarkably rigid relations among response, spontaneous fluctuations, dissipation, activity, information, and network structure. They also sharpen a foundational question: which parts of this structure truly require Markovianity, and which reflect a deeper trajectory-level principle?

Beyond Markovianity, important progress exists for particular model classes. Response has been studied for general non-Markovian stochastic processes \cite{hanggi1982stochastic} and nonexponential renewal processes \cite{barbi2005linear}; current fluctuation theorems, dynamical fluctuations, thermodynamics, and uncertainty relations have been developed for semi-Markov dynamics \cite{andrieux2008fluctuation,maes2009dynamical,esposito2008continuous,ertel2022operationally}; and large-deviation and uncertainty structures have been established for self-interacting jump processes \cite{coghi2026level}. Computational mechanics also constructs predictive representations from observed histories through causal states and $\epsilon$-machines \cite{crutchfield1989inferring,shalizi2001computational}. Markov embedding is particularly powerful and has produced master-equation descriptions for broad classes of history-dependent processes \cite{kanazawa2024standard}; an exact embedding of the observed trajectory density preserves all unperturbed multi-time statistics. When the embedding coordinates are experimentally resolved and the action of the control on them is known, fluctuation-response relations are indeed recovered on the embedded dynamics \cite{goerlich2026fluctuation}. For response, however, embedding faces a basic obstruction. An embedding of the unperturbed path distribution does not specify how a laboratory control acts on the hidden or auxiliary coordinates. Physical response, however, is a property of a parameterized family of perturbed dynamics. The response problem is therefore undefined precisely where the embedding is silent. Conversely, a complete predictive embedding typically retains far more information than a given observable and perturbation require. Its auxiliary coordinates need not be unique nor identifiable with microscopic configurations \cite{siegle2010markovian,alexandrovich2016nonparametric,flomenbom2005what}.

These obstructions reflect a deeper change: beyond Markovianity, a perturbation may reshape the memory functional itself, so the elementary response object changes identity. Three questions follow that existing tools cannot address. First, when the observed-state probabilities no longer close, what carries the exact fluctuation-response structure? Markovian identities are built from the generator, its pseudo-inverse, or the propagator--objects that no longer exist at the observed level--while general information bounds constrain response without identifying what fluctuations are made of. Second, can one certify that a reduced description preserves the response of a chosen observable? Matching occupations, mean fluxes, dynamical activity, and every single-state residence-time distribution cannot supply such a certificate: we exhibit a surrogate that reproduces all of them and still fails. Third, when a system admits observation but not clean intervention, as in spontaneous single-channel recordings, is response experimentally accessible at all? Each question must be answered on the observed event history, where both the memory and the experimental control reside.

Here, we answer these questions by constructing the fluctuation-response theory of history-dependent jump processes directly on the observed trajectory, for the broad class of causal, nonexplosive dynamics whose transition rates are determined by the realized past. The mathematical ingredients are classical tools of point-process theory--conditional intensities, the Doob-Meyer decomposition, and martingale representation theorems \cite{bremaud1981point,jacod2003limit,daley2008introduction}--and martingale methods have recently entered stochastic thermodynamics \cite{neri2017statistics,chetrite2011two,roldan2023martingales,stutzer2026stochastic}; what is new is the physics constructed on this foundation. The structural statement is simple. The Doob-Meyer decomposition separates every observed jump into two parts with distinct physical roles. One part is an event tendency that the realized history already determines, carried by the history-dependent rate. The other is a martingale increment that no knowledge of the history can anticipate: it has zero conditional mean, and it is uncorrelated across times. All memory resides in the tendency; the martingale increments remain as simple as in a memoryless process. Non-Markovianity destroys closure at the level of instantaneous state probabilities, so the observed-state probabilities obey no closed master equation, but it leaves this trajectory-level structure intact. The organizing principle of jump dynamics is not the absence of memory, but the separation of each event into predictable tendency and unpredictable increments.

The central result of this construction is that response is the stochastic coordinate of trajectory fluctuations. The elementary response object is the response conditioned on the realized pre-perturbation history. It is how the expected future of an observable changes when a single transition rate is briefly perturbed, given the particular past the system has actually traversed. The core identity then states that every trajectory observable fluctuates as a sum of martingale increments, each weighted by the marginal consequence of that event for the observable--the intensity-normalized, history-conditioned response. In physical terms, each unanticipated event shifts the observable by exactly the consequence that event carries. The same quantity, multiplied by the event rate, is the response to perturbing that transition. Watching spontaneous fluctuations is watching responses, one event at a time. Response is therefore not merely related to fluctuations through a posterior inequality; it is the coordinate through which the martingale increments of the trajectory generate the fluctuation itself.

From this single representation, the fluctuation-response theory unfolds as a hierarchy. Because martingale increments at different times are uncorrelated, squaring the representation--the martingale isometry--gives exact finite-time and finite-frequency FRRs: the variance and the power spectrum of an observable are the intensity-weighted second moments of its history-conditioned responses. An ordinary susceptibility measurement averages the response over histories, so it retains only the first moment. The exact identities then become inequalities (FRIs). Their slack is the response-heterogeneity gap: the dispersion of the causal consequence of the same transition across the histories on which it occurs. Projecting onto physical perturbation directions yields trajectory Fisher-information bounds, and bounding the logarithmic sensitivity of the rates produces response-kinetic uncertainty relations in which the dynamical activity, the expected number of events, is the resource limiting response precision. For Markov dynamics with local additive observables and history-independent (open-loop) perturbations, trajectories sharing the same present state respond identically. The gap then closes. The exact covariance relation then reduces to their dynamical FRRs \cite{aslyamov2026dynamical}; the correspondence is made explicit in \cref{sec: markov_dfrr}. The hierarchy collapses onto the known Markovian identities and inequalities \cite{dechant2020fluctuation,aslyamov2025nonequilibrium,zheng2025unified,dechant2026finite,stutzer2026stochastic}, which the present theory contains as its memoryless boundary.

The gap is not a universal scalar measure of memory duration or non-Markovianity. It quantifies precisely those distinctions among histories that alter the causal consequence of the perturbed transitions for a chosen observable, perturbation family, and observation timescale, and this task specificity is what makes it operational. Coarse-graining is a generic source of observed memory \cite{esposito2012stochastic,hartich2021emergent}: a state-only Markov surrogate may reproduce occupations, mean fluxes, dynamical activity, and all single-state residence-time distributions while failing to preserve a directional response and its fluctuations. Conditioning the theory on a proposed memory coordinate interpolates between the ordinary FRI and the exact FRR. The residual gap decreases monotonically under refinement of the coordinate. Its vanishing defines response sufficiency: a test of whether histories merged by a reduced description carry identical marginal consequences for the response at hand. No unique reconstruction of the hidden network is required. This is the certificate that occupations, fluxes, and waiting-time statistics cannot supply. Crucially, the normalized response kernel is an intrinsic property of the spontaneous trajectory ensemble. It is the change in the conditional prediction of a future observable produced by appending a jump to a given history, relative to continuing without one. The entire hierarchy can therefore be evaluated from unperturbed trajectories alone, without applying any external perturbation.

We demonstrate this program on experimental data. Analyzing single-channel patch-clamp trajectories of cell-membrane and mitochondrial BK channels--more than half a million resolved dwell events across six membrane voltages--we find that the two channel classes order oppositely under statistical and response-relevant measures of memory. Cell-membrane BK channels carry the stronger conventional dwell-time correlations, yet their present state and the time spent on the state ({\it i.e.,} residence age) nearly suffice to determine the consequence of a gating event. Mitochondrial BK channels show weaker dwell correlations but retain a far larger beyond-age dependence of the future consequence of individual gating events. This dependence persists across voltages and resolved frequencies. Statistical memory and response-relevant memory are therefore distinct and can order oppositely within the same family of molecular systems. Memory is revealed not merely by how long the past persists, but by how the past diversifies the consequences of present events.

\section{Summary of Main Results}
\label{sec:summary_main_results}

The main results form a single chain. First, non-Markovianity may destroy closure at the level of instantaneous state probabilities. It does not destroy the exact stochastic closure of the observed trajectory. Second, the martingale representation theorem shows that all fluctuations generated after the initial condition arise from one space--time orthogonal family of martingale increments. The coefficients of this expansion are event-consequence kernels. Third, a weak perturbation is carried by the same martingale increments. This identifies the event-consequence kernels with intensity-normalized history-conditioned responses.

The consequences follow directly. Full history resolution gives exact fluctuation-response relations. Averaging over histories produces a response-heterogeneity gap. Projection onto a physical control gives Fisher-information bounds. Bounding the rate sensitivity gives activity-controlled response-kinetic uncertainty relations. The same gap also tests whether a proposed memory coordinate preserves a specified response. We apply this test to BK-channel recordings. Statistical memory and response-relevant memory then give opposite orderings of the two channel classes. This section states each result and what it means. \Cref{sec:theory,sec:linear_response,sec:time-domain_frr_bounds,sec:frequency_domain_frr,sec:response_sufficient_coarse_graining} develop the theory and its applications, while \cref{sec: markov_dfrr} makes its relation to the Markovian dynamical fluctuation-response relations explicit.

\begin{figure*}[t]
  \centering
  \includegraphics[width=\textwidth]{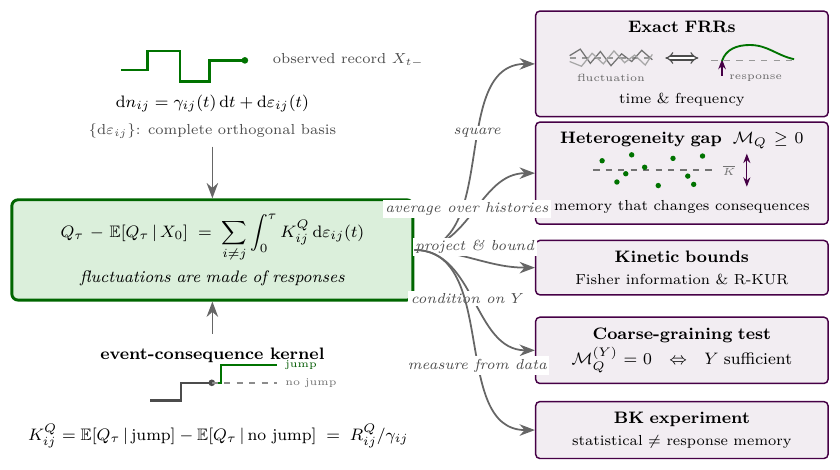}
  \caption{Logical structure of the main results.}
  \label{fig:summary_map}
\end{figure*}

\subsection{Trajectory-Level Closure and the Martingale Basis}
\label{subsec:summary_martingale}

We consider regular, causal, simple, and nonexplosive multichannel jump processes. Their transition intensities are determined by the observed history. For a directed transition $j\to i$, let $n_{ij}(t)$ be the counting process and let
$\gamma_{ij}(t)=r_{ij}^{X_{t-}}\mathds{1}_{x_{t-}=j}$ be its stochastic intensity. The Doob--Meyer decomposition states that every jump increment splits into two parts (\cref{sec:theory}):
\begin{subequations}
\label{eq:summary_doob_meyer}
\begin{align}
  \rmd n_{ij}(t) &= \gamma_{ij}(t)\,\rmd t+\rmd\varepsilon_{ij}(t), \\
  \mathbb{E}\left[ \rmd\varepsilon_{ij}(t) \!\mid\! X_{t-} \right] &= 0, \\
  \mathbb{E}\left[ \rmd\varepsilon_{ij}(t)\rmd\varepsilon_{kl}(t) \!\mid\! X_{t-} \right] &= \delta_{ik}\delta_{jl}\gamma_{ij}(t)\,\rmd t.
\end{align}
\end{subequations}

The first term is the event tendency determined by the realized past. The second is the martingale increment that remains after this tendency is removed. All dynamical memory that controls the next event is contained in $\gamma_{ij}(t)$. Memory may also enter separately through a history-dependent physical readout.

The martingale increments are orthogonal across different transition channels and different times. Their amplitude depends on history. Their orthogonal structure does not. They need not be Gaussian or statistically independent.

The state probabilities obey a gain--loss balance. In general, however, the mean flux $\langle\gamma_{ij}(t)\rangle$ depends on the joint statistics of state and history. The instantaneous state probabilities therefore need not satisfy a closed master equation. The trajectory equation remains closed. Once the history and the intensity functional are specified, every coefficient in \cref{eq:summary_doob_meyer} is known. This is the first central result: state-level closure may fail, while trajectory-level closure survives (\cref{sec:theory}).

The second structural result is completeness. The increments $\{\rmd\varepsilon_{ij}(t)\}$ form a complete orthogonal basis for every fluctuation generated after time zero. This is the martingale representation theorem, stated for the filtration generated by the complete initial information and the resolved jump channels (\cref{sec:theory}). Here $X_0$ denotes the complete initial information, including any prehistory needed to initialize the memory. Uncertainty in $X_0$ gives the only additional source of fluctuation.

The corresponding expansion coefficient has a simple physical meaning. For a fixed pre-event history $X_{t-}$,
\begin{equation}
  K_{ij}^{Q}(\tau,t \!\mid\! X_{t-}) = \mathbb{E}\left[ Q_\tau \,\Big|\, X_t^{ij} \right] - \mathbb{E}\left[ Q_\tau \,\Big|\, X_t^{0} \right].
  \label{eq:summary_event_consequence}
\end{equation}
The history $X_t^{ij}$ contains an appended transition $j\to i$ at time $t$. The history $X_t^{0}$ continues without that transition. Thus, $K_{ij}^{Q}$---the \emph{event-consequence kernel}---is the marginal consequence of one event for the future observable. It contains any immediate contribution of the event. It also contains its complete later propagation through the history-dependent dynamics.

\subsection{History-Conditioned Response as a Stochastic Coordinate}
\label{subsec:summary_response_coordinate}

A weak perturbation changes the same transition intensities. Its infinitesimal score is therefore carried by the same martingale increments,
\begin{equation}
  \rmd\Lambda_\alpha(t) = \sum_{i\neq j}\alpha_{ij}^{X_{t-}}\rmd\varepsilon_{ij}(t),
  \label{eq:summary_score}
\end{equation}
where $\alpha_{ij}^{X_{t-}}$ is the logarithmic rate sensitivity (\cref{sec:linear_response}). Response is consequently a correlation with a martingale score. Causality follows because future score increments are orthogonal to observables already determined by the past.

In a non-Markovian process, the response depends on the pre-perturbation history. The elementary response object is therefore the edge-wise history-conditioned response
$R_{ij}^{Q}(\tau,t \,|\, X_{t-})$. Its exact relation to the event-consequence kernel (\cref{sec:linear_response}) is
\begin{equation}
  R_{ij}^{Q}(\tau,t \!\mid\! X_{t-}) = \gamma_{ij}(t)K_{ij}^{Q}(\tau,t \!\mid\! X_{t-}).
  \label{eq:summary_response_kernel}
\end{equation}
The two factors have different meanings. The intensity $\gamma_{ij}(t)$ describes how available the event is on the realized history. The kernel $K_{ij}^{Q}$ describes what that event does to the future observable.

The same kernel therefore plays three roles. It is the consequence of one event. It is the coefficient of that event's increment in the observable fluctuation. And, multiplied by the intensity, it is the edge response.

The central equation of the theory expresses every fluctuation in response coordinates (\cref{sec:time-domain_frr_bounds}),
\begin{equation}
  Q_\tau-\mathbb{E}\left[ Q_\tau \!\mid\! X_0 \right] = \sum_{i\neq j}\int_0^\tau \frac{R_{ij}^{Q}(\tau,t \!\mid\! X_{t-})}{\gamma_{ij}(t)}\rmd\varepsilon_{ij}(t).
  \label{eq:summary_response_representation}
\end{equation}
Ratios involving $\gamma_{ij}(t)$ are defined on the support of that intensity and are set to zero outside it. Read it from right to left. Each event contributes its increment. Each increment is weighted by the consequence of that event. The sum is the fluctuation. Response therefore does not enter through a later bound. It is the stochastic coordinate of the fluctuation itself.

Any two trajectory observables then obey an exact covariance relation (\cref{sec:time-domain_frr_bounds}),
\begin{subequations}
\label{eq:summary_time_frr}
\begin{align}
  \operatorname{Cov}&(Q_\tau,Q_\tau') = C_0(Q,Q') \\
  &+\int_0^\tau\sum_{i\neq j}\left\langle \frac{R_{ij}^{Q}(\tau,t \!\mid\! X_{t-}) R_{ij}^{Q'}(\tau,t \!\mid\! X_{t-})}{\gamma_{ij}(t)} \right\rangle\rmd t, \nonumber \\
  C_0&(Q,Q') = \operatorname{Cov}\Big(\mathbb{E}\left[ Q_\tau \!\mid\! X_0 \right], \mathbb{E}\left[ Q_\tau' \!\mid\! X_0 \right]\Big).
\end{align}
\end{subequations}
The term $C_0(Q,Q')$ comes from uncertainty already present at the initial time. It vanishes when the complete initial history is fixed. The remaining covariance is generated during the observation interval. It is exactly the overlap of the two sets of history-resolved response coordinates.

For $Q_\tau'=Q_\tau$, the variance is the total history-resolved response power. At full history resolution, there is no inequality. There is also no unexplained residual.

\subsection{Exact Relations, Inequalities, and Kinetic Bounds}
\label{subsec:summary_hierarchy}

The exact FRR resolves the consequence of an event on every active history. An ordinary edge response instead averages over those histories,
\begin{equation}
  R_{ij}^{Q}(\tau,t) = \left\langle\gamma_{ij}(t)K_{ij}^{Q}(\tau,t \!\mid\! X_{t-})\right\rangle.
  \label{eq:summary_ordinary_edge_response}
\end{equation}
The natural reference is the event-weighted mean consequence,
\begin{equation}
  \overline K_{ij}^{Q}(\tau,t) \equiv \frac{R_{ij}^{Q}(\tau,t)}{\left\langle\gamma_{ij}(t)\right\rangle}.
  \label{eq:summary_event_weighted_kernel}
\end{equation}
Histories contribute in proportion to the intensity of the event $j\to i$. Edges with vanishing mean intensity do not contribute.

The exact variance relation separates into two terms,
\begin{subequations}
\label{eq:summary_time_gap}
\begin{align}
  \operatorname{Var}&(Q_\tau) - C_0(Q,Q) = \mathcal V_Q^{\mathrm{av}}+\mathcal M_Q, \\
  \mathcal V_Q^{\mathrm{av}} &= \int_0^\tau\sum_{i\neq j}\frac{\left[R_{ij}^{Q}(\tau,t)\right]^2}{\left\langle\gamma_{ij}(t)\right\rangle}\rmd t, \\
  \mathcal M_Q &= \int_0^\tau\sum_{i\neq j}\left\langle \gamma_{ij}(t)\left[K_{ij}^{Q}(\tau,t \!\mid\! X_{t-}) - \overline K_{ij}^{Q}(\tau,t)\right]^2 \right\rangle\rmd t \nonumber\\
  &\geq0.
\end{align}
\end{subequations}
The term $\mathcal V_Q^{\mathrm{av}}$ is the part reconstructible from ordinary edge responses. The term $\mathcal M_Q$ is the response-heterogeneity gap. It is the intensity-weighted dispersion of the event consequence over the histories merged by ordinary response averaging.

Since $\mathcal M_Q\geq0$, the ordinary responses can only underestimate the variance. This is the edge-wise fluctuation-response inequality. The difference between the exact FRR and the ordinary-response reconstruction therefore has a definite origin. It is not an unspecified loss of tightness.

The gap is task-specific. It depends on the observable, the perturbed edge family, and the observation horizon. It is neither a universal scalar measure of memory duration nor a universal measure of non-Markovianity. It records only those distinctions among histories that change the consequence of the selected events.

For Markov jump dynamics with local readouts and history-independent open-loop edge perturbations, the present state and time determine the event consequence. The additional full-history gap then vanishes. The converse does not hold. A zero gap for one observable and one perturbation family does not imply that the complete process is Markovian. A history-dependent readout can also produce a positive gap even when the underlying dynamics is Markovian.

Response precision is also bounded. For a physical perturbation with direction $\alpha_{ij}^{X_{t-}}$, temporal protocol $h(t)$, and bounded sensitivity $|\alpha_{ij}^{X_{t-}}|\leq\alpha_{\max}$ (\cref{sec:time-domain_frr_bounds}),
\begin{equation}
  \frac{\left|\left.\partial_\epsilon\left\langle Q_\tau\right\rangle_{\epsilon h}\right|_{\epsilon=0}\right|^2}{\operatorname{Var}(Q_\tau)-C_0(Q,Q)} \leq \mathcal I_{\alpha,h} \leq \alpha_{\max}^2\mathcal A_{h,\tau}.
  \label{eq:summary_time_rkur}
\end{equation}
Here $\mathcal I_{\alpha,h}$ is the path-space Fisher information of the perturbation, and $\mathcal A_{h,\tau}$ is the protocol-weighted dynamical activity. The first inequality measures the mismatch between the observable fluctuation and the chosen physical score direction. The second replaces the actual sensitivities by the envelope $\alpha_{\max}$. This produces a separate sensitivity-bound loss. Neither is the history-averaging gap $\mathcal M_Q$.

Activity appears because it measures the martingale variance available to carry the trajectory score. No global rescaling of time is required. For a constant protocol, $\mathcal A_{h,\tau}$ becomes the expected number of resolved events during the observation interval.

The same structure persists at finite frequency (\cref{sec:frequency_domain_frr}). The frequency domain adds a new feature: phase. Consider a stationary regime with finite spectra and no persistent random component fixed by the remote past. Each history then carries a complex response spectrum $\mathcal R_{ij}^{Q}(\omega \!\mid\! X_{0-})$. Here $X_{0-}$ is the stationary history preceding the perturbation. The magnitude of the spectrum gives the response gain. Its phase gives the temporal lag.

The power spectrum obeys
\begin{subequations}
\label{eq:summary_frequency_frr}
\begin{align}
  \mathcal S_Q(\omega) &= \sum_{i\neq j}\left\langle \frac{\left|\mathcal R_{ij}^{Q}(\omega \!\mid\! X_{0-})\right|^2}{\gamma_{ij}(0)} \right\rangle_{\mathrm{ss}}, \\
  \mathcal S_Q(\omega) &= \sum_{i\neq j}\frac{\left|\mathcal R_{ij}^{Q}(\omega)\right|^2}{\left\langle\gamma_{ij}(0)\right\rangle_{\mathrm{ss}}}+\mathcal M_Q(\omega), \\
  \mathcal M_Q(\omega) &\geq0.
\end{align}
\end{subequations}
The ordinary susceptibility is a coherent complex average. It can be small for two different reasons. The conditioned responses may all be weak. Alternatively, strong conditioned responses may cancel because their phases differ.

The spectral gap $\mathcal M_Q(\omega)$ measures the intensity-weighted dispersion of the complex event consequence. It therefore contains both gain heterogeneity and phase-lag heterogeneity. Equal-time jump responses are included in the response spectrum and generate the corresponding shot-noise contribution.

Without time-translation invariance, frequency is no longer diagonal. An input at frequency $\nu$ may generate an output at frequency $\omega$. For a fixed unit-modulus complex harmonic convention,
\begin{equation}
  \frac{\left|\mathcal R_{\alpha}^{Q,\tau}(\omega,\nu)\right|^2}{\mathcal S_Q^\tau(\omega,\omega)-\mathcal S_{0,Q}^\tau(\omega,\omega)} \leq \mathcal I_\alpha^\tau \leq \alpha_{\max}^2\mathcal A_\tau.
  \label{eq:summary_two_frequency}
\end{equation}
The initial spectrum $\mathcal S_{0,Q}^\tau$ contains fluctuations inherited from the initial state and initial memory. It does not come from martingale increments generated during the protocol.

A finite observation window broadens even a stationary diagonal response. Genuine frequency conversion is the additional input--output mixing beyond this window-induced structure. The physical real protocol is obtained from the real part of the complex representation. Its Fisher information is determined by the squared real protocol.

The time-domain, stationary, and two-frequency relations are not separate theories. They are different resolutions of the same response representation.

\subsection{Response-Sufficient Coarse-Graining}
\label{subsec:summary_coarse_graining}

The response-heterogeneity gap can test a proposed memory coordinate. Let
$Y_{t-}=\Phi_t(X_{t-})$ and require
\begin{equation}
  \sigma(x_{t-})\subseteq\sigma(Y_{t-})\subseteq\mathcal F_{t-}.
  \label{eq:summary_candidate_information_order}
\end{equation}
Here $\sigma(Y_{t-})$ is the information retained by the candidate state.

For any such candidate, the variance splits again (\cref{sec:response_sufficient_coarse_graining}),
\begin{subequations}
\label{eq:summary_coarse_graining}
\begin{align}
  \operatorname{Var}(Q_\tau) &- C_0(Q,Q) = \mathcal V_Q^{(Y)}+\mathcal M_Q^{(Y)}, \\
  \mathcal V_Q^{(Y)} ={}& \int_0^\tau\sum_{i\neq j}\left\langle \gamma_{ij}(t \!\mid\! Y_{t-}) \left[K_{ij}^{Q}(Y_{t-})\right]^2 \right\rangle\rmd t, \\
  \mathcal M_Q^{(Y)} ={}& \int_0^\tau\sum_{i\neq j}\left\langle \gamma_{ij}(t)\left[K_{ij}^{Q}(X_{t-}) - K_{ij}^{Q}(Y_{t-})\right]^2 \right\rangle\rmd t \nonumber\\
  \geq{}& 0, \\
  \sigma(Y_{t-}) & \subseteq \sigma(Y'_{t-}) \Longrightarrow \mathcal M_Q^{(Y')} \leq \mathcal M_Q^{(Y)}.
\end{align}
\end{subequations}
Here $\gamma_{ij}(t \,|\, Y_{t-})$ and $K_{ij}^{Q}(Y_{t-})$ are the intensity and the kernel conditioned on the retained information. The common arguments $(\tau,t)$ of the kernels are suppressed. Ratios are set to zero outside the support of the corresponding conditioned intensity.

Adding information cannot increase the residual. We call $Y_{t-}$ response-sufficient when $\mathcal M_Q^{(Y)}=0$ for the chosen observable and resolved edge-perturbation family. If the physical logarithmic sensitivity $\alpha_{ij}^{X_{t-}}$ is history dependent, the candidate state must also determine the relevant sensitivity.

Response sufficiency is deliberately task specific. It does not require $Y_{t-}$ to predict the full future process. It does not require a unique hidden representation. It asks a narrower question: do histories merged by $Y_{t-}$ assign the same consequence to the events relevant for the chosen response?

We illustrate this point with a microscopic Markov network. Two hidden states, $A_+$ and $A_-$, are observed as one state $A$. Their total escape rates are equal. The coarse-grained residence time in $A$ is therefore exactly exponential. A state-only Markov surrogate can match all single-state residence-time distributions. It also matches stationary occupations, mean fluxes, and total activity.

Directional information nevertheless survives in the hidden polarization,
\begin{subequations}
\label{eq:summary_polarization}
\begin{align}
  m_t &= \sigma_{\mathrm{pre}}e^{-2ua_t}, \\
  r_{BA}^{X_{t-}} &= \frac{\Gamma}{2}+\frac{\delta}{2}m_t, \\
  r_{CA}^{X_{t-}} &= \frac{\Gamma}{2}-\frac{\delta}{2}m_t.
\end{align}
\end{subequations}
Here $a_t$ is the residence age. The sign $\sigma_{\mathrm{pre}}$ records the state from which $A$ was entered. We also define $\Gamma=b+c$ and $\delta=b-c$. The sum of the two escape rates is history independent. Their difference is not.

We compare two observables,
\begin{subequations}
\label{eq:summary_activity_directional_observables}
\begin{align}
  Q_\tau^{\mathrm{act}} &= n_{BA}(\tau) + n_{CA}(\tau), \\ Q_\tau^{\mathrm{dir}} &= n_{BA}(\tau) - n_{CA}(\tau).
\end{align}
\end{subequations}
The present observed state is response-sufficient for $Q_\tau^{\mathrm{act}}$. It is not sufficient for $Q_\tau^{\mathrm{dir}}$. Residence age preserves the magnitude of the polarization but not its sign. The previous state preserves the sign but not its decay. Both pieces are needed.

Retaining $(x_{t-},m_t)$ restores the microscopic directional response and the exact variance reconstruction. In terms of
$\eta_Q^{(Y)}\equiv\mathcal M_Q^{(Y)}/[\operatorname{Var}(Q_\tau)-C_0(Q,Q)]$,
\begin{subequations}
\label{eq:summary_coarse_grained_result}
\begin{align}
  \eta_{Q_{\mathrm{act}}}^{(x)} = 0, &\qquad \eta_{Q_{\mathrm{dir}}}^{(x)} > 0, \\
  \eta_{Q_{\mathrm{dir}}}^{(Y^\star)} = 0, &\qquad Y^\star\simeq(x_{t-},m_t).
\end{align}
\end{subequations}
The lesson is direct. Matching occupations, fluxes, activities, and waiting-time distributions does not guarantee response equivalence. The required memory depends on the observable that must be preserved.

\subsection{Statistical versus Response-Relevant Memory in BK Channels}
\label{subsec:summary_bk}

The event-consequence kernel is defined by conditional expectations under the unperturbed trajectories. At a fixed pre-event history, it compares a transition at the present instant with continuation without a transition at that instant; subsequent evolution follows the same unperturbed dynamics on both branches. This construction motivates finite-history observational diagnostics that can be estimated from spontaneous recordings without an externally imposed perturbation (\cref{sec:response_sufficient_coarse_graining}).

We apply this construction to single-channel recordings of cell-membrane and mitochondrial BK channels. The data contain more than half a million resolved dwells across six membrane voltages. We compare two measures of memory. The first is the conventional correlation between successive dwell durations \cite{colquhoun1987note,magleby1992dependency}. The second is the residual fraction $\widehat\eta_0$ left by the state-and-age candidate $Y=(x_{t-},a_t)$.

Both quantities are reported relative to order-shuffled controls. The shuffling preserves the marginal open- and closed-dwell distributions, the open fraction, and the total recording time. It removes serial order. The two measures give opposite orderings,
\begin{subequations}
\label{eq:summary_bk_dissociation}
\begin{align}
  \operatorname{median} C_{\mathrm{dwell}}^{\mathrm{ex}}:
  \quad 0.2077 \; \text{(cell-BK)} > 0.1290 \; \text{(mitoBK)}, \nonumber \\
  \operatorname{median}\widehat\eta_0^{\mathrm{ex}}:
  \quad 0.0373 \; \text{(cell-BK)} < 0.1954 \; \text{(mitoBK)}. \nonumber
\end{align}
\end{subequations}
Cell-membrane channels have stronger conventional dwell correlations but a smaller and heterogeneous excess dependence of the predictive event consequence on preceding dwells. Mitochondrial channels have weaker dwell correlations and a substantially larger beyond-age excess residual. Adding preceding dwell durations reduces the mitochondrial residual,
\begin{equation}
  \widehat\eta_0^{\mathrm{ex}} = 0.1954 \; \longrightarrow \; \widehat\eta_1^{\mathrm{ex}} = 0.1529 \; \longrightarrow \; \widehat\eta_2^{\mathrm{ex}} = 0.1038.
  \label{eq:summary_bk_history_hierarchy}
\end{equation}
The residual decreases, but it does not close. Within the finite-window analysis, the excess frequency-resolved residual remains positive throughout the resolved band. It is largest at low frequency.

These are finite-history observational estimates. They are not direct measurements of the full-filtration gap. Their message is nevertheless clear. Statistical memory and response-relevant memory are distinct. They can even order oppositely within the same molecular family.

Memory is revealed not only by how long the past persists. It is also revealed by how the past changes the consequences of present events.

\section{Trajectory-level closure and martingale representation}
\label{sec:theory}

In this section, we establish two trajectory-level structures that survive beyond Markovianity. The first is a closed stochastic equation for the observed trajectory. Although history dependence generally prevents the instantaneous state probabilities from obeying a closed master equation, every observed counting process still admits an exact stochastic equation once its history-dependent intensity is specified. The second is a complete representation of trajectory fluctuations. The martingale increments associated with different transition channels and different times are mutually orthogonal, and the martingale representation theorem shows that they form a complete stochastic basis for fluctuations generated during the observation interval. The corresponding expansion coefficient is a history-conditioned causal kernel with a direct physical meaning: it quantifies how an individual transition changes the predicted future of an observable. Technical details are provided in \cref{app:trajectory_martingale_structure}.

\subsection{Non-Markovian Jump Processes}
\label{subsec:NMJP}

\begin{figure}[t]
  \centering
  \includegraphics[width=0.99\linewidth]{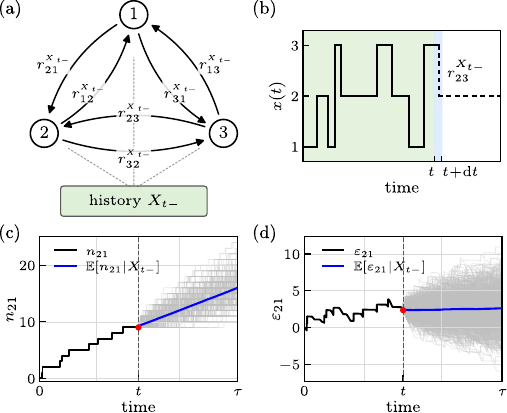}
  \caption{\textbf{Non-Markovian jump trajectories and martingale structure.}
  (a) Schematic illustration of a three-state non-Markovian jump process. The transition rate $r_{ij}^{X_{t-}}$ depends on the past trajectory $X_{t-}$. In the illustrative model, the memory is encoded by $m_j(t) = \int_0^t e^{-(t-s)/\tau_m} \mathds{1}_{x(s)=j} \,\rmd s$, and the rate is chosen as $r_{ij}^{X_{t-}} = k_{ij} \exp[\beta m_j(t)/\tau_m]$, with $\beta$ controlling the memory strength. (b) A representative trajectory. The green region indicates the history $X_{t-}$, the blue region indicates the infinitesimal interval $[t,t+\rmd t)$ in which a jump occurs, and the dashed line indicates the future after time $t$. (c) Several realizations of the counting process $n_{21}$ continued from the same history, together with their conditional ensemble mean. (d) The corresponding compensated processes $\varepsilon_{21}$. Individual realizations fluctuate, whereas their conditional ensemble mean remains equal to its value at the conditioning time, illustrating the martingale property.}
  \label{fig:non-Markovian_and_martingale}
\end{figure}

We consider a jump process on a finite discrete state space $\{1, 2, \cdots, N\}$, as schematically shown in \cref{fig:non-Markovian_and_martingale} (a), (b). The state of the system at time $t$ is denoted by $x_t \equiv x(t) \in \{1, 2, \cdots, N\}$. A stochastic trajectory over the time interval $[0,\tau]$ is the piecewise-constant path
\begin{equation}
  X_\tau = \{ x_t : 0 \leq t \leq \tau \}.
  \label{eq: trajectory_definition}
\end{equation}
The probability density of observing the trajectory $X_\tau$ is denoted by $\mathcal{P}[X_\tau]$.

Physically, conditioning on $X_{t-}$ below means conditioning on everything an observer who has watched the trajectory up to, but not including, time $t$ knows. Formally, we work on a filtered probability space $(\Omega,\mathcal{F},(\mathcal{F}_{t}^{X})_{t \geq 0},\mathbb{P})$, where $\mathcal{F}_{t}^{X}$ is the natural filtration generated by the observed trajectory \cite{bremaud1981point,jacod2003limit,daley2008introduction}. The symbol $X_{t-}$ denotes the realized history immediately before time $t$, whereas $\mathcal{F}_{t-}^{X}$ denotes all information contained in that history. With a slight abuse of notation, we write conditional expectations as $\mathbb{E}[\,\cdot\, \,|\, X_{t-}\,]$ instead of the mathematically concrete notation $\mathbb{E}[\,\cdot\, \,|\, \mathcal{F}^X_{t-}\,]$. Because the transition intensities may depend on events preceding the observation window, the initial information may contain a fixed or random prehistory $X_{0-}$. Equivalently, all memory variables may be initialized at $t=0$ according to a prescribed initial-history distribution. The uncertainty carried by this initial condition will later generate an initial contribution to trajectory fluctuations. The separation of a trajectory into the realized pre-jump history $X_{t-}$, the infinitesimal interval at time $t$, and the subsequent future is illustrated in \cref{fig:non-Markovian_and_martingale} (b).

For each directed transition from state $j$ to state $i$, we introduce a history-dependent transition rate $r_{ij}^{X_{t-}}$. For simplicity, each ordered state pair is assumed to correspond to one resolved transition channel. Multiple physical channels connecting the same pair of states can be included by adding a channel label. The rate is a functional of the trajectory already realized and is therefore a stochastic quantity. It determines the conditional probability of a transition during the infinitesimal time interval $[t,t+\rmd t)$,
\begin{equation}
  P\left[\, x_{t+\rmd t}=i \!\mid\! x_{t-}=j,X_{t-}\, \right] = r_{ij}^{X_{t-}} \,\rmd t + o(\rmd t).
  \label{eq: conditional_jump_probability}
\end{equation}
The probability of two or more jumps during the same infinitesimal interval is assumed to be $o(\rmd t)$. As illustrated in \cref{fig:non-Markovian_and_martingale} (a), each directed edge carries a transition rate $r_{ij}^{X_{t-}}$ determined by the same realized pre-jump history.

The notation $r_{ij}^{X_{t-}}$ emphasizes that the local jump tendency is determined by the observed past rather than only by the present state. We consider regular, causal, and nonexplosive dynamics. Thus, infinitely many jumps cannot accumulate within a finite interval. More than one jump cannot occur with non-negligible probability during an infinitesimal interval. The rate at time $t$ depends only on $X_{t-}$, not on future information. Singular events that occur with unit probability at a prescribed deterministic time are not included in the present rate description.

Many familiar non-Markovian models admit this history-dependent intensity representation. In a semi-Markov process, the rate may depend on the residence age $a_t$ since the most recent transition,
\begin{equation}
  r_{ij}^{X_{t-}} \propto h_j(a_t),
  \label{eq: semimarkov_example}
\end{equation}
where $h_j(a)$ is the age-dependent escape hazard from state $j$. In neural firing dynamics, the activation rate may depend on the previous number of activation events,
\begin{equation}
  r_{ij}^{X_{t-}} = f_{ij}\big( n_{\mathrm{act}}(t-) \big),
  \label{eq: activation_count_example}
\end{equation}
where $n_{\mathrm{act}}(t-)$ counts activation events in the trajectory $X_{t-}$ and $f_{ij}$ is a nonnegative function. In self-interacting processes, the rate may depend on empirical occupations and currents,
\begin{subequations}
\label{eq: self_interacting_example}
\begin{align}
  \mu_k(t) &= \frac{1}{t} \int_0^t \mathds{1}_{x_{s-}=k} \,\rmd s, \\
  J_{ij}(t) &= \frac{1}{t} \big[ n_{ij}(t) - n_{ji}(t) \big], \\
  r_{ij}^{X_{t-}} &= f_{ij}\big( \bm{\mu}(t),\bm{J}(t) \big).
\end{align}
\end{subequations}
These examples share one structure. Non-markovianity enters through the dependence of the instantaneous jump tendency on the realized past. The stochastic dynamics itself is formulated directly on the observed trajectory.

\subsection{Closed Trajectory Stochastic Equations}
\label{subsec:doob_meyer}

For each directed transition from state $j$ to state $i$, let $n_{ij}(t)$ denote the total number of such jumps up to time $t$. Its increment $\rmd n_{ij}(t)$ equals one when a transition $j\to i$ occurs during $[t,t+\rmd t)$ and is otherwise zero. From \cref{eq: conditional_jump_probability},
\begin{equation}
  \mathbb{E}\left[\, \rmd n_{ij}(t) \!\mid\! X_{t-}\, \right] = r_{ij}^{X_{t-}} \mathds{1}_{x_{t-}=j} \,\rmd t.
  \label{eq: conditional_mean_counting_increment}
\end{equation}
Several realizations of the counting process continued from the same history, together with their conditional ensemble mean, are illustrated in \cref{fig:non-Markovian_and_martingale} (c). Although the individual counting trajectories fluctuate, their conditional mean growth is determined by the history-dependent event tendency in \cref{eq: conditional_mean_counting_increment}.

Since $n_{ij}(t)$ is an increasing adapted process, the Doob--Meyer decomposition \cite{bremaud1981point,jacod2003limit} separates it uniquely into a history-determined compensator and a martingale. The compensator is
\begin{equation}
  \int_0^t r_{ij}^{X_{s-}} \mathds{1}_{x_{s-}=j} \,\rmd s,
  \label{eq: compensator_nonmarkov}
\end{equation}
and the compensated counting process
\begin{equation}
  \varepsilon_{ij}(t) = n_{ij}(t) - \int_0^t r_{ij}^{X_{s-}} \mathds{1}_{x_{s-}=j} \,\rmd s
  \label{eq: compensated_counting_martingale}
\end{equation}
is a martingale. As illustrated in \cref{fig:non-Markovian_and_martingale} (d), the compensated trajectories continue to fluctuate after the conditioning time, while their conditional ensemble mean remains fixed at its value determined by the common prehistory. Equivalently,
\begin{equation}
  \underbrace{\rmd n_{ij}(t)}_{\substack{\text{jump}\\\text{increment}}} = \, \underbrace{r_{ij}^{X_{t-}} \mathds{1}_{x_{t-}=j} \,\rmd t}_{\substack{\text{history-dependent}\\\text{tendency}}} \,\, + \underbrace{\rmd\varepsilon_{ij}(t)}_{\substack{\text{martingale}\\\text{increment}}}.
  \label{eq: nonmarkov_jump_sde}
\end{equation}
We define the stochastic intensity
\begin{equation}
  \gamma_{ij}(t) \equiv r_{ij}^{X_{t-}} \mathds{1}_{x_{t-}=j},
  \label{eq: stochastic_intensity_gamma}
\end{equation}
so that
\begin{equation}
  \rmd n_{ij}(t) = \gamma_{ij}(t) \,\rmd t + \rmd\varepsilon_{ij}(t).
  \label{eq: compact_count_sde}
\end{equation}

\Cref{eq: compact_count_sde} is an exact stochastic equation of motion for the observed non-Markovian trajectory. Once the intensity functional and the realized history are specified, every coefficient on the right-hand side is determined by information already available at time $t$. The dynamics therefore closes exactly on the trajectory history even when it does not close on the instantaneous state probabilities. This trajectory-level closure is the first central structure of the present theory.

In the Markovian setting, the same stochastic equation has recently been used to construct a stochastic calculus for pathwise observables of Markov jump processes \cite{stutzer2026stochastic}. In the Markovian limit, $r_{ij}^{X_{t-}} \to r_{ij}(t,x_{t-})$, the intensity is determined by the present state and time. Markovian and non-Markovian jump dynamics therefore do not differ in the form of the martingale increment. They differ in the information required to determine the event tendency: the complete observed history is generally required in the non-Markovian case.

The martingale increments--the \emph{innovations}, in the language of filtering theory \cite{bremaud1981point}--have simple conditional statistics. By construction,
\begin{equation}
  \mathbb{E}\left[\, \rmd\varepsilon_{ij}(t) \!\mid\! X_{s-}\, \right] = 0, \quad t\geq s,
  \label{eq: martingale_noise_mean}
\end{equation}
and their conditional equal-time correlations are
\begin{equation}
  \mathbb{E}\left[\, \rmd\varepsilon_{ij}(t)\rmd\varepsilon_{kl}(t) \!\mid\! X_{t-}\, \right] = \delta_{ik}\delta_{jl}\gamma_{ij}(t) \,\rmd t.
  \label{eq: martingale_noise_equal_time}
\end{equation}
After averaging over histories,
\begin{equation}
  \left\langle \rmd\varepsilon_{ij}(t)\rmd\varepsilon_{kl}(t') \right\rangle = \delta_{ik}\delta_{jl}\delta(t-t')\left\langle\gamma_{ij}(t)\right\rangle \rmd t\rmd t'.
  \label{eq: martingale_noise_autocorrelation}
\end{equation}
Thus, martingale increments belonging to different directed edges or different times are orthogonal in the second-moment sense. Their amplitude is history dependent through $\gamma_{ij}(t)$, but their space--time orthogonality remains as simple as in a memoryless process. This statement is solely the martingale property and does not require the increments to be Gaussian or statistically independent.

The same stochastic equation determines the change of the state indicator. Along each trajectory,
\begin{equation}
  \rmd\mathds{1}_{x_t=i} = \sum_{j\neq i}\rmd n_{ij}(t) - \sum_{j\neq i}\rmd n_{ji}(t).
  \label{eq: state_indicator_increment}
\end{equation}
Substituting \cref{eq: nonmarkov_jump_sde} gives
\begin{align}
  \rmd\mathds{1}_{x_t=i} ={}& \left[ \sum_{j\neq i} r_{ij}^{X_{t-}}\mathds{1}_{x_{t-}=j} - \sum_{j\neq i} r_{ji}^{X_{t-}}\mathds{1}_{x_{t-}=i} \right]\rmd t \nonumber\\
  &+ \sum_{j\neq i}\rmd\varepsilon_{ij}(t) - \sum_{j\neq i}\rmd\varepsilon_{ji}(t).
  \label{eq: state_indicator_sde}
\end{align}
Taking the ensemble average yields
\begin{equation}
  \frac{\rmd p_i(t)}{\rmd t} = \sum_{j\neq i} \Big[ \left\langle\gamma_{ij}(t)\right\rangle - \left\langle\gamma_{ji}(t)\right\rangle \Big],
  \label{eq: nonmarkov_probability_evolution}
\end{equation}
where $p_i(t)=\left\langle\mathds{1}_{x_t=i}\right\rangle$. \Cref{eq: nonmarkov_probability_evolution} retains the usual gain--loss form, but it is generally not a closed equation for $\{p_i(t)\}$. The mean flux $\left\langle\gamma_{ij}(t)\right\rangle$ depends on the joint statistics of the present state and the realized history. In the Markovian limit, $\left\langle\gamma_{ij}(t)\right\rangle=r_{ij}(t)p_j(t)$ and the ordinary master equation is recovered.

The contrast between \cref{eq: compact_count_sde,eq: nonmarkov_probability_evolution} is fundamental. Non-Markovianity can destroy finite-dimensional closure at the level of instantaneous state probabilities, but it does not destroy the exact stochastic closure of the trajectory. The closed object is no longer the probability vector alone; it is the observed history together with its history-dependent intensity.

\subsection{Martingale Representation Theorem}
\label{subsec:martingale_representation}

We now ask how the fluctuations of an arbitrary trajectory observable are generated by the martingale increments introduced above. Let $Q_\tau$ be any square-integrable observable determined by the observed trajectory up to time $\tau$, and define its history-conditioned prediction
\begin{equation}
  M_t^Q \equiv \mathbb{E}\left[\, Q_\tau \!\mid\! X_t\, \right].
  \label{eq: observable_doob_martingale}
\end{equation}
The quantity $M_t^Q$, the Doob martingale of $Q_\tau$, is the prediction of the final observable using all trajectory information available at time $t$. As time advances, each observed event, and each interval during which no event occurs, updates this prediction. By construction, $M_t^Q$ is a martingale. For completeness, we verify the martingale property of $M_t^Q$ in \cref{app:doob_martingale_property}.

The conditions are simple. The filtration is generated jointly by the initial condition and the resolved jump channels, so the observed events are the only source of randomness entering after time zero. The dynamics is nonexplosive with integrable intensities. Under these conditions, the classical martingale representation theorem for multivariate point processes \cite{davis1976representation,bremaud1981point,jacod2003limit,daley2008introduction} gives
\begin{equation}
  \rmd M_t^Q = \sum_{i\neq j} K_{ij}^Q(\tau,t \!\mid\! X_{t-}) \,\rmd\varepsilon_{ij}(t).
  \label{eq: differential_martingale_representation}
\end{equation}
Here $K_{ij}^Q(\tau,t \!\mid\! X_{t-})$ is the history-dependent expansion coefficient associated with the martingale increment on the transition $j\to i$ at time $t$; it is predictable, and it is unique on the support of the corresponding intensity (\cref{app:martingale_representation_proof}). Integrating over the observation interval and using $M_\tau^Q=Q_\tau$ gives
\begin{equation}
  Q_\tau-\mathbb{E}\left[\, Q_\tau \!\mid\! X_0\, \right] = \sum_{i\neq j}\int_0^\tau K_{ij}^Q(\tau,t \!\mid\! X_{t-}) \,\rmd\varepsilon_{ij}(t).
  \label{eq: martingale_representation}
\end{equation}

\Cref{eq: martingale_noise_autocorrelation,eq: martingale_representation} have complementary meanings. The space--time orthogonality of the martingale increments makes the family $\{\rmd\varepsilon_{ij}(t)\}$ a natural orthogonal stochastic basis indexed by transition channel and time. The martingale representation theorem establishes that this basis is complete for trajectory fluctuations generated after the initial condition. When the complete initial history is fixed, $\mathbb{E}[\,Q_\tau \,|\, X_0\,]=\mathbb{E}[Q_\tau]$, and every fluctuation of $Q_\tau$ around its mean is generated by the martingale increments in \cref{eq: martingale_representation}. If the initial history is random, the only additional contribution is the fluctuation already contained in the initial conditional mean $\mathbb{E}[\,Q_\tau \,|\, X_0\,]$. Memory changes the stochastic intensities and the coefficients multiplying the martingale increments, but it does not introduce an additional source of trajectory randomness.

The coefficient $K_{ij}^Q(\tau,t \!\mid\! X_{t-})$ has a direct causal interpretation. Consider a fixed pre-event history $X_{t-}$. Let $X_t^{ij}$ denote the history obtained by appending a transition $j\to i$ during the infinitesimal interval at time $t$, and let $X_t^0$ denote the corresponding continuation in which no transition is appended. Then
\begin{equation}
  K_{ij}^Q(\tau,t \!\mid\! X_{t-}) = \mathbb{E}\left[ Q_\tau \Big| X_t^{ij} \right] - \mathbb{E}\left[ Q_\tau \Big| X_t^0 \right].
  \label{eq: event_consequence_kernel}
\end{equation}
captures the infinitesimal event--continuation comparison as detailed in \cref{app:event_consequence_kernel}. Both conditional expectations refer to the unperturbed trajectories. The event/no-jump distinction applies only to the infinitesimal interval at the comparison time; it neither constrains subsequent jumps nor changes the dynamics over the remaining observation window. Thus, $K_{ij}^Q$ is the marginal consequence of one transition for the future observable; we call it the \emph{event-consequence kernel}. It contains both an immediate contribution, when $Q_\tau$ directly counts or weights that transition, and the complete subsequent propagation of the event through the history-dependent dynamics.

Combining the martingale representation with the space--time orthogonality of the martingale increments gives the local identity
\begin{equation}
  \mathbb{E}\left[\, Q_\tau\rmd\varepsilon_{ij}(t) \!\mid\! X_{t-}\, \right] = \gamma_{ij}(t) \, K_{ij}^Q(\tau,t \!\mid\! X_{t-}) \,\rmd t.
  \label{eq: kernel_martingale_correlation}
\end{equation}
The correlation of a future observable with a martingale increment equals the event rate times the future consequence of that event. The relation between this causal kernel and the response to a perturbation of the corresponding transition rate will be established in the next section.

\subsection{Trajectory Observables}
\label{subsec:trajectory_observables}

We now specialize this general representation to additive trajectory observables, which constitute the principal class used throughout the remainder of this work. The first elementary object is the occupation-time increment
\begin{equation}
  \rmd\tau_i(t) = \mathds{1}_{x_t=i} \,\rmd t,
  \label{eq: waiting_increment}
\end{equation}
which records the time spent in state $i$ during $[t,t+\rmd t)$. The second is the jump increment $\rmd n_{ij}(t)$, which records whether a transition from $j$ to $i$ occurs during the same interval. These two objects form the natural building blocks of additive observables\footnote{Here ``additive'' means that the observable is accumulated along the trajectory as a time integral of state-dependent contributions and a sum of jump-dependent contributions.}.

A general additive trajectory observable can be written as
\begin{equation}
  Q_\tau[X_\tau] = \int_{t=0}^{t=\tau} \sum_i g_i^{X_{t-}}\rmd\tau_i(t) + \int_0^\tau \sum_{i\neq j} c_{ij}^{X_{t-}}\rmd n_{ij}(t).
  \label{eq: general_additive_observable}
\end{equation}
Here $g_i^{X_{t-}}$ and $c_{ij}^{X_{t-}}$ are weights determined by the history available immediately before time $t$.

History dependence may enter the dynamics and the observable in two logically distinct ways. Dynamical memory means that the transition intensity $\gamma_{ij}(t)$ depends on $X_{t-}$. Readout memory means that the physical signal generated by an occupation or event depends on the preceding trajectory through $g_i^{X_{t-}}$ or $c_{ij}^{X_{t-}}$. The latter may occur even when the underlying jump dynamics is Markovian. Standard local additive observables use only state-, edge-, and time-dependent weights, whereas \cref{eq: general_additive_observable} also allows the measured output itself to carry adaptation, depletion, or desensitization.

\begin{figure}
  \centering
  \includegraphics[width=0.80\linewidth]{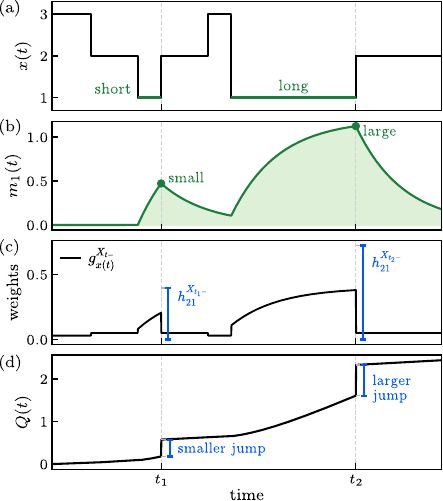}
  \caption{\textbf{History-dependent additive observable.}
  (a) A representative trajectory of a three-state non-Markovian jump process. The green segments indicate residence in state $1$. (b) The corresponding memory variable $m_1(t) = \int_0^t e^{-(t-s)/\tau_m}\mathds{1}_{x(s)=1} \,\rmd s$. (c) History-dependent weights used to construct the additive observable. The waiting-time weight is $g_1^{X_{t-}} = g_1^0+\eta_gm_1(t)/\tau_m$ in state $1$, while $g_2$ and $g_3$ are constants. The jump weight for the edge $1\to2$ is $c_{21}^{X_{t-}} = c_{21}^0+\eta_cm_1(t)/\tau_m$. The blue brackets indicate the two values of $c_{21}^{X_{t-}}$ at the highlighted jumps. (d) The accumulated additive observable $Q(t) = \int_0^t \sum_i g_i^{X_{s-}}\mathds{1}_{x(s-)=i} \,\rmd s+\int_0^t c_{21}^{X_{s-}}\rmd n_{21}(s)$. The two $1\to2$ jumps have different increments because their pre-jump histories, and hence their weights $c_{21}^{X_{t-}}$, are different.}
  \label{fig:additive_observable}
\end{figure}

For example, in short-term synaptic plasticity \cite{markram1996redistribution}, the output induced by a presynaptic spike depends on the recent spike history. If
\begin{equation}
  m_t = \int_{[0,t)} \phi(t-s) \,\rmd n_{\mathrm{sp}}(s)
  \label{eq: synaptic_memory_variable}
\end{equation}
encodes previous spikes, an integrated postsynaptic output can be written as
\begin{equation}
  Q_\tau^{\mathrm{syn}} = \int_0^\tau A(m_{t-}) \,\rmd n_{\mathrm{sp}}(t),
  \label{eq: synaptic_history_dependent_output}
\end{equation}
where $A(m_{t-})$ is the spike-triggered output amplitude after the realized stimulation history. The history-dependent coefficient is therefore part of the physical readout rather than an artificial mathematical probe. A representative construction is illustrated in \cref{fig:additive_observable} (a)--(d), where the trajectory generates a memory variable that modulates the occupation and jump weights and thereby determines the accumulated observable.

For additive observables of the form in \cref{eq: general_additive_observable}, the general identity \cref{eq: kernel_martingale_correlation} can be evaluated through the elementary occupation and jump increments. Choosing $Q_\tau=\mathds{1}_{x_t=k}$ defines the history-conditioned state-influence kernel
\begin{equation}
  I_{ij}^{k}(t,s \!\mid\! X_{s-}) = \mathbb{E}\left[\, \mathds{1}_{x_t=k} \!\mid\! X_s^{ij}\, \right] - \mathbb{E}\left[\, \mathds{1}_{x_t=k} \!\mid\! X_s^0\, \right], \quad t>s.
  \label{eq: state_influence_kernel}
\end{equation}
The corresponding occupation--martingale correlation is
\begin{equation}
  \left\langle \rmd\tau_k(t)\rmd\varepsilon_{ij}(s) \right\rangle = \mathds{1}_{t>s}\left\langle \gamma_{ij}(s) \, I_{ij}^{k}(t,s \!\mid\! X_{s-}) \right\rangle \rmd t\rmd s.
  \label{eq: occupation_martingale_correlation}
\end{equation}
This formula is the non-Markovian analogue of the occupation--martingale correlation in Markov jump processes \cite{stutzer2026stochastic}. The causal indicator $\mathds{1}_{t>s}$ follows from the martingale property rather than from Markovianity. In the Markovian limit,
\begin{align}
  \left\langle \rmd\tau_k(t)\rmd\varepsilon_{ij}(s) \right\rangle ={}& \mathds{1}_{t>s}r_{ij}(s)p_j(s) \nonumber\\
  &\times\left[ P(t,k \!\mid\! s,i)-P(t,k \!\mid\! s,j) \right]\rmd t\rmd s,
  \label{eq: markov_occupation_martingale_correlation}
\end{align}
where $P(t,k \!\mid\! s,i)$ is the Markov propagator from state $i$ at time $s$ to state $k$ at time $t$.

Similarly, choosing the future intensity $Q_\tau=\gamma_{kl}(t)$ defines the history-conditioned intensity-influence kernel
\begin{equation}
  J_{ij}^{kl}(t,s \!\mid\! X_{s-}) = \mathbb{E}\left[\, \gamma_{kl}(t) \!\mid\! X_s^{ij}\, \right] - \mathbb{E}\left[\, \gamma_{kl}(t) \!\mid\! X_s^0\, \right], \quad t>s.
  \label{eq: intensity_influence_kernel}
\end{equation}
For $t>s$,
\begin{equation}
  \left\langle \rmd n_{kl}(t)\rmd\varepsilon_{ij}(s) \right\rangle = \left\langle \gamma_{ij}(s) \, J_{ij}^{kl}(t,s \!\mid\! X_{s-}) \right\rangle \rmd t\rmd s.
  \label{eq: future_jump_martingale_correlation}
\end{equation}
Including the equal-time contribution, the jump--martingale correlation is
\begin{align}
  \left\langle \rmd n_{kl}(t)\rmd\varepsilon_{ij}(s) \right\rangle ={}& \delta_{ik}\delta_{jl}\delta(t-s)\left\langle\gamma_{ij}(s)\right\rangle \rmd t\rmd s \nonumber\\
  &+ \mathds{1}_{t>s}\left\langle \gamma_{ij}(s) \, J_{ij}^{kl}(t,s \!\mid\! X_{s-}) \right\rangle \rmd t\rmd s.
  \label{eq: jump_martingale_correlation}
\end{align}
The corresponding time--time, time--jump, and jump--jump correlations are collected in \cref{app:elementary_correlations}. For an observable with predictable history-dependent weights, the weights are included directly in the future functional entering $K_{ij}^Q$.

The trajectory equation \cref{eq: compact_count_sde} and the martingale representation \cref{eq: martingale_representation} provide two complementary forms of closure. The former states that the stochastic evolution closes on the observed history even when the state probabilities do not. The latter states that all trajectory fluctuations generated during the observation interval close on the same orthogonal family of martingale increments. In the next section, we show that a local perturbation of a transition intensity probes precisely the event-consequence kernel $K_{ij}^Q$.

\section{Trajectory score and history-conditioned response}
\label{sec:linear_response}

\Cref{sec:theory} established that the fluctuations of an arbitrary trajectory observable are generated by the martingale increments, with the event-consequence kernel $K_{ij}^{Q}(\tau,t \,|\, X_{t-})$ as the corresponding expansion coefficient. We now identify the physical meaning of this coefficient through response. A weak perturbation changes the history-dependent transition intensity, and its trajectory score is carried by the same martingale increments that generate spontaneous fluctuations. The elementary response object is therefore the response to a unit logarithmic perturbation of one transition, conditioned on the realized pre-perturbation history. We show that this edge-wise response is the stochastic intensity multiplied by $K_{ij}^{Q}$. General physical perturbations and ordinary responses then follow by combining edges and averaging over histories. Detailed derivations of the trajectory-score identities are provided in \cref{appsec:trajectory_response_theory}.

\subsection{Perturbations of History-Dependent Transition Rates}
\label{subsec:parameter_dependent_rate}

We first specify how an external parameter enters a non-Markovian jump process. In standard open-loop perturbations of Markov jump dynamics, the logarithmic change of a transition rate is determined by the edge, the present state, and time \cite{maes2020response,stutzer2026stochastic}. Here, the rate is a functional of the past trajectory. A perturbation may therefore change not only the instantaneous magnitude of the rate, but also how the realized history affects the present jump tendency.

We write the parameter-dependent transition rate as
\begin{equation}
  r_{ij}^{X_{t-}}(\lambda),
  \label{eq: parameter_dependent_rate}
\end{equation}
where $\lambda$ is an external control parameter. The reference dynamics corresponds to $\lambda=0$, with
\begin{equation}
  r_{ij}^{X_{t-}} \equiv r_{ij}^{X_{t-}}(0).
  \label{eq: reference_transition_rate}
\end{equation}
The logarithmic sensitivity of the history-dependent rate is defined as
\begin{equation}
  \alpha_{ij}^{X_{t-}} \equiv \left.\partial_\lambda\ln r_{ij}^{X_{t-}}(\lambda)\right|_{\lambda=0}.
  \label{eq: logarithmic_rate_sensitivity}
\end{equation}
Equivalently,
\begin{equation}
  r_{ij}^{X_{t-}}(\lambda) = r_{ij}^{X_{t-}}\exp\!\left[\lambda\alpha_{ij}^{X_{t-}}+o(\lambda)\right].
  \label{eq: linearized_parameter_dependent_rate}
\end{equation}
The derivative is taken at fixed realized history $X_{t-}$. Thus, if $\lambda$ changes a memory strength, a memory timescale, or a feedback rule, $\alpha_{ij}^{X_{t-}}$ is obtained by differentiating the rate functional evaluated on the same trajectory history. The redistribution of histories under the perturbation is accounted for separately by the perturbed trajectory probability density. Since $\alpha_{ij}^{X_{t-}}$ depends only on the trajectory already realized, it is determined before the jump at time $t$, but it can vary from one history to another.

\subsection{Trajectory Score and Ordinary Response}
\label{subsec:trajectory_score}

When the instantaneous state probabilities do not close, response cannot in general be derived from a finite-dimensional generator on the observed state space. The trajectory remains exactly specified by its history-dependent intensities, so we formulate response by differentiating the trajectory probability density. For the parameter-dependent rates in \cref{eq: parameter_dependent_rate}, and assuming that the initial density $p_{\mathrm{ini}}$ is independent of $\lambda$, the trajectory density is \cite{bremaud1981point,jacod2003limit,daley2008introduction}
\begin{align}
  \mathcal{P}_{\lambda}[X_\tau] = p_{\mathrm{ini}}\exp\Bigg\{&\sum_{i\neq j}\int_0^\tau \ln r_{ij}^{X_{t-}}(\lambda)\,\rmd n_{ij}(t) \nonumber\\
  &-\int_0^\tau\sum_{i\neq j}r_{ij}^{X_{t-}}(\lambda)\mathds{1}_{x_{t-}=j}\,\rmd t\Bigg\}.
  \label{eq: parameterized_trajectory_density}
\end{align}
The first term assigns the history-dependent rate to every realized transition. The second is the escape contribution accumulated during the waiting periods. The derivation is given in \cref{app:trajectory_density}.

The trajectory score associated with the perturbation is
\begin{equation}
  \Lambda_\tau[X_\tau] = \left.\partial_\lambda\ln\mathcal{P}_{\lambda}[X_\tau]\right|_{\lambda=0}.
  \label{eq: trajectory_score_definition}
\end{equation}
Using \cref{eq: logarithmic_rate_sensitivity}, we obtain
\begin{align}
  \Lambda_\tau[X_\tau] &= \sum_{i\neq j}\int_0^\tau \alpha_{ij}^{X_{t-}} \Big[\rmd n_{ij}(t)-\gamma_{ij}(t)\,\rmd t\Big] \nonumber\\
  &= \sum_{i\neq j}\int_0^\tau \alpha_{ij}^{X_{t-}}\,\rmd\varepsilon_{ij}(t).
  \label{eq: martingale_score}
\end{align}
Thus, even when the perturbation changes a history-dependent rate functional, its trajectory score is carried by the same martingale increments that generate spontaneous trajectory fluctuations.

For a trajectory observable $Q_\tau$ with no explicit dependence on $\lambda$, differentiation of the trajectory density gives
\begin{equation}
  \left.\partial_\lambda\left\langle Q_\tau\right\rangle_\lambda\right|_{\lambda=0} = \left\langle Q_\tau\Lambda_\tau\right\rangle.
  \label{eq: score_response_identity}
\end{equation}
Since $\left\langle\Lambda_\tau\right\rangle=0$,
\begin{equation}
  \left.\partial_\lambda\left\langle Q_\tau\right\rangle_\lambda\right|_{\lambda=0} = \operatorname{Cov}(Q_\tau,\Lambda_\tau).
  \label{eq: score_response_covariance}
\end{equation}
If the observable itself depends explicitly on the parameter,
\begin{align}
  \left.\partial_\lambda\left\langle Q_\tau(\lambda)\right\rangle_\lambda\right|_{\lambda=0} ={}& \left\langle\left.\partial_\lambda Q_\tau(\lambda)\right|_{\lambda=0}\right\rangle \nonumber\\
  &+\operatorname{Cov}(Q_\tau,\Lambda_\tau).
  \label{eq: explicitly_perturbed_observable}
\end{align}
Unless otherwise stated, we focus below on the dynamical response and assume that $Q_\tau$ has no explicit parameter dependence.

For a time-dependent perturbation, we introduce an external protocol $h(t)$ through
\begin{equation}
  \ln r_{ij}^{X_{t-},\epsilon h} = \ln r_{ij}^{X_{t-}}+\epsilon h(t)\alpha_{ij}^{X_{t-}}+O(\epsilon^2),
  \label{eq: time_dependent_rate_perturbation}
\end{equation}
where $\epsilon\ll1$. The corresponding trajectory score is
\begin{equation}
  \Lambda_{\alpha,h}[X_\tau] = \left.\partial_\epsilon\ln\mathcal{P}_{\epsilon h}[X_\tau]\right|_{\epsilon=0} = \int_0^\tau h(t)\,\rmd\Lambda_\alpha(t),
  \label{eq: protocol_score}
\end{equation}
with
\begin{equation}
  \rmd\Lambda_\alpha(t) = \sum_{i\neq j}\alpha_{ij}^{X_{t-}}\,\rmd\varepsilon_{ij}(t).
  \label{eq: infinitesimal_score}
\end{equation}

For an observable measured over $[0,\tau]$, the ordinary response kernel is defined by
\begin{equation}
  R_\alpha^Q(\tau,t) \equiv \left.\frac{\partial}{\partial\epsilon}\frac{\delta\left\langle Q_\tau\right\rangle_{\epsilon h}}{\delta h(t)}\right|_{\epsilon=0}.
  \label{eq: ordinary_response_definition}
\end{equation}
The response to the complete protocol is
\begin{equation}
  \left.\partial_\epsilon\left\langle Q_\tau\right\rangle_{\epsilon h}\right|_{\epsilon=0} = \int_0^\tau h(t) \, R_\alpha^Q(\tau,t)\,\rmd t.
  \label{eq: protocol_response}
\end{equation}
The time-resolved trajectory-score identity gives
\begin{equation}
  R_\alpha^Q(\tau,t) = \left\langle Q_\tau\frac{\rmd\Lambda_\alpha(t)}{\rmd t}\right\rangle.
  \label{eq: response_score_correlation}
\end{equation}
The ordinary response averages over all histories present immediately before the perturbation. To identify the coefficient $K_{ij}^Q$ appearing in the fluctuation representation, we now resolve the response first by transition edge and then by pre-perturbation history.

Girsanov-type trajectory scores of this form underlie likelihood-ratio sensitivity estimators for stochastic reaction networks \cite{plyasunov2007efficient,warren2012steady}. Here the score is further resolved on the realized history and identified, through the martingale representation, with the coefficient of the observable fluctuation.

\subsection{History-Conditioned Edge Response}
\label{subsec:history_conditioned_response}

For Markov jump dynamics with local observables and history-independent open-loop perturbations, the present state and time suffice to determine the conditioned future response. In a non-Markovian process, the same present state may correspond to different histories, and the same edge perturbation can consequently have different future effects. The elementary response object is therefore the response to a unit logarithmic perturbation of one transition, conditioned on the realized history immediately before the perturbation.

For a fixed transition $j\to i$, consider the edge perturbation
\begin{equation}
  \ln r_{kl}^{X_{s-},\epsilon h} = \ln r_{kl}^{X_{s-}}+\epsilon h(s)\delta_{ki}\delta_{lj}+O(\epsilon^2).
  \label{eq: edge_log_rate_perturbation}
\end{equation}
Holding the pre-perturbation history $X_{t-}$ fixed, we define the edge-wise history-conditioned response as
\begin{equation}
  R_{ij}^{Q}(\tau,t \!\mid\! X_{t-}) \equiv \left.\frac{\partial}{\partial\epsilon}\frac{\delta}{\delta h(t)}\mathbb{E}_{\epsilon h}\left[Q_\tau \!\mid\! X_{t-}\right]\right|_{\epsilon=0}.
  \label{eq: edge_wise_history_conditioned_response}
\end{equation}
The perturbation changes the conditional density of the future trajectory but does not alter the history already realized. The conditional trajectory-score identity derived in Appendix~\ref{app:conditional_score_identity} gives
\begin{equation}
  R_{ij}^{Q}(\tau,t \!\mid\! X_{t-}) = \mathbb{E}\left[Q_\tau\frac{\rmd\varepsilon_{ij}(t)}{\rmd t} \,\Bigg|\, X_{t-}\right].
  \label{eq: conditioned_edge_response_correlation}
\end{equation}
Using the martingale--kernel identity \cref{eq: kernel_martingale_correlation}, we obtain the central result of this section,
\begin{equation}
  R_{ij}^{Q}(\tau,t \!\mid\! X_{t-}) = \gamma_{ij}(t) \, K_{ij}^{Q}(\tau,t \!\mid\! X_{t-}).
  \label{eq: edge_response_kernel}
\end{equation}
The event-consequence kernel introduced from spontaneous fluctuations therefore has a direct response interpretation,
\begin{equation}
  K_{ij}^{Q}(\tau,t \!\mid\! X_{t-}) = \frac{R_{ij}^{Q}(\tau,t \!\mid\! X_{t-})}{\gamma_{ij}(t)},
  \label{eq: kernel_as_normalized_response}
\end{equation}
where the ratio is understood on histories with nonzero intensity. The stochastic intensity $\gamma_{ij}(t)$ specifies how available the transition $j\to i$ is on the realized history, while $K_{ij}^{Q}$ specifies the marginal consequence carried by one such transition for the future observable. For Markov dynamics and local additive observables, $K_{ij}^{Q}$ reduces to a state-conditioned jump consequence. \Cref{sec: markov_dfrr} shows that this quantity is precisely the edge amplitude $\varphi_e$ entering the Markovian dynamical FRRs of \cite{aslyamov2026dynamical}.

For a general physical perturbation direction $\alpha_{ij}^{X_{t-}}$, the history-conditioned response is
\begin{subequations}
\label{eq: history_conditioned_parameter_response}
\begin{align}
  R_{\alpha}^{Q}(\tau,t \!\mid\! X_{t-}) &= \sum_{i\neq j}\alpha_{ij}^{X_{t-}}R_{ij}^{Q}(\tau,t \!\mid\! X_{t-}), \\
  &= \sum_{i\neq j}\alpha_{ij}^{X_{t-}}\gamma_{ij}(t) \, K_{ij}^{Q}(\tau,t \!\mid\! X_{t-}).
\end{align}
\end{subequations}
The three factors have distinct physical roles. The logarithmic sensitivity $\alpha_{ij}^{X_{t-}}$ describes how the control parameter couples to the edge. The intensity $\gamma_{ij}(t)$ describes the event availability on the realized history. The kernel $K_{ij}^{Q}$ describes the event consequence for the subsequent trajectory. The first two factors determine the local change of the present event tendency, while the third determines how that local change propagates into the future.

\begin{figure}[t]
  \centering
  \includegraphics[width=0.80\linewidth]{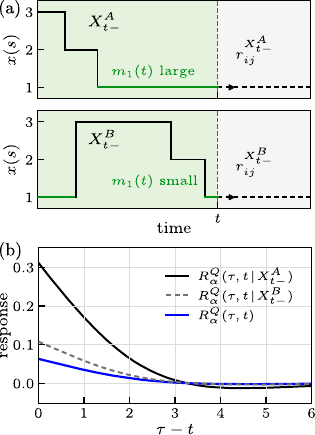}
  \caption{\textbf{History-conditioned and ordinary edge responses.}
  (a) Two representative histories $X_{t-}^{A}$ and $X_{t-}^{B}$ that reach the same state at time $t$. The dark-green segments mark the past residence in state $1$, which determines the memory variable $m_1(t)$ and affects the history-dependent transition rates. At time $t$, we apply the same unit logarithmic perturbation to the edge $1\to2$.
  (b) Schematic comparison between the edge responses $R_{21}^{Q}(\tau,t \,|\, X_{t-}^{A})$ and $R_{21}^{Q}(\tau,t \,|\, X_{t-}^{B})$, and the ordinary edge response $R_{21}^{Q}(\tau,t)$ obtained by averaging over histories. The horizontal axis $\tau-t$ varies the future observation horizon at fixed perturbation time and fixed prehistory; no time-translation invariance is assumed. Although the present state and the perturbation are identical, the two histories generally carry different responses.}
  \label{fig:conditional_response}
\end{figure}

The model used in \cref{fig:non-Markovian_and_martingale,fig:additive_observable,fig:conditional_response} provides a direct illustration. Its history-dependent rate satisfies
\begin{equation}
  \ln r_{ij}^{X_{t-}} = \ln k_{ij}+\beta\frac{m_j(t)}{\tau_m}.
  \label{eq: illustrative_log_rate}
\end{equation}
For the same transition edge, perturbing the baseline logarithmic rate or the memory strength gives
\begin{subequations}
\label{eq: illustrative_logarithmic_sensitivities}
\begin{align}
  \lambda &= \ln k_{ij}, & \alpha_{ij}^{X_{t-}} &= 1 \\
  \lambda &= \beta, & \alpha_{ij}^{X_{t-}} &= \frac{m_j(t)}{\tau_m}.
\end{align}
\end{subequations}
The first perturbation has a history-independent local logarithmic coupling, although its absolute response remains history-dependent through $\gamma_{ij}(t)$ and $K_{ij}^{Q}$. The second perturbation reads the memory variable already through the local coupling $\alpha_{ij}^{X_{t-}}$.

The ordinary response is recovered by averaging over the reference distribution of pre-perturbation histories,
\begin{subequations}
\label{eq: averaged_response_from_history_conditioned}
\begin{align}
  R_{\alpha}^{Q}(\tau,t) &= \left\langle R_{\alpha}^{Q}(\tau,t \!\mid\! X_{t-})\right\rangle, \\
  &= \sum_{i\neq j}\left\langle\alpha_{ij}^{X_{t-}}\gamma_{ij}(t) \, K_{ij}^{Q}(\tau,t \!\mid\! X_{t-})\right\rangle.
\end{align}
\end{subequations}
The ordinary response therefore cannot in general be factorized into separately averaged control couplings, intensities, and event consequences.

As illustrated in \cref{fig:conditional_response}, the two histories reach the same state and are subjected to the same logarithmic perturbation of the transition $1\to2$, but their different past residence in state $1$ produces different memory variables and different conditional future dynamics. The responses $R_{21}^{Q}(\tau,t \!\mid\! X_{t-}^{A})$ and $R_{21}^{Q}(\tau,t \!\mid\! X_{t-}^{B})$ can therefore differ even though neither the present state nor the perturbation has changed. The ordinary response $R_{21}^{Q}(\tau,t)$ retains only their history average.

Causality follows directly from the martingale property. If $Q_t$ is determined by the trajectory up to time $t$, then for $s>t$,
\begin{equation}
  \mathbb{E}\left[\rmd\varepsilon_{ij}(s) \!\mid\! X_t\right] = 0, \quad s>t,
  \label{eq: future_martingale_zero_mean}
\end{equation}
and consequently
\begin{equation}
  \left\langle Q_t\,\rmd\varepsilon_{ij}(s)\right\rangle = 0, \quad s>t.
  \label{eq: causal_score_orthogonality}
\end{equation}
A future score increment therefore cannot change an observable already determined by the past.

The distinction between conditioned and ordinary response is essential beyond Markovianity. For Markov jump dynamics with local readouts and history-independent perturbations, conditioning on the present state and time is sufficient. In the non-Markovian setting, the response remains a random functional of the realized history. The ordinary response retains only its first moment, whereas the exact fluctuation-response relations developed in the next section depend on its history-resolved structure.

\subsection{Responses of Elementary Observables}
\label{subsec:responses_trajectory_observables}

The response formulas become especially transparent for the elementary occupation and jump observables introduced in \cref{subsec:trajectory_observables}. For the occupation-time increment $\rmd\tau_k(s)$, using \cref{eq: occupation_martingale_correlation},
\begin{align}
  \frac{\partial}{\partial\epsilon}\!&\left.\frac{\delta\left\langle\rmd\tau_k(s)\right\rangle_{\epsilon h}}{\delta h(t)}\right|_{\epsilon=0} \nonumber\\
  &= \mathds{1}_{s>t}\sum_{i\neq j}\left\langle\alpha_{ij}^{X_{t-}}\gamma_{ij}(t)I_{ij}^{k}(s,t \!\mid\! X_{t-})\right\rangle\,\rmd s.
  \label{eq: response_waiting_increment}
\end{align}
The occupation response contains only the causal propagation of the perturbation after time $t$.

For the jump increment $\rmd n_{kl}(s)$, using \cref{eq: jump_martingale_correlation},
\begin{align}
  \frac{\partial}{\partial\epsilon}\!&\left.\frac{\delta\left\langle\rmd n_{kl}(s)\right\rangle_{\epsilon h}}{\delta h(t)}\right|_{\epsilon=0} = \delta(s-t)\left\langle\alpha_{kl}^{X_{t-}}\gamma_{kl}(t)\right\rangle\,\rmd s \nonumber\\
  &+\mathds{1}_{s>t}\sum_{i\neq j}\left\langle\alpha_{ij}^{X_{t-}}\gamma_{ij}(t)J_{ij}^{kl}(s,t \!\mid\! X_{t-})\right\rangle\,\rmd s.
  \label{eq: response_jump_increment}
\end{align}
The first term is the direct equal-time response of the perturbed jump itself, while the second is the delayed contribution propagated through the subsequent history-dependent dynamics. The response of a general additive observable follows by weighting and integrating these elementary contributions; its explicit score expansion is given in Appendix~\ref{app:additive_observable_response}.

\Cref{sec:theory} identifies $K_{ij}^{Q}$ as the coefficient through which the orthogonal martingale increments generate an observable fluctuation. The present section identifies the same coefficient as the intensity-normalized history-conditioned edge response. Substituting \cref{eq: kernel_as_normalized_response} into the martingale representation therefore produces the exact fluctuation-response representation developed in the next section.

\section{Time-domain fluctuation-response relations and inequalities}
\label{sec:time-domain_frr_bounds}

The results of this section form a hierarchy. Resolving every transition channel, perturbation time, and pre-perturbation history yields exact fluctuation-response relations (FRRs). Averaging over histories converts these identities into edge-wise fluctuation-response inequalities (FRIs) and leaves a nonnegative response-heterogeneity gap. Combining the edge-wise response coordinates into a physical perturbation direction gives a path-space Fisher-information FRI. Finally, bounding the logarithmic rate sensitivity replaces the Fisher information by the dynamical activity and yields a response-kinetic uncertainty relation (R-KUR). Each step retains less of the full history-resolved response structure. Detailed derivations are provided in Appendix~\ref{app:time_domain_frr_bounds}.

\subsection{Response Representation and Exact Time-Domain FRRs}
\label{subsec:exact_time_domain_frr}

Combining the martingale representation in \cref{eq: martingale_representation} with the response identity \cref{eq: edge_response_kernel} gives
\begin{equation}
  Q_\tau-\mathbb{E}\left[ Q_\tau \!\mid\! X_0 \right] = \sum_{i\neq j}\int_0^\tau \frac{R_{ij}^Q(\tau,t \!\mid\! X_{t-})}{\gamma_{ij}(t)}\rmd\varepsilon_{ij}(t).
  \label{eq: response_representation}
\end{equation}
Here $X_0$ denotes the complete initial information, including the initial state and any initial prehistory required to initialize the memory. All ratios involving $\gamma_{ij}(t)$ are understood on the support of the corresponding intensity and are set to zero outside that support.

\Cref{eq: response_representation} identifies the intensity-normalized history-conditioned response as the coefficient of each orthogonal martingale increment. The fluctuation of $Q_\tau$ generated after the initial condition is therefore expanded directly in response coordinates: each transition channel and time contributes according to the future consequence carried by that event on the realized history.

For two trajectory observables $Q_\tau$ and $Q_\tau'$, the space--time orthogonality of the martingale increments gives
\begin{subequations}
\label{eq: exact_covariance_frr}
\begin{align}
  \operatorname{Cov}(Q_\tau, {}&Q_\tau') = C_0(Q,Q') \\
  &+\int_0^\tau\sum_{i\neq j}\left\langle \frac{R_{ij}^Q(\tau,t \!\mid\! X_{t-}) R_{ij}^{Q'}(\tau,t \!\mid\! X_{t-})}{\gamma_{ij}(t)} \right\rangle \rmd t, \nonumber \\
  C_0(Q,Q') {}& = \operatorname{Cov}\Big(\mathbb{E}\left[ Q_\tau \!\mid\! X_0 \right], \mathbb{E}\left[ Q_\tau' \!\mid\! X_0 \right]\Big).
\end{align}
\end{subequations}
The term $C_0(Q,Q')$ is inherited from uncertainty in the initial state and initial memory. It is orthogonal to the covariance generated subsequently by the martingale increments and vanishes when the complete initial condition is fixed. The remaining covariance is exactly the overlap of the full history-resolved response coordinates of the two observables. At this level, no response information has been discarded and no unexplained residual remains.

For $Q_\tau'=Q_\tau$,
\begin{equation}
  \operatorname{Var}(Q_\tau) = C_0(Q,Q)+\int_0^\tau\sum_{i\neq j}\left\langle \frac{\left[R_{ij}^Q(\tau,t \!\mid\! X_{t-})\right]^2}{\gamma_{ij}(t)} \right\rangle\rmd t.
  \label{eq: exact_variance_frr}
\end{equation}
Equivalently,
\begin{equation}
  \operatorname{Var}(Q_\tau)-C_0(Q,Q) = \mathbb{E}\left[\operatorname{Var}(Q_\tau \!\mid\! X_0)\right].
  \label{eq: martingale_generated_variance}
\end{equation}
Thus, the variance remaining after the initial contribution is removed is precisely the fluctuation generated by the martingale increments during $[0,\tau]$, and the exact FRR reconstructs it as the total history-resolved response power.

\subsection{History Averaging, Time-Domain FRIs, and the Response-Heterogeneity Gap}
\label{subsec:from_frr_to_fri}

The exact FRRs above retain the response on every pre-perturbation history. The ordinary edge response instead averages over those histories,
\begin{equation}
  R_{ij}^{Q}(\tau,t) = \left\langle R_{ij}^{Q}(\tau,t \!\mid\! X_{t-}) \right\rangle = \left\langle \gamma_{ij}(t)K_{ij}^{Q}(\tau,t \!\mid\! X_{t-}) \right\rangle.
  \label{eq: averaged_edge_response_fri}
\end{equation}
For $\left\langle\gamma_{ij}(t)\right\rangle>0$, we define the event-weighted mean consequence
\begin{equation}
  \overline{K}_{ij}^{Q}(\tau,t) \equiv \frac{R_{ij}^{Q}(\tau,t)}{\left\langle\gamma_{ij}(t)\right\rangle}.
  \label{eq: event_weighted_mean_consequence}
\end{equation}
This is not an unweighted average over all histories. Histories contribute in proportion to the intensity of the event $j\to i$ at time $t$, so $\overline{K}_{ij}^{Q}$ is the mean consequence within the ensemble of histories on which that event is available. If $\left\langle\gamma_{ij}(t)\right\rangle=0$, the corresponding edge contribution is defined to be zero.

Completing the square in the exact variance FRR gives the exact decomposition
\begin{subequations}
\label{eq: frr_fri_decomposition}
\begin{align}
  \operatorname{Var}&(Q_\tau) - C_0(Q,Q) = \mathcal{V}_Q^{\mathrm{av}} + \mathcal{M}_Q, \\
  \mathcal{V}_Q^{\mathrm{av}} &= \int_0^\tau\sum_{i\neq j}\frac{\left[R_{ij}^{Q}(\tau,t)\right]^2}{\left\langle\gamma_{ij}(t)\right\rangle}\rmd t,
  \label{eq: averaged_response_variance}\\
  \mathcal{M}_Q &= \int_0^\tau\sum_{i\neq j}\Bigg\langle \gamma_{ij}(t)\Bigg[K_{ij}^{Q}(\tau,t \!\mid\! X_{t-}) - \overline{K}_{ij}^{Q}(\tau,t)\Bigg]^2 \Bigg\rangle\rmd t \nonumber\\
  &\geq 0.
  \label{eq: memory_response_fluctuation_gap}
\end{align}
\end{subequations}
The first term is the part of the variance reconstructible from ordinary edge responses. The second term $\mathcal{M}_Q$ is the response-heterogeneity gap. It is the intensity-weighted variation of the event consequence over active histories that have been merged by ordinary response averaging. The gap therefore has a specific physical meaning: it records how much fluctuation power comes from the same edge event carrying different future consequences on different histories.

Since $\mathcal{M}_Q\geq0$, \cref{eq: frr_fri_decomposition} immediately gives
\begin{equation}
  \operatorname{Var}(Q_\tau)-C_0(Q,Q) \geq \int_0^\tau\sum_{i\neq j}\frac{\left[R_{ij}^{Q}(\tau,t)\right]^2}{\left\langle\gamma_{ij}(t)\right\rangle}\rmd t.
  \label{eq: time_domain_fri_edge}
\end{equation}
This is the edge-wise time-domain FRI for non-Markovian jump processes. The quantity $\left\langle\gamma_{ij}(t)\right\rangle$ is the instantaneous mean event rate on the directed edge $j\to i$,
\begin{equation}
  \left\langle\gamma_{ij}(t)\right\rangle = \frac{\rmd}{\rmd t}\left\langle n_{ij}(t)\right\rangle.
  \label{eq: mean_edge_event_rate}
\end{equation}
Its time integral is the expected number of transitions on that edge during the observation interval. The inverse mean event rate in \cref{eq: time_domain_fri_edge} therefore compares the squared response with the kinetic events available to generate it.

The edge-wise FRI is saturated when all active histories assign the same consequence to the corresponding edge event,
\begin{equation}
  K_{ij}^{Q}(\tau,t \!\mid\! X_{t-}) = \overline{K}_{ij}^{Q}(\tau,t) = \frac{R_{ij}^{Q}(\tau,t)}{\left\langle\gamma_{ij}(t)\right\rangle}
  \label{eq: time_domain_fri_equality_condition}
\end{equation}
for histories with nonzero $\gamma_{ij}(t)$. In this case, history averaging loses no response information and $\mathcal{M}_Q=0$. For Markov jump dynamics with local observables and history-independent open-loop edge perturbations, the present state and time determine the event consequence, and the additional full-history gap vanishes. The converse is not true: $\mathcal{M}_Q=0$ for a particular observable, edge family, and observation horizon does not imply that the complete process is Markovian. And a history-dependent readout may generate a positive gap even when the underlying jump dynamics is Markovian. The gap is therefore observable-dependent, perturbation-dependent, and timescale-dependent rather than a universal scalar measure of non-Markovianity.

\subsection{General Time-Domain FRIs and Path-Space Fisher Information}
\label{subsec:general_time_domain_fri}

We next combine the elementary edge coordinates into an arbitrary physical perturbation direction. The ordinary response kernel is
\begin{equation}
  R_{\alpha}^{Q}(\tau,t) = \sum_{i\neq j}\left\langle \alpha_{ij}^{X_{t-}}R_{ij}^{Q}(\tau,t \!\mid\! X_{t-}) \right\rangle.
  \label{eq: alpha_response_fri}
\end{equation}
The observable fluctuation generated after the initial condition and the trajectory score are expanded in the same orthogonal family of martingale increments,
\begin{subequations}
\label{eq: observable_score_common_basis}
\begin{align}
  \widetilde{Q}_\tau &\equiv Q_\tau - \mathbb{E}\left[ Q_\tau \!\mid\! X_0 \right] \\
  &= \sum_{i\neq j}\int_0^\tau K_{ij}^{Q}(\tau,t \!\mid\! X_{t-}) \, \rmd\varepsilon_{ij}(t), \nonumber \\
  \Lambda_{\alpha,h} &= \sum_{i\neq j}\int_0^\tau h(t)\alpha_{ij}^{X_{t-}}\rmd\varepsilon_{ij}(t).
\end{align}
\end{subequations}
The response to the complete protocol is the overlap of these two sets of coordinates,
\begin{equation}
  \left.\partial_\epsilon\left\langle Q_\tau\right\rangle_{\epsilon h}\right|_{\epsilon=0} = \left\langle \widetilde{Q}_\tau\Lambda_{\alpha,h}\right\rangle.
  \label{eq: response_as_coordinate_overlap}
\end{equation}
The initial conditional mean does not contribute because it is determined before the martingale score is generated and is orthogonal to that score.

The squared norm of the score is the path-space Fisher information,
\begin{equation}
  \mathcal{I}_{\alpha,h} = \left\langle\Lambda_{\alpha,h}^2\right\rangle = \int_0^\tau\sum_{i\neq j}\left\langle \gamma_{ij}(t)\left[h(t)\alpha_{ij}^{X_{t-}}\right]^2 \right\rangle\rmd t.
  \label{eq: fisher_activity_general}
\end{equation}
It measures the local path-space sensitivity of the trajectory density along the chosen perturbation direction. Applying Cauchy--Schwarz to \cref{eq: response_as_coordinate_overlap} gives
\begin{equation}
  \frac{\left|\left.\partial_\epsilon\left\langle Q_\tau\right\rangle_{\epsilon h}\right|_{\epsilon=0}\right|^2}{\operatorname{Var}(Q_\tau)-C_0(Q,Q)} \leq \mathcal{I}_{\alpha,h}.
  \label{eq: general_time_domain_fri}
\end{equation}
This is the general time-domain FRI for a physical perturbation of history-dependent rates. It is saturated when
\begin{equation}
  Q_\tau-\mathbb{E}\left[ Q_\tau \!\mid\! X_0 \right] = c\Lambda_{\alpha,h}
  \label{eq: fisher_fri_equality_condition}
\end{equation}
for a history-independent constant $c$ on the reference ensemble. This condition measures the alignment between the observable fluctuation and the selected physical score direction. It is distinct from the response-heterogeneity gap $\mathcal{M}_Q$. The gap is produced by averaging over histories, whereas the slack of \cref{eq: general_time_domain_fri} is produced by an imperfect match between one physical perturbation direction and the complete fluctuation coordinates of the observable. This mismatch can remain nonzero even for a Markov process.

For $h(t)=1$, \cref{eq: general_time_domain_fri} reduces to the corresponding static-parameter response bound. In the Markovian limit, it recovers the activity-related FRI for Markov jump dynamics \cite{stutzer2026stochastic}.

For a unit logarithmic perturbation of a single transition $j\to i$, with $h(t)=1$ and $\alpha_{kl}^{X_{t-}}=\delta_{ki}\delta_{lj}$, the Fisher information is exactly the expected activity of that edge,
\begin{equation}
  \mathcal{I}_{ij}(\tau) = \mathcal{A}_{ij}(\tau) \equiv \int_0^\tau\left\langle\gamma_{ij}(t)\right\rangle\rmd t.
  \label{eq: local_edge_fisher_activity}
\end{equation}
The corresponding local response-kinetic bound is
\begin{equation}
  \frac{\left|\left.\partial_\epsilon\left\langle Q_\tau\right\rangle_{\epsilon,ij}\right|_{\epsilon=0}\right|^2}{\operatorname{Var}(Q_\tau)-C_0(Q,Q)} \leq \mathcal{A}_{ij}(\tau).
  \label{eq: local_edge_response_kur}
\end{equation}

\begin{figure}[t]
  \centering
  \includegraphics[width=0.98\linewidth]{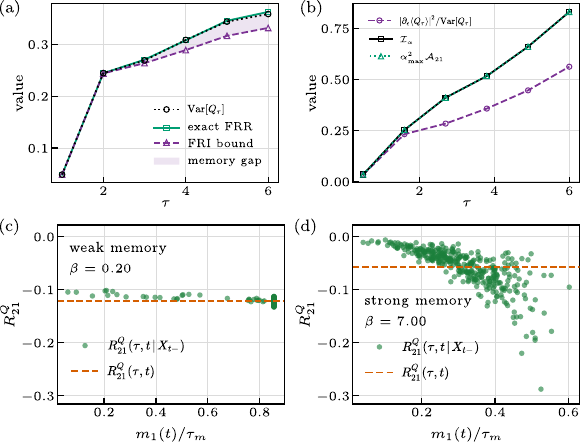}
  \caption{\textbf{Time-domain FRRs, FRIs, and response-KUR.}
  (a) The variance, exact FRR reconstruction, and FRI lower bound as functions of $\tau$ for the observable $Q_\tau = \int_0^\tau \mathds{1}_{x(s)=1} \,\rmd s$; the shaded region indicates the response-heterogeneity gap generated by history averaging. (b) Time-domain response-KUR for a local perturbation of the edge $1 \to 2$. We choose the logarithmic rate perturbation $\ln r_{21}^{X_{t-},\epsilon} = \ln r_{21}^{X_{t-}} + \epsilon$, so that $\alpha_{21}^{X_{t-}} = 1$ and the Fisher information reduces exactly to the local activity of the perturbed edge, $\mathcal{I}_\alpha(\tau) = \mathcal{A}_{21}(\tau)$. (c,d) History-conditioned responses $R_{21}^{Q}(\tau, t \!\mid\! X_{t-})$ for the perturbed edge $1 \to 2$ under weak and strong memory, respectively. Here $m_j(t)$ encodes the accumulated occupation history of state $j$, while $\beta$ controls how strongly this memory modulates the transition rate through $\ln r_{ij}^{X_{t-}} = \ln k_{ij} + \beta m_j(t) / \tau_m$. Each point corresponds to one history with $x(t-) = 1$, and the dashed line denotes the ordinary response obtained by averaging over histories.}
  \label{fig:time_domain_frr_fri}
\end{figure}

\subsection{Non-Markovian Response-Kinetic Uncertainty Relations}
\label{subsec:non_markovian_response_kur}

The general time-domain FRI becomes an activity-controlled kinetic bound when the logarithmic sensitivity is bounded,
\begin{equation}
  \left|\alpha_{ij}^{X_{t-}}\right| \leq \alpha_{\max}
  \label{eq: time_domain_bounded_logarithmic_sensitivity}
\end{equation}
for all transitions and histories contributing to the ensemble. The parameter $\epsilon$ and the temporal protocol $h(t)$ are understood to have a fixed normalization; otherwise an overall rescaling could be shifted arbitrarily between $\epsilon$, $h$, and $\alpha$.

The path-space Fisher information is then bounded by the protocol-weighted dynamical activity,
\begin{subequations}
\begin{align}
  \mathcal{I}_{\alpha,h} &\leq \alpha_{\max}^2\mathcal{A}_{h,\tau},
  \label{eq: time_domain_fisher_activity_activity_bound}\\
  \mathcal{A}_{h,\tau} &= \int_0^\tau h(t)^2\sum_{i\neq j}\left\langle\gamma_{ij}(t)\right\rangle\rmd t.
  \label{eq: protocol_weighted_activity}
\end{align}
\end{subequations}
Combining this result with \cref{eq: general_time_domain_fri} gives
\begin{equation}
  \frac{\left|\left.\partial_\epsilon\left\langle Q_\tau\right\rangle_{\epsilon h}\right|_{\epsilon=0}\right|^2}{\operatorname{Var}(Q_\tau)-C_0(Q,Q)} \leq \alpha_{\max}^2\mathcal{A}_{h,\tau}.
  \label{eq: time_domain_response_kur}
\end{equation}
\label{eq: time_domain_response_precision_kur}
This is the time-domain non-Markovian R-KUR. It states that the precision of the response of a trajectory observable to a bounded kinetic modulation is controlled by the dynamical activity available during the observation interval. The result does not require Markovianity, stationarity, local detailed balance, a thermodynamic interpretation of the event channels, or an identification of the perturbation with a rescaling of time. It follows from the martingale fluctuation-response structure and the boundedness of the logarithmic rate sensitivity.

Saturation requires two independent conditions. First, the martingale-generated fluctuation of $Q_\tau$ must be proportional to the trajectory score, as in \cref{eq: fisher_fri_equality_condition}. Second, $\left|\alpha_{ij}^{X_{t-}}\right|=\alpha_{\max}$ must hold on the edge--history--time support weighted by $h(t)^2\gamma_{ij}(t)$. The difference introduced by replacing the Fisher information with $\alpha_{\max}^2\mathcal{A}_{h,\tau}$ is therefore a sensitivity-bound loss, not a response-heterogeneity gap.

For a constant protocol $h(t)=1$, the protocol-weighted activity reduces to the total expected number of transitions over the observation interval,
\begin{equation}
  \mathcal{A}_\tau = \int_0^\tau\sum_{i\neq j}\left\langle\gamma_{ij}(t)\right\rangle\rmd t.
  \label{eq: total_activity_observation_time}
\end{equation}
The corresponding R-KUR is
\begin{equation}
  \frac{\left|\left.\partial_\epsilon\left\langle Q_\tau\right\rangle_{\epsilon}\right|_{\epsilon=0}\right|^2}{\operatorname{Var}(Q_\tau)-C_0(Q,Q)} \leq \alpha_{\max}^2\mathcal{A}_\tau.
  \label{eq: constant_protocol_time_domain_response_kur}
\end{equation}
Thus, over a fixed observation time, the total activity is the kinetic resource limiting response precision.

As illustrated in \cref{fig:time_domain_frr_fri}, the exact FRR reconstructs the full martingale-generated variance, while the ordinary edge responses reconstruct only the FRI contribution. Panels (c) and (d) display the spread of the raw history-conditioned responses. The response-heterogeneity gap itself is determined by the intensity-normalized coefficients $K_{21}^{Q}=R_{21}^{Q}/\gamma_{21}$ and their intensity-weighted variation; a broad distribution of raw responses alone does not imply a positive gap. For the illustrative model shown, increasing the memory strength also increases the variation of these normalized event consequences and therefore enlarges $\mathcal{M}_Q$. For the unit logarithmic perturbation in panel (b), the Fisher information equals the local edge activity, as in \cref{eq: local_edge_fisher_activity}.

The hierarchy contains three physically different losses of resolution. History averaging removes distinctions among event consequences and produces the response-heterogeneity gap $\mathcal{M}_Q$. Selecting one physical perturbation direction retains only the component of the observable fluctuation aligned with its score and produces the Fisher-information mismatch. Replacing the actual sensitivities by the uniform envelope $\alpha_{\max}$ produces the sensitivity-bound loss. Only the first quantity is a diagnostic of response-relevant memory. The same hierarchy will reappear in the finite-frequency theory.

\section{Finite-frequency fluctuation-response relations and inequalities}
\label{sec:frequency_domain_frr}

The frequency domain adds two physical resolutions that are not explicit in the finite-time relations of the previous section. In a stationary process, each history carries a complex event consequence. Its magnitude gives the response gain. Its phase gives the temporal lag. The power spectrum is the full intensity-weighted power of these complex consequences. An ordinary susceptibility retains only their coherent complex average. Without time-translation invariance, frequency no longer diagonalizes the response: an input at frequency $\nu$ can generate an output at a different frequency $\omega$. We first develop the stationary spectral hierarchy and then extend it to finite-window two-frequency relations. Detailed derivations are provided in \cref{appsec: stationary_finite_frequency_derivations,appsec: dynamical_two_frequency_derivations}.

\subsection{Stationary Edge Responses and Spectra}
\label{subsec:stationary_finite_frequency_response}

We first consider autonomous non-Markovian jump processes in a stationary regime, where correlation and response functions are invariant under time translation. We restrict to well-behaved stationary regimes. No additional random component of the observable rate remains fixed by the remote past. The relevant causal response kernels and spectral densities have finite Fourier transforms. The more general form, including a possible remote-past contribution, is given in \cref{appsec: stationary_spectral_assumptions}.

Let $\rmd Q(t)$ be the infinitesimal increment of an additive trajectory observable. It may contain both occupation-time and jump contributions,
\begin{equation}
  \rmd Q(t) = \sum_i g_i^{X_{t-}} \,\rmd \tau_i(t) + \sum_{i \neq j} c_{ij}^{X_{t-}} \,\rmd n_{ij}(t).
  \label{eq: stationary_additive_increment}
\end{equation}
The history-dependent weights are assumed to be stationary under a common shift of the trajectory and observation time. When $\rmd Q(t)$ contains jump contributions, its time derivative $\dot{Q}(t)$ is understood as a generalized process with delta peaks at the jump times.

We specialize the edge-wise logarithmic perturbation introduced in the previous section to the stationary setting,
\begin{equation}
  \ln r_{ij}^{X_{s-},\epsilon h} = \ln r_{ij}^{X_{s-}} + \epsilon h(s) + O(\epsilon^2).
  \label{eq: edge_log_rate_perturbation_frequency}
\end{equation}
The history-conditioned response kernel is defined by the derivative of the corresponding conditional mean,
\begin{equation}
  R_{ij}^{Q}(t-s \!\mid\! X_{s-}) \equiv \left. \frac{\partial}{\partial \epsilon} \frac{\delta}{\delta h(s)} \mathbb{E}_{\epsilon h}\!\left[ \dot{Q}(t) \!\mid\! X_{s-} \right] \right|_{\epsilon=0}, \quad t \geq s.
  \label{eq: stationary_history_conditioned_response_kernel}
\end{equation}
By causality, this response vanishes for $t<s$. Stationarity implies that its functional form depends on $t$ and $s$ only through $t-s$. Its realized value remains a random functional of the history at the perturbation time. We therefore set the perturbation time to the origin and write $R_{ij}^{Q}(t \!\mid\! X_{0-})$ for $t\geq0$. Here $X_{0-}$ denotes the stationary prehistory viewed from the perturbation time.

The ordinary time-domain response kernel is obtained by averaging over the stationary history ensemble,
\begin{equation}
  R_{ij}^{Q}(t) = \left\langle R_{ij}^{Q}(t \!\mid\! X_{0-}) \right\rangle_{\mathrm{ss}}.
  \label{eq: averaged_stationary_response_kernel}
\end{equation}
Recalling the event-consequence representation derived above, the history-conditioned edge response satisfies
\begin{equation}
  R_{ij}^{Q}(t \!\mid\! X_{0-}) = \gamma_{ij}(0) \, K_{ij}^{Q}(t \!\mid\! X_{0-}).
  \label{eq: stationary_response_causal_influence}
\end{equation}
The frequency-domain history-conditioned response and event-consequence kernel are defined by the Fourier transforms of their causal time-domain counterparts,
\begin{subequations}
\label{eq: stationary_frequency_response_and_kernel}
\begin{align}
  \mathcal{R}_{ij}^{Q}(\omega \!\mid\! X_{0-}) &= \int_{[0,\infty)} e^{\mathrm{i}\omega t} \, R_{ij}^{Q}(t \!\mid\! X_{0-}) \, \rmd t,
  \label{eq: history_conditioned_frequency_response} \\
  \mathcal{K}_{ij}^{Q}(\omega \!\mid\! X_{0-}) &= \int_{[0,\infty)} e^{\mathrm{i}\omega t} \, K_{ij}^{Q}(t \!\mid\! X_{0-}) \, \rmd t,
  \label{eq: history_conditioned_frequency_causal_kernel} \\
  \mathcal{R}_{ij}^{Q}(\omega \!\mid\! X_{0-}) &= \gamma_{ij}(0) \, \mathcal{K}_{ij}^{Q}(\omega \!\mid\! X_{0-}).
  \label{eq: frequency_response_causal_influence}
\end{align}
\end{subequations}
The notation $[0,\infty)$ includes the complete equal-time contribution at the origin. The quantity $\mathcal{K}_{ij}^{Q}(\omega \!\mid\! X_{0-})$ is the frequency-resolved event consequence. Its magnitude measures how strongly one event contributes at frequency $\omega$, while its phase measures when that consequence arrives relative to the perturbation.

The ordinary finite-frequency edge response is the coherent complex average
\begin{subequations}
\label{eq: averaged_frequency_response}
\begin{align}
  \mathcal{R}_{ij}^{Q}(\omega) &= \left\langle \mathcal{R}_{ij}^{Q}(\omega \!\mid\! X_{0-}) \right\rangle_{\mathrm{ss}} \\
  &= \int_{[0,\infty)} e^{\mathrm{i}\omega t} \, R_{ij}^{Q}(t) \,\rmd t.
\end{align}
\end{subequations}
Histories with comparable response magnitudes but different phase lags can therefore partially cancel. A small ordinary susceptibility does not necessarily imply that the individual history-conditioned responses are weak.

We define the stationary cross spectrum of two real generalized observable rates by
\begin{equation}
  \mathcal{S}_{QQ'}(\omega) = \int_{-\infty}^{\infty} e^{\mathrm{i}\omega t} \operatorname{Cov}\!\left( \dot{Q}(t),\dot{Q}'(0) \right) \,\rmd t.
  \label{eq: stationary_cross_spectrum_definition}
\end{equation}
For $Q'=Q$, this reduces to the power spectral density
\begin{equation}
  \mathcal{S}_Q(\omega) = \int_{-\infty}^{\infty} e^{\mathrm{i}\omega t} \operatorname{Cov}\!\left( \dot{Q}(t),\dot{Q}(0) \right) \,\rmd t.
  \label{eq: observable_power_spectrum}
\end{equation}
The spectrum is associated with the generalized observable rate or increment process, rather than with the accumulated variable $Q(t)$ itself.

\subsection{Exact Stationary Finite-frequency FRRs}
\label{subsec:exact_finite_frequency_frr}

The stationary martingale representation and its Fourier transformation are derived in \cref{appsec: stationary_finite_frequency_frr_derivation}. Under the stationary assumptions stated above, the exact cross-spectrum FRR is
\begin{subequations}
\label{eq: exact_finite_frequency_cross_frr}
\begin{align}
  \mathcal{S}_{QQ'}(\omega) &= \sum_{i \neq j} \left\langle \gamma_{ij}(0) \, \mathcal{K}_{ij}^{Q}(\omega \!\mid\! X_{0-}) \, \mathcal{K}_{ij}^{Q'}(-\omega \!\mid\! X_{0-}) \right\rangle_{\mathrm{ss}} \\
  &= \sum_{i \neq j} \left\langle \frac{\mathcal{R}_{ij}^{Q}(\omega \!\mid\! X_{0-}) \, \mathcal{R}_{ij}^{Q'}(-\omega \!\mid\! X_{0-})}{\gamma_{ij}(0)} \right\rangle_{\mathrm{ss}}.
\end{align}
\end{subequations}
For real time-domain response kernels,
\begin{equation}
  \mathcal{R}_{ij}^{Q'}(-\omega \!\mid\! X_{0-}) = \left[ \mathcal{R}_{ij}^{Q'}(\omega \!\mid\! X_{0-}) \right]^*.
  \label{eq: negative_frequency_response_conjugate}
\end{equation}
The cross spectrum consequently satisfies $\mathcal{S}_{QQ'}(-\omega)=\mathcal{S}_{Q'Q}(\omega)=\mathcal{S}_{QQ'}(\omega)^*$. For $Q'=Q$, the exact power-spectrum FRR becomes
\begin{subequations}
\label{eq: exact_finite_frequency_frr}
\begin{align}
  \mathcal{S}_Q(\omega) &= \sum_{i \neq j} \left\langle \gamma_{ij}(0) \left| \mathcal{K}_{ij}^{Q}(\omega \!\mid\! X_{0-}) \right|^2 \right\rangle_{\mathrm{ss}} \\
  &= \sum_{i \neq j} \left\langle \frac{\left| \mathcal{R}_{ij}^{Q}(\omega \!\mid\! X_{0-}) \right|^2}{\gamma_{ij}(0)} \right\rangle_{\mathrm{ss}}.
\end{align}
\end{subequations}
Thus, the stationary power spectrum is exactly the total intensity-weighted power of the complex event consequences. At full history resolution, the spontaneous spectrum and the response spectra are two descriptions of the same martingale coordinates.

If the observable contains jump contributions, the equal-time response at $t=0$ is included in $\mathcal{R}_{ij}^{Q}(\omega \!\mid\! X_{0-})$. Consequently, \cref{eq: exact_finite_frequency_frr} includes the full shot-noise contribution of jump observables.

The zero-frequency limit connects this stationary theory to the finite-time relations of Section~V. Under integrable stationary correlations,
\begin{equation}
  \lim_{\tau\to\infty}\frac{\operatorname{Var}(Q_\tau)}{\tau} = \mathcal{S}_Q(0),
  \label{eq: zero_frequency_variance_rate}
\end{equation}
and, for a constant stationary perturbation direction,
\begin{equation}
  \lim_{\tau\to\infty}\frac{1}{\tau}\left.\partial_\epsilon\left\langle Q_\tau\right\rangle_\epsilon\right|_{\epsilon=0} = \mathcal{R}_{\alpha}^{Q}(0).
  \label{eq: zero_frequency_response_rate}
\end{equation}
The finite-time and stationary spectral relations are therefore different resolutions of the same response-coordinate representation.

\subsection{History Projection and the Frequency-Resolved Response-Heterogeneity Gap}
\label{subsec:finite_frequency_fri}

The ordinary edge response retains only the coherent history average of the complex event consequence. For $\left\langle\gamma_{ij}(0)\right\rangle_{\mathrm{ss}}>0$, we define the event-weighted mean consequence
\begin{equation}
  \overline{\mathcal{K}}_{ij}^{Q}(\omega) \equiv \frac{\mathcal{R}_{ij}^{Q}(\omega)}{\left\langle \gamma_{ij}(0) \right\rangle_{\mathrm{ss}}}.
  \label{eq: stationary_event_weighted_mean_consequence}
\end{equation}
If $\left\langle\gamma_{ij}(0)\right\rangle_{\mathrm{ss}}=0$, the corresponding edge contribution is defined to be zero.

Completing the square in the exact power-spectrum FRR gives the exact decomposition
\begin{subequations}
\label{eq: finite_frequency_frr_fri_decomposition}
\begin{align}
  \mathcal{S}_Q(\omega) &= \sum_{i \neq j} \frac{\left| \mathcal{R}_{ij}^{Q}(\omega) \right|^2}{\left\langle \gamma_{ij}(0) \right\rangle_{\mathrm{ss}}} + \mathcal{M}_Q(\omega), \\
  \mathcal{M}_Q(\omega) &= \sum_{i \neq j} \left\langle \gamma_{ij}(0) \left| \mathcal{K}_{ij}^{Q}(\omega \!\mid\! X_{0-})-\overline{\mathcal{K}}_{ij}^{Q}(\omega) \right|^2 \right\rangle_{\mathrm{ss}} \\
  &\geq 0.
  \label{eq: finite_frequency_memory_gap}
\end{align}
\end{subequations}
The nonnegative term $\mathcal{M}_Q(\omega)$ is the frequency-resolved response-heterogeneity gap. It is the intensity-weighted dispersion of the complex event consequence over active pre-perturbation histories. It therefore retains both gain dispersion and phase-lag dispersion. The ordinary susceptibility is a coherent complex average, so a small value may reflect either weak responses on every history or cancellation among strong responses with different phases.

Since $\mathcal{M}_Q(\omega)\geq0$, \cref{eq: finite_frequency_frr_fri_decomposition} immediately gives the edge-wise finite-frequency FRI,
\begin{equation}
  \mathcal{S}_Q(\omega) \geq \sum_{i \neq j} \frac{\left| \mathcal{R}_{ij}^{Q}(\omega) \right|^2}{\left\langle \gamma_{ij}(0) \right\rangle_{\mathrm{ss}}}.
  \label{eq: finite_frequency_edge_fri}
\end{equation}
The FRI is saturated when
\begin{equation}
  \mathcal{K}_{ij}^{Q}(\omega \!\mid\! X_{0-}) = \overline{\mathcal{K}}_{ij}^{Q}(\omega) = \frac{\mathcal{R}_{ij}^{Q}(\omega)}{\left\langle \gamma_{ij}(0) \right\rangle_{\mathrm{ss}}}
  \label{eq: finite_frequency_edge_fri_equality_condition}
\end{equation}
for all active histories. Saturation requires agreement of both the response magnitude and the response phase across histories.

For Markov jump dynamics with local readouts and history-independent open-loop edge perturbations, the frequency-resolved event consequence is fixed by the present state, edge, and frequency, so the additional full-history gap vanishes. The converse is not true: $\mathcal{M}_Q(\omega)=0$ for one observable and frequency does not imply that the complete process is Markovian. A history-dependent readout may also generate a positive spectral gap even when the underlying jump dynamics is Markovian. The gap is therefore observable dependent, perturbation dependent, and frequency dependent rather than a universal measure of memory duration.

\begin{figure}[htbp]
  \centering
  \includegraphics[width=0.98\linewidth]{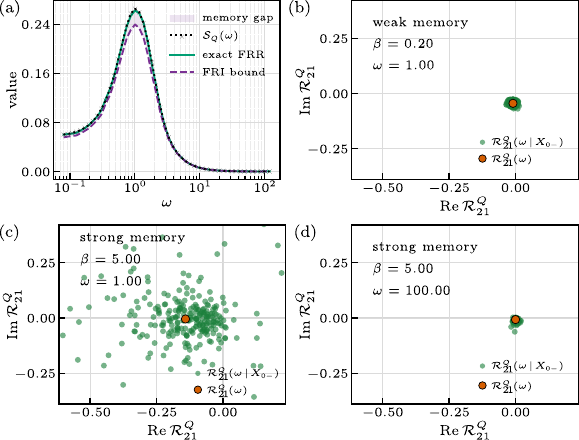}
  \caption{\textbf{Steady-state finite-frequency response.}
  (a) Power spectrum $\mathcal{S}_Q(\omega)$, exact FRR reconstruction, and FRI lower bound for the stationary observable $Q$. The shaded region indicates the frequency-resolved response-heterogeneity gap between the exact FRR and the ordinary-response FRI. (b,c) Complex-plane representation of the history-conditioned response $\mathcal{R}_{21}^{Q}(\omega \!\mid\! X_{0-})$ for weak and strong memory, respectively. Green points denote responses conditioned on individual histories, while the blue point denotes the ordinary response $\mathcal{R}_{21}^{Q}(\omega)$ obtained by averaging over histories. Here $m_j(t)$ encodes the accumulated occupation history of state $j$, while $\beta$ controls how strongly this memory modulates the transition rate through $\ln r_{ij}^{X_{t-}} = \ln k_{ij} + \beta m_j(t) / \tau_m$. (d) The same strong-memory response at a higher frequency. The reduced spread illustrates that history-induced response heterogeneity is frequency-dependent.}
  \label{fig:frequency_domain_response}
\end{figure}

As illustrated in \cref{fig:frequency_domain_response}, history dependence can broaden both the magnitude and the phase of the conditioned response spectrum. The ordinary response is the coherent complex average of these points. It can therefore be suppressed not only because individual histories respond weakly, but also because substantial conditioned responses acquire different phase lags and partially cancel.

Panels (b)--(d) display the raw history-conditioned responses $\mathcal{R}_{21}^{Q}$. The spectral gap itself is determined by the intensity-normalized coefficients $\mathcal{K}_{21}^{Q}=\mathcal{R}_{21}^{Q}/\gamma_{21}$ and their intensity-weighted dispersion. A broad raw-response cloud alone is therefore not sufficient to imply a positive gap. In the illustrative model, the corresponding normalized event consequences also become more concentrated at high frequency.

\subsection{Stationary Fisher-Information and Kinetic Bounds}
\label{subsec:stationary_fisher_information_fri}
\label{subsec:finite_frequency_response_kur}

For a stationary perturbation direction whose only explicit time dependence is carried by the external protocol $h(t)$,
\begin{equation}
  \ln r_{ij}^{X_{t-},\epsilon h} = \ln r_{ij}^{X_{t-}} + \epsilon h(t) \alpha_{ij}^{X_{t-}} + O(\epsilon^2),
  \label{eq: general_log_rate_perturbation_frequency}
\end{equation}
the corresponding finite-frequency response is
\begin{equation}
  \mathcal{R}_{\alpha}^{Q}(\omega) = \sum_{i \neq j} \left\langle \alpha_{ij}^{X_{0-}} \mathcal{R}_{ij}^{Q}(\omega \!\mid\! X_{0-}) \right\rangle_{\mathrm{ss}}.
  \label{eq: general_frequency_response_alpha}
\end{equation}
For a weak harmonic perturbation $h(t)=h_{\omega}e^{-\mathrm{i}\omega t}$, the induced oscillation of the mean observable rate at the same frequency is proportional to $\mathcal{R}_{\alpha}^{Q}(\omega)$. The complex notation represents the two real quadratures. The physical perturbation is the real part, with a fixed amplitude convention.

The stationary path-space Fisher-information rate per unit protocol power is
\begin{equation}
  \mathcal{I}_{\alpha} = \sum_{i \neq j} \left\langle \gamma_{ij}(0) \left( \alpha_{ij}^{X_{0-}} \right)^2 \right\rangle_{\mathrm{ss}}.
  \label{eq: stationary_frequency_fisher_activity_alpha}
\end{equation}
The observable spectrum and the perturbation direction are two vectors in the same complex martingale-coordinate space. Their overlap is $\mathcal{R}_{\alpha}^{Q}(\omega)$, the squared norm of the observable coordinates is $\mathcal{S}_Q(\omega)$, and the squared norm of the perturbation direction is $\mathcal{I}_{\alpha}$. Complex Cauchy--Schwarz therefore gives
\begin{equation}
  \frac{\left| \mathcal{R}_{\alpha}^{Q}(\omega) \right|^2}{\mathcal{S}_Q(\omega)} \leq \mathcal{I}_{\alpha}.
  \label{eq: general_stationary_frequency_fri}
\end{equation}
The bound is saturated when
\begin{equation}
  \mathcal{K}_{ij}^{Q}(\omega \!\mid\! X_{0-}) = c(\omega)\alpha_{ij}^{X_{0-}}
  \label{eq: stationary_fisher_fri_equality_condition}
\end{equation}
for a history-independent complex constant $c(\omega)$ on the active edge--history support. This is an alignment condition between the observable mode and the selected perturbation direction. It is distinct from the history-averaging gap.

When the logarithmic sensitivity is bounded,
\begin{equation}
  \left| \alpha_{ij}^{X_{0-}} \right| \leq \alpha_{\max},
  \label{eq: bounded_logarithmic_sensitivity}
\end{equation}
the Fisher-information rate is bounded by the stationary dynamical activity rate,
\begin{equation}
  \mathcal{I}_{\alpha} \leq \alpha_{\max}^2 \mathcal{A}, \qquad \mathcal{A} = \sum_{i \neq j} \left\langle \gamma_{ij}(0) \right\rangle_{\mathrm{ss}}.
  \label{eq: stationary_fisher_activity_bound}
\end{equation}
Combining this result with \cref{eq: general_stationary_frequency_fri} gives
\begin{equation}
  \frac{\left| \mathcal{R}_{\alpha}^{Q}(\omega) \right|^2}{\mathcal{S}_Q(\omega)} \leq \alpha_{\max}^2 \mathcal{A}.
  \label{eq: stationary_finite_frequency_response_kur}
\end{equation}
This is the stationary finite-frequency R-KUR. For a fixed normalization of the control parameter and harmonic protocol, the activity rate provides the kinetic resource controlling response precision at frequency $\omega$. Saturation requires both the alignment condition in \cref{eq: stationary_fisher_fri_equality_condition} and saturation of $|\alpha_{ij}^{X_{0-}}|\leq\alpha_{\max}$ on the active support.

\subsection{Dynamical Two-frequency Relations and Frequency Conversion}
\label{subsec:dynamical_two_frequency_frr}

The stationary relations rely on time-translation invariance. For explicitly time-dependent dynamics or a finite-time driving protocol, the response retains its full two-time structure. We use $\omega$ for the output frequency of $Q$, $\omega'$ for the output frequency of a second observable $Q'$, and $\nu$ for the input frequency of an external perturbation. The resulting two-frequency theory is the finite-window extension of the stationary spectral hierarchy.

\subsubsection{Finite-window response and covariance spectra}

We consider a finite observation window $[0,\tau]$. The two-frequency spectral covariance of two additive observable rates is
\begin{align}
  \mathcal{S}_{QQ'}^{\tau}(\omega,\omega') ={}& \int_0^\tau \rmd t \int_0^\tau \rmd t' \, e^{\mathrm{i}\omega t} e^{-\mathrm{i}\omega' t'} \nonumber\\
  &\times \operatorname{Cov}\!\left( \dot{Q}(t),\dot{Q}'(t') \right).
  \label{eq: dynamical_two_frequency_spectrum}
\end{align}
For $Q'=Q$ and $\omega'=\omega$, this becomes the nonnegative finite-window spectral power
\begin{equation}
  \mathcal{S}_{Q}^{\tau}(\omega,\omega) = \int_0^\tau \rmd t \int_0^\tau \rmd t' \, e^{\mathrm{i}\omega(t-t')} \operatorname{Cov}\!\left( \dot{Q}(t),\dot{Q}(t') \right).
  \label{eq: dynamical_power_spectrum}
\end{equation}
This quantity is not divided by $\tau$ and should be distinguished from the stationary power spectral density.

Without time-translation invariance, the history-conditioned response depends separately on the observation time $t$ and perturbation time $s$,
\begin{equation}
  R_{ij}^{Q}(t,s \!\mid\! X_{s-}) \equiv \left. \frac{\partial}{\partial \epsilon} \frac{\delta}{\delta h(s)} \mathbb{E}_{\epsilon h}\!\left[ \dot{Q}(t) \!\mid\! X_{s-} \right] \right|_{\epsilon=0}, \quad t \geq s.
  \label{eq: dynamical_history_conditioned_response_kernel}
\end{equation}
The corresponding event-consequence kernel satisfies
\begin{equation}
  R_{ij}^{Q}(t,s \!\mid\! X_{s-}) = \gamma_{ij}(s) \, K_{ij}^{Q}(t,s \!\mid\! X_{s-}).
  \label{eq: dynamical_response_causal_influence}
\end{equation}
The finite-window output-frequency response and event-consequence kernel are
\begin{subequations}
\label{eq: dynamical_frequency_response_and_kernel}
\begin{align}
  \mathcal{R}_{ij}^{Q,\tau}(\omega,s \!\mid\! X_{s-}) &= \int_{[s,\tau]} e^{\mathrm{i}\omega t} \, R_{ij}^{Q}(t,s \!\mid\! X_{s-}) \,\rmd t,
  \label{eq: dynamical_history_conditioned_frequency_response} \\
  \mathcal{K}_{ij}^{Q,\tau}(\omega,s \!\mid\! X_{s-}) &= \int_{[s,\tau]} e^{\mathrm{i}\omega t} \, K_{ij}^{Q}(t,s \!\mid\! X_{s-}) \,\rmd t, \\
  \mathcal{R}_{ij}^{Q,\tau}(\omega,s \!\mid\! X_{s-}) &= \gamma_{ij}(s) \, \mathcal{K}_{ij}^{Q,\tau}(\omega,s \!\mid\! X_{s-}).
\end{align}
\end{subequations}
The averaged output-frequency response at perturbation time $s$ is
\begin{equation}
  \mathcal{R}_{ij}^{Q,\tau}(\omega,s) = \left\langle \mathcal{R}_{ij}^{Q,\tau}(\omega,s \!\mid\! X_{s-}) \right\rangle.
  \label{eq: dynamical_averaged_frequency_response}
\end{equation}

\subsubsection{Exact two-frequency FRR and history projection}

The contribution inherited from uncertainty in the initial state and initial memory is
\begin{equation}
  C_{0,QQ'}(t,t') = \operatorname{Cov}\!\left( \mathbb{E}\!\left[ \dot{Q}(t) \!\mid\! X_0 \right],\mathbb{E}\!\left[ \dot{Q}'(t') \!\mid\! X_0 \right] \right).
  \label{eq: dynamical_initial_covariance}
\end{equation}
Its finite-window transform is
\begin{equation}
  \mathcal{S}_{0,QQ'}^{\tau}(\omega,\omega') = \int_0^\tau \rmd t \int_0^\tau \rmd t' \, e^{\mathrm{i}\omega t} e^{-\mathrm{i}\omega' t'} C_{0,QQ'}(t,t').
  \label{eq: dynamical_initial_spectrum}
\end{equation}
The derivation in \cref{appsec: dynamical_two_frequency_frr_derivation} gives the exact two-frequency FRR,
\begin{widetext}
\begin{subequations}
\label{eq: dynamical_two_frequency_frr}
\begin{align}
  \mathcal{S}_{QQ'}^{\tau}(\omega,\omega') &= \mathcal{S}_{0,QQ'}^{\tau}(\omega,\omega') + \int_0^\tau \sum_{i \neq j} \left\langle \gamma_{ij}(s) \, \mathcal{K}_{ij}^{Q,\tau}(\omega,s \!\mid\! X_{s-}) \, \mathcal{K}_{ij}^{Q',\tau}(-\omega',s \!\mid\! X_{s-}) \right\rangle \,\rmd s \\
  &= \mathcal{S}_{0,QQ'}^{\tau}(\omega,\omega') + \int_0^\tau \sum_{i \neq j} \left\langle \frac{\mathcal{R}_{ij}^{Q,\tau}(\omega,s \!\mid\! X_{s-}) \, \mathcal{R}_{ij}^{Q',\tau}(-\omega',s \!\mid\! X_{s-})}{\gamma_{ij}(s)} \right\rangle \,\rmd s.
\end{align}
\end{subequations}
\end{widetext}
For real observables, $\mathcal{S}_{QQ'}^{\tau}(\omega,\omega')=\left[\mathcal{S}_{Q'Q}^{\tau}(\omega',\omega)\right]^*$. For $Q'=Q$ and $\omega'=\omega$, the dynamical power-spectrum FRR is
\begin{align}
  \mathcal{S}_{Q}^{\tau}(\omega,\omega) ={}& \mathcal{S}_{0,Q}^{\tau}(\omega,\omega) \nonumber \\
  &+ \int_0^\tau \sum_{i \neq j} \left\langle \frac{\left| \mathcal{R}_{ij}^{Q,\tau}(\omega,s \!\mid\! X_{s-}) \right|^2}{\gamma_{ij}(s)} \right\rangle \,\rmd s.
  \label{eq: dynamical_power_spectrum_frr}
\end{align}
For stationary dynamics and sufficiently long observation windows,
\begin{equation}
  \lim_{\tau \to \infty} \frac{1}{\tau} \mathcal{S}_{Q}^{\tau}(\omega,\omega) = \mathcal{S}_Q(\omega).
  \label{eq: dynamical_stationary_spectrum_limit}
\end{equation}

For $\left\langle\gamma_{ij}(s)\right\rangle>0$, define the event-weighted finite-window consequence
\begin{equation}
  \overline{\mathcal{K}}_{ij}^{Q,\tau}(\omega,s) \equiv \frac{\mathcal{R}_{ij}^{Q,\tau}(\omega,s)}{\left\langle \gamma_{ij}(s) \right\rangle}.
  \label{eq: dynamical_event_weighted_mean_consequence}
\end{equation}
The exact dynamical FRR then decomposes as
\begin{widetext}
\begin{subequations}
\label{eq: dynamical_frr_fri_decomposition}
\begin{align}
  \mathcal{S}_{Q}^{\tau}(\omega,\omega)-\mathcal{S}_{0,Q}^{\tau}(\omega,\omega) ={}& \int_0^\tau \sum_{i \neq j} \frac{\left| \mathcal{R}_{ij}^{Q,\tau}(\omega,s) \right|^2}{\left\langle \gamma_{ij}(s) \right\rangle} \,\rmd s + \mathcal{M}_{Q}^{\tau}(\omega), \\
  \mathcal{M}_{Q}^{\tau}(\omega) ={}& \int_0^\tau \sum_{i \neq j} \left\langle \gamma_{ij}(s)\left|\mathcal{K}_{ij}^{Q,\tau}(\omega,s \!\mid\! X_{s-})-\overline{\mathcal{K}}_{ij}^{Q,\tau}(\omega,s)\right|^2 \right\rangle \,\rmd s \geq 0.
  \label{eq: dynamical_frequency_memory_gap}
\end{align}
\end{subequations}
\end{widetext}
The gap resolves how history-dependent response heterogeneity is distributed over perturbation times and output frequencies. Since it is nonnegative, the finite-window edge-wise FRI follows directly,
\begin{equation}
  \mathcal{S}_{Q}^{\tau}(\omega,\omega) - \mathcal{S}_{0,Q}^{\tau}(\omega,\omega) \geq \int_0^\tau \sum_{i \neq j} \frac{\left| \mathcal{R}_{ij}^{Q,\tau}(\omega,s) \right|^2}{\left\langle \gamma_{ij}(s) \right\rangle} \,\rmd s.
  \label{eq: dynamical_edge_fri}
\end{equation}

\subsubsection{Frequency-conversion response and kinetic bounds}

Consider a time-dependent perturbation direction
\begin{equation}
  \ln r_{ij}^{X_{s-},\epsilon h}(s) = \ln r_{ij}^{X_{s-}}(s) + \epsilon h(s) \alpha_{ij}^{X_{s-}}(s) + O(\epsilon^2).
  \label{eq: dynamical_general_log_rate_perturbation}
\end{equation}
For a general real protocol, or its complex representation, the output-frequency response is
\begin{equation}
  \mathcal{R}_{\alpha,h}^{Q,\tau}(\omega) = \int_0^\tau h(s) \sum_{i \neq j} \left\langle \alpha_{ij}^{X_{s-}}(s) \mathcal{R}_{ij}^{Q,\tau}(\omega,s \!\mid\! X_{s-}) \right\rangle \,\rmd s.
  \label{eq: dynamical_general_protocol_response}
\end{equation}
The corresponding protocol-dependent Fisher information, or complexified score norm for a complex protocol representation, is
\begin{equation}
  \mathcal{I}_{\alpha,h}^{\tau} = \int_0^\tau |h(s)|^2 \sum_{i \neq j} \left\langle \gamma_{ij}(s) \left[ \alpha_{ij}^{X_{s-}}(s) \right]^2 \right\rangle \,\rmd s.
  \label{eq: dynamical_protocol_fisher_information}
\end{equation}
Complex Cauchy--Schwarz gives
\begin{equation}
  \frac{\left| \mathcal{R}_{\alpha,h}^{Q,\tau}(\omega) \right|^2}{\mathcal{S}_{Q}^{\tau}(\omega,\omega) - \mathcal{S}_{0,Q}^{\tau}(\omega,\omega)} \leq \mathcal{I}_{\alpha,h}^{\tau}.
  \label{eq: dynamical_general_protocol_fri}
\end{equation}
Equality requires the output-frequency event consequence to be proportional to $h(s)^*\alpha_{ij}^{X_{s-}}(s)$ over the active edge--history--time support.

For the unit-modulus harmonic convention $h(s)=e^{-\mathrm{i}\nu s}$, the response at output frequency $\omega$ is
\begin{align}
  \mathcal{R}_{\alpha}^{Q,\tau}(\omega,\nu) ={}& \int_0^\tau e^{-\mathrm{i}\nu s} \sum_{i \neq j} \left\langle \alpha_{ij}^{X_{s-}}(s) \mathcal{R}_{ij}^{Q,\tau}(\omega,s \!\mid\! X_{s-}) \right\rangle \,\rmd s.
  \label{eq: dynamical_two_frequency_response_alpha}
\end{align}
Since $|e^{-\mathrm{i}\nu s}|=1$, the corresponding score norm is
\begin{equation}
  \mathcal{I}_{\alpha}^{\tau} = \int_0^\tau \sum_{i \neq j} \left\langle \gamma_{ij}(s) \left[ \alpha_{ij}^{X_{s-}}(s) \right]^2 \right\rangle \,\rmd s.
  \label{eq: dynamical_fisher_activity_alpha}
\end{equation}
The two-frequency FRI is therefore
\begin{equation}
  \frac{\left| \mathcal{R}_{\alpha}^{Q,\tau}(\omega,\nu) \right|^2}{\mathcal{S}_{Q}^{\tau}(\omega,\omega) - \mathcal{S}_{0,Q}^{\tau}(\omega,\omega)} \leq \mathcal{I}_{\alpha}^{\tau}.
  \label{eq: dynamical_general_fri}
\end{equation}
A physical cosine or sine drive is the corresponding real quadrature. Its Fisher information contains the actual squared real protocol and therefore differs by the chosen amplitude normalization.

If the logarithmic sensitivity is bounded during the protocol,
\begin{equation}
  \left| \alpha_{ij}^{X_{s-}}(s) \right| \leq \alpha_{\max},
  \label{eq: dynamical_bounded_logarithmic_sensitivity}
\end{equation}
then
\begin{subequations}
\label{eq: dynamical_fisher_activity_activity_bound}
\begin{align}
  \mathcal{I}_{\alpha,h}^{\tau} &\leq \alpha_{\max}^2 \mathcal{A}_{h,\tau}, \\
  \mathcal{A}_{h,\tau} &= \int_0^\tau |h(s)|^2 \sum_{i \neq j} \left\langle \gamma_{ij}(s) \right\rangle \,\rmd s.
  \label{eq: dynamical_protocol_weighted_activity}
\end{align}
\end{subequations}
For the unit-modulus harmonic convention, $\mathcal{A}_{h,\tau}$ reduces to the total activity
\begin{equation}
  \mathcal{A}_{\tau} = \int_0^\tau \sum_{i \neq j} \left\langle \gamma_{ij}(s) \right\rangle \,\rmd s.
  \label{eq: dynamical_total_activity}
\end{equation}
Combining these results gives the dynamical two-frequency R-KUR,
\begin{equation}
  \frac{\left| \mathcal{R}_{\alpha}^{Q,\tau}(\omega,\nu) \right|^2}{\mathcal{S}_{Q}^{\tau}(\omega,\omega) - \mathcal{S}_{0,Q}^{\tau}(\omega,\omega)} \leq \alpha_{\max}^2 \mathcal{A}_{\tau}.
  \label{eq: dynamical_two_frequency_response_kur}
\end{equation}

In a stationary long-time experiment, time-translation invariance makes the infinite-window response diagonal in frequency. A finite window broadens this diagonal through a known window kernel and can therefore produce off-diagonal leakage even for stationary dynamics. Genuine frequency conversion is the additional input--output mixing that cannot be accounted for by this stationary finite-window benchmark. The explicit window formula is derived in \cref{appsec: finite_window_leakage}.

The initial spectrum $\mathcal{S}_{0,Q}^{\tau}(\omega,\omega)$ is subtracted because it originates from uncertainty in the initial state and prehistory rather than from martingale fluctuations generated during the protocol. If the complete initial condition is fixed, this term vanishes.

The frequency-domain theory therefore extends the time-domain hierarchy in two directions. In stationary dynamics, it resolves the gain and phase of the event consequence at each frequency. In nonstationary dynamics, it resolves how an input frequency is converted into output frequencies. The same frequency-resolved gap will be used in the next section to test whether a retained memory state is sufficient on fast and slow timescales separately.

\section{Application: Response-Sufficient Coarse-Graining and Experimental Memory Separation}
\label{sec:response_sufficient_coarse_graining}

Coarse-graining is a common source of non-Markovianity. Even when the microscopic dynamics is Markovian, an observed state may collect several unresolved configurations. Two trajectories that reach the same observed state can then have different future jump statistics and different responses because they imply different distributions over the hidden configurations. Matching stationary occupations, mean fluxes, or residence-time distributions is therefore not sufficient to guarantee that a reduced state preserves the response and fluctuations of a trajectory observable.

Here we use the FRR-FRI gap as a diagnostic of a proposed memory state. For a chosen observable $Q_\tau$, we specify a physically motivated history variable $Y_{t-}=\Phi_t(X_{t-})$, calculate the response conditioned on $Y_{t-}$, and compare the resulting FRI reconstruction with the measured variance. The remaining gap quantifies the response-relevant memory not retained by that candidate state. This is deliberately a hypothesis-driven procedure: we test and rank proposed memory coordinates rather than infer an optimal $Y_{t-}$ directly from an arbitrary trajectory.

We develop this application in three stages. We first formulate the general response-sufficiency test and its monotone hierarchy under refinement of the retained history. We then illustrate the construction in a solvable coarse-grained Markov network, where the exact memory coordinate is analytically known. Finally, we extract the intrinsic event-consequence kernel from spontaneous cell-membrane and mitochondrial BK-channel trajectories.

The martingale representation has a key practical consequence. The normalized kernel $K_{ij}^{Q}$ needs no externally imposed perturbation. It is the change in the conditional prediction of a future observable when a jump is appended to a given pre-jump history. The kernel can therefore be inferred directly from unperturbed trajectories by comparing the event branch with the corresponding no-jump continuation branch. The experimental trajectories reveal an opposite ordering. Cell-membrane BK channels exhibit the stronger conventional dwell-time memory. Mitochondrial BK channels retain a much larger beyond-age dependence of the future consequence of individual gating events. The frequency-resolved analysis shows that this mitoBK residual is enhanced at low frequency but remains appreciable throughout the resolved band. Cell-BK approaches the shuffled baseline outside a weak and heterogeneous slow component.

\subsection{The FRR-FRI Gap as a Test of Candidate Memory States}
\label{subsec:response_sufficient_states}

To see how memory arises under coarse-graining, let $z_t$ be a microscopic Markov state with filtration $\mathcal{F}_{t-}$ and let $x_t=c(z_t)$ be the observed state. We denote the microscopic transition rate from $\nu$ to $\mu$ by $k_{\mu\nu}$ and define the total microscopic rate from $\nu$ into the observed block $\Omega_i$ as
\begin{equation}
  \kappa_{i\nu} \equiv \sum_{\mu \in \Omega_i} k_{\mu\nu}.
  \label{eq: microscopic_rate_into_observed_block}
\end{equation}
If $\Omega_j$ is the microscopic block associated with $x_t=j$, the observed rate for a jump $j\to i$ is
\begin{subequations}
\label{eq: coarse_grained_history_dependent_rate}
\begin{align}
  r_{ij}^{X_{t-}} &= \sum_{\nu \in \Omega_j} \pi_\nu(t \!\mid\! X_{t-}) \, \kappa_{i\nu}, \\
  \pi_\nu(t \!\mid\! X_{t-}) &= P(z_{t-}=\nu \!\mid\! X_{t-}).
\end{align}
\end{subequations}
The hidden posterior $\pi_\nu(t \!\mid\! X_{t-})$ is sufficient to predict the full observed future, but it is generally larger than needed for a particular observable and perturbation.

We therefore consider a candidate memory state satisfying
\begin{equation}
  \sigma(x_{t-}) \subseteq \sigma(Y_{t-}) \subseteq \mathcal{F}_{t-}.
  \label{eq: candidate_memory_information_order}
\end{equation}
Here $\sigma(Z)$ denotes the sigma algebra generated by a random variable $Z$. Physically, it represents all information accessible from knowing $Z$; hence, \cref{eq: candidate_memory_information_order} states that $Y_{t-}$ retains at least the current-state information contained in $x_{t-}$, but no information beyond the complete observed history available immediately before time $t$. Conditioning the intensity and the edge-wise response on this state gives
\begin{subequations}
\label{eq: reduced_state_conditioned_quantities}
\begin{align}
  \gamma_{ij}(t \!\mid\! Y_{t-}) &= \mathbb{E}\left[ \, \gamma_{ij}(t) \!\mid\! Y_{t-} \, \right], \\
  R_{ij}^{Q}(\tau,t \!\mid\! Y_{t-}) &= \mathbb{E}\left[ \, R_{ij}^{Q}(\tau,t \!\mid\! X_{t-}) \,\Big|\, Y_{t-} \, \right].
\end{align}
\end{subequations}
The fluctuation reconstructed from these conditioned responses is
\begin{equation}
  \mathcal{V}_Q^{(Y)} \equiv \int_0^\tau \sum_{i \neq j} \left\langle \frac{\left[ R_{ij}^{Q}(\tau,t \!\mid\! Y_{t-}) \right]^2}{\gamma_{ij}(t \!\mid\! Y_{t-})} \right\rangle \,\rmd t,
  \label{eq: reduced_state_frr_reconstruction}
\end{equation}
where ratios are set to zero off the support of the corresponding intensity. The exact FRR then gives
\begin{equation}
  \operatorname{Var}(Q_\tau)-C_0(Q,Q) = \mathcal{V}_Q^{(Y)}+\mathcal{M}_Q^{(Y)}, \qquad \mathcal{M}_Q^{(Y)}\geq0.
  \label{eq: reduced_state_response_memory_decomposition}
\end{equation}
Writing
\begin{subequations}
\label{eq: reduced_state_normalized_responses}
\begin{align}
  K_{ij}^{Q}(\tau,t \!\mid\! X_{t-}) &\equiv \frac{R_{ij}^{Q}(\tau,t \!\mid\! X_{t-})}{\gamma_{ij}(t)}, \\
  K_{ij}^{Q}(\tau,t \!\mid\! Y_{t-}) &\equiv \frac{R_{ij}^{Q}(\tau,t \!\mid\! Y_{t-})}{\gamma_{ij}(t \!\mid\! Y_{t-})},
\end{align}
\end{subequations}
the reduced coefficient is the activity-weighted projection of the full-history coefficient,
\begin{equation}
  K_{ij}^{Q}(\tau,t \!\mid\! Y_{t-}) = \frac{\mathbb{E}\left[ \, \gamma_{ij}(t) \, K_{ij}^{Q}(\tau,t \!\mid\! X_{t-}) \,\Big|\, Y_{t-} \, \right]}{\mathbb{E}\left[ \, \gamma_{ij}(t) \!\mid\! Y_{t-} \, \right]}.
  \label{eq: reduced_state_activity_weighted_projection}
\end{equation}
The residual takes the explicit form
\begin{align}
\label{eq: reduced_state_residual_memory_gap}
  \mathcal{M}_Q^{(Y)} = \int_0^\tau \sum_{i \neq j} \Bigg\langle \gamma_{ij}(t) &\Big[ K_{ij}^{Q}(\tau,t \!\mid\! X_{t-}) \\
  &- K_{ij}^{Q}(\tau,t \!\mid\! Y_{t-}) \Big]^2 \Bigg\rangle \, \rmd t. \nonumber
\end{align}
Thus, $\mathcal{M}_Q^{(Y)}$ measures the part of the fluctuation arising from histories that are merged by $Y_{t-}$ but for which the same jump has different future consequences. We call $Y_{t-}$ \emph{response-sufficient} for the observable $Q_\tau$, the resolved edge-perturbation family, and the observation horizon $\tau$ when $\mathcal{M}_Q^{(Y)}=0$. Physically, response sufficiency means the following. Once $Y_{t-}$ is specified, the merged histories become indistinguishable for the chosen response problem. For every resolved transition and perturbation time, the same jump has the same marginal consequence for $Q_\tau$. The unresolved past may still affect other observables or other aspects of the future trajectory. But it carries no information needed to reconstruct the response and the fluctuations of $Q_\tau$. In this sense, $Y_{t-}$ provides a task-specific closure at the level of response. It acts as an effective state for the chosen observable and perturbation. It need not be a complete predictive state \cite{shalizi2001computational,marzen2017structure}, a Markovian embedding, or a unique reconstruction of the hidden dynamics. For a physical perturbation whose logarithmic sensitivity $\alpha_{ij}^{X_{t-}}$ is itself history-dependent, response sufficiency additionally requires the relevant sensitivity to be determined by the retained state $Y_{t-}$.

When $Y_{t-}=x_{t-}$, the edge label already fixes the pre-jump state, and $\mathcal{V}_Q^{(x)}$ reduces to the ordinary edge-wise FRI. At the opposite extreme, retaining the full history recovers the exact FRR. Hence,
\begin{equation}
  \underbrace{\int_0^\tau \sum_{i \neq j} \frac{\left[ R_{ij}^{Q}(\tau,t) \right]^2}{\left\langle \gamma_{ij}(t) \right\rangle} \,\rmd t}_{\mathcal{V}_Q^{(x)}\;\text{(ordinary FRI)}} \leq \mathcal{V}_Q^{(Y)} \leq \underbrace{\operatorname{Var}(Q_\tau)-C_0(Q,Q)}_{\mathcal{V}_Q^{(X)}\;\text{(exact FRR)}}.
  \label{eq: frr_fri_candidate_state_hierarchy}
\end{equation}
If $Y'$ retains all information in $Y$, i.e., $\sigma(Y'_{t-}) \supseteq \sigma(Y_{t-})$, then we have the hierarchical relation
\begin{equation}
  \mathcal{M}_Q^{(Y')} \leq \mathcal{M}_Q^{(Y)}.
  \label{eq: candidate_state_gap_monotonicity}
\end{equation}
The corresponding orthogonal decomposition and proof are given in \cref{appsec: response_sufficiency_monotonicity}. Adding a useful history coordinate therefore closes a definite portion of the FRR-FRI gap. For comparison across observables, we use the residual fraction
\begin{equation}
  \eta_Q^{(Y)} \equiv \frac{\mathcal{M}_Q^{(Y)}}{\operatorname{Var}(Q_\tau)-C_0(Q,Q)}.
  \label{eq: residual_response_memory_fraction}
\end{equation}

Operationally, one first proposes a small set of interpretable candidates, such as the present state, residence age, previous state, or a filtered history coordinate. A central advantage of the martingale formulation is that the normalized kernel $K_{ij}^{Q}(\tau,t \!\mid\! X_{t-})$ is an intrinsic property of the spontaneous trajectory distribution. For a fixed pre-jump history, it is the change in the conditional prediction of $Q_\tau$ produced by appending the jump $j\to i$ at time $t$, relative to continuing the trajectory without a jump at that time. It can therefore be evaluated from unperturbed trajectories, either directly in a known model or by estimating the corresponding conditional event and continuation ensembles from data. Multiplication by the stochastic intensity gives the response to the elementary edge log-rate perturbation, \begin{equation}
  R_{ij}^{Q}(\tau,t \!\mid\! X_{t-}) = \gamma_{ij}(t) \, K_{ij}^{Q}(\tau,t \!\mid\! X_{t-}).
  \label{eq: intrinsic_kernel_edge_response}
\end{equation}
Therefore, the hierarchy of candidate memory states can be tested directly from spontaneous trajectories even without applying a controlled perturbation. A response to a specific physical control parameter additionally requires its logarithmic sensitivity $\alpha_{ij}^{X_{t-}}$. An analogous hierarchy can be formulated frequency by frequency using the finite-frequency FRR.

\begin{figure*}[t]
  \centering
  \includegraphics[width=0.95\textwidth]{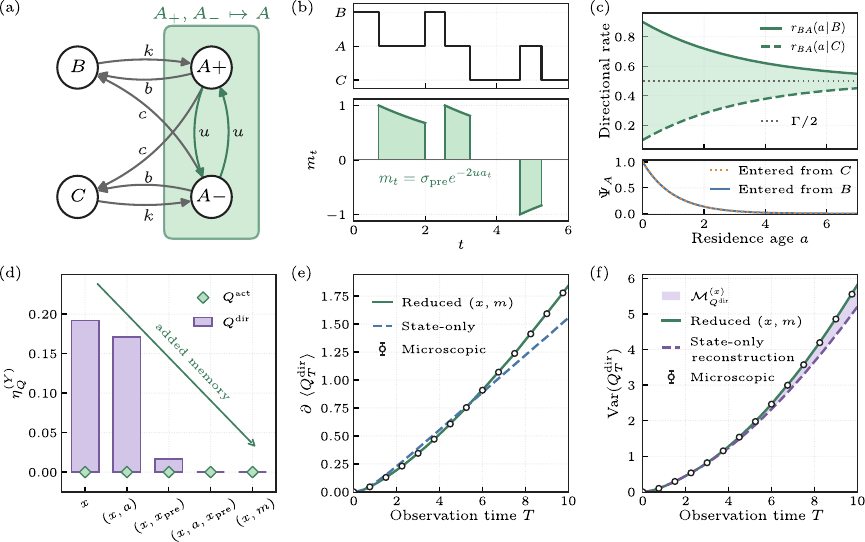}
  \caption{\textbf{Testing candidate memory states by the FRR-FRI gap.}
  (a) Microscopic four-state Markov network and the projection $A_+, A_- \mapsto A$.
  (b) An observed trajectory and the polarization $m_t=\sigma_{\mathrm{pre}}e^{-2ua_t}$ reconstructed from the previous state and residence age.
  (c) The directional rates out of $A$ retain entry-history dependence even though the survival probability is exactly $e^{-\Gamma a}$.
  (d) Residual fractions for several candidate states. The current state is sufficient for the outgoing-activity observable $Q^{\mathrm{act}}$, whereas $Q^{\mathrm{dir}}$ requires both the previous state and the residence age, or equivalently $m_t$.
  (e) Directional response of the microscopic process, the response-sufficient dynamics $(x_t,m_t)$, and the state-only Markov surrogate.
  (f) Variance of $Q_\tau^{\mathrm{dir}}$ and its response reconstruction. The shaded region is the FRR-FRI gap left by the current-state projection of the true coarse-grained dynamics.
  Parameters are $k=0.5$, $b=0.9$, $c=0.1$, and $u=0.15$.}
  \label{fig:response_sufficient_coarse_graining}
\end{figure*}

\subsection{A Solvable Coarse-Grained Network Model}
\label{subsec:minimal_memory_solvable_network}

We illustrate the test with the microscopic Markov network in \cref{fig:response_sufficient_coarse_graining}(a):
\begin{subequations}
\label{eq: solvable_microscopic_network}
\begin{align}
  &B \underset{b}{\overset{k}{\rightleftarrows}} A_+, \qquad C \underset{b}{\overset{k}{\rightleftarrows}} A_-, \qquad A_+ \underset{u}{\overset{u}{\rightleftarrows}} A_-, \nonumber \\
  &A_+ \xrightarrow{c} C, \qquad A_- \xrightarrow{c} B. \nonumber
\end{align}
\end{subequations}
We assume $b>c>0$. The states $A_+$ and $A_-$ are unresolved and are observed as a single state $A$. Let $\Gamma=b+c$ and $\delta=b-c$. Since both hidden states have the same total escape rate $\Gamma$,
\begin{equation}
  P(\tau_A>s \!\mid\! \text{entry history}) = e^{-\Gamma s}.
  \label{eq: exponential_residence_time_A}
\end{equation}
The residence times in $B$ and $C$ are also exponential with rate $k$. A state-only Markov surrogate with rates $r_{AB}^{\mathrm{M}}=r_{AC}^{\mathrm{M}}=k$ and $r_{BA}^{\mathrm{M}}=r_{CA}^{\mathrm{M}}=\Gamma/2$ therefore reproduces all single-state residence-time distributions, stationary occupations, mean fluxes, and total activity.

The unresolved state is summarized by the polarization
\begin{subequations}
\label{eq: hidden_polarization_observed_history}
\begin{align}
  m_t &= P(z_{t-}=A_+ \!\mid\! X_{t-})-P(z_{t-}=A_- \!\mid\! X_{t-}) \\
  &= \sigma_{\mathrm{pre}}(t)e^{-2ua_t}, \qquad x_{t-}=A,
\end{align}
\end{subequations}
where $a_t$ is the residence age and $\sigma_{\mathrm{pre}}=+1$ or $-1$ depending on whether $A$ was entered from $B$ or $C$. The directional rates are
\begin{equation}
  r_{BA}^{X_{t-}} = \frac{\Gamma}{2}+\frac{\delta}{2}m_t, \qquad r_{CA}^{X_{t-}} = \frac{\Gamma}{2}-\frac{\delta}{2}m_t.
  \label{eq: solvable_coarse_grained_intensities}
\end{equation}
Their sum is history independent, whereas their difference is $\delta m_t$. Hence, exponential waiting times hide a memory of the next jump direction.

We compare four candidate states,
\begin{subequations}
\label{eq: solvable_candidate_memory_states}
\begin{align}
  Y^{(0)} &= x_{t-}, \\
  Y^{(a)} &= (x_{t-},a_t), \\
  Y^{(\mathrm{pre})} &= (x_{t-},x_{\mathrm{pre}}), \\
  Y^\star &= (x_{t-},a_t,x_{\mathrm{pre}}),
\end{align}
\end{subequations}
and two observables,
\begin{subequations}
\label{eq: activity_and_directional_observables}
\begin{align}
  Q_\tau^{\mathrm{act}} &= n_{BA}(\tau)+n_{CA}(\tau), \\ Q_\tau^{\mathrm{dir}} &= n_{BA}(\tau)-n_{CA}(\tau).
\end{align}
\end{subequations}
Within the observed state $A$ and for $u>0$, the pair $(a_t,x_{\mathrm{pre}})$ is equivalent to the polarization $m_t$, so that $Y^\star\simeq(x_{t-},m_t)$. The outgoing activity from $A$, $Q_\tau^{\mathrm{act}}$, is insensitive to $m_t$, so the current state already closes the FRR-FRI gap. The directional count is different: the age determines the magnitude of the remaining polarization but not its sign, while the previous state determines the sign but not its decay. For the nondegenerate parameters used in \cref{fig:response_sufficient_coarse_graining}, direct evaluation gives
\begin{subequations}
\label{eq: candidate_state_gap_comparison}
\begin{align}
  &\eta_{Q^{\mathrm{act}}}^{(x)}=0, \\
  &\eta_{Q^{\mathrm{dir}}}^{(Y^{(0)})} > \eta_{Q^{\mathrm{dir}}}^{(Y^{(a)})} > \eta_{Q^{\mathrm{dir}}}^{(Y^{(\mathrm{pre})})} > \eta_{Q^{\mathrm{dir}}}^{(Y^\star)} = 0.
\end{align}
\end{subequations}
Thus, the gap ranks the proposed memory variables and identifies $(x_t,m_t)$ as the first response-sufficient state within this candidate family for $Q_\tau^{\mathrm{dir}}$ and the chosen directional perturbation.

To probe the directional response, we perturb the two rates out of $A$ by opposite logarithmic fields,
\begin{subequations}
\label{eq: directional_field_perturbation}
\begin{align}
  \ln r_{BA}^{X_{t-},\epsilon} &= \ln r_{BA}^{X_{t-}}+\frac{\epsilon}{2}, \\
  \ln r_{CA}^{X_{t-},\epsilon} &= \ln r_{CA}^{X_{t-}}-\frac{\epsilon}{2}.
\end{align}
\end{subequations}
The corresponding martingale score is
\begin{equation}
  \Lambda_\tau^{\mathrm{dir}} = \frac{1}{2}Q_\tau^{\mathrm{dir}}-\frac{\delta}{2}\int_0^\tau m_t\mathds{1}_{x_{t-}=A} \,\rmd t.
  \label{eq: directional_field_score}
\end{equation}
The second term is the memory correction missed by the state-only Markov surrogate. As shown in \cref{fig:response_sufficient_coarse_graining}(e,f), the microscopic dynamics and the reduced dynamics $(x_t,m_t)$ give the same response and variance reconstruction. The state-only projection of the true coarse-grained dynamics leaves a positive FRR-FRI gap, while the state-only Markov surrogate fails to reproduce the true directional response and variance reconstruction.

This example shows that a candidate state can reproduce standard kinetic statistics and still fail a response-based test, and that the required state depends on the observable and perturbation. We next apply the same organizing principle to experimental open-closed trajectories without reconstructing a unique microscopic network.

\subsection{Statistical and Response-Relevant Memory in BK-Channel Gating}
\label{subsec:experimental_bk_response_memory}

Do conventional dwell-time correlations identify the history needed to predict what a gating event changes? We address this question using single-channel patch-clamp trajectories of cell-membrane BK and mitochondrial BK channels \cite{borys2022new,agata_wawrzkiewicz_jalowiecka_2022_6340407}. The two datasets give a direct negative answer. Cell-BK channels exhibit stronger conventional dwell-time correlations, whereas mitoBK channels retain a much larger dependence of event consequences on history beyond the present state and residence age.

\begin{figure*}[htbp]
  \centering
  \includegraphics[width=0.95\textwidth]{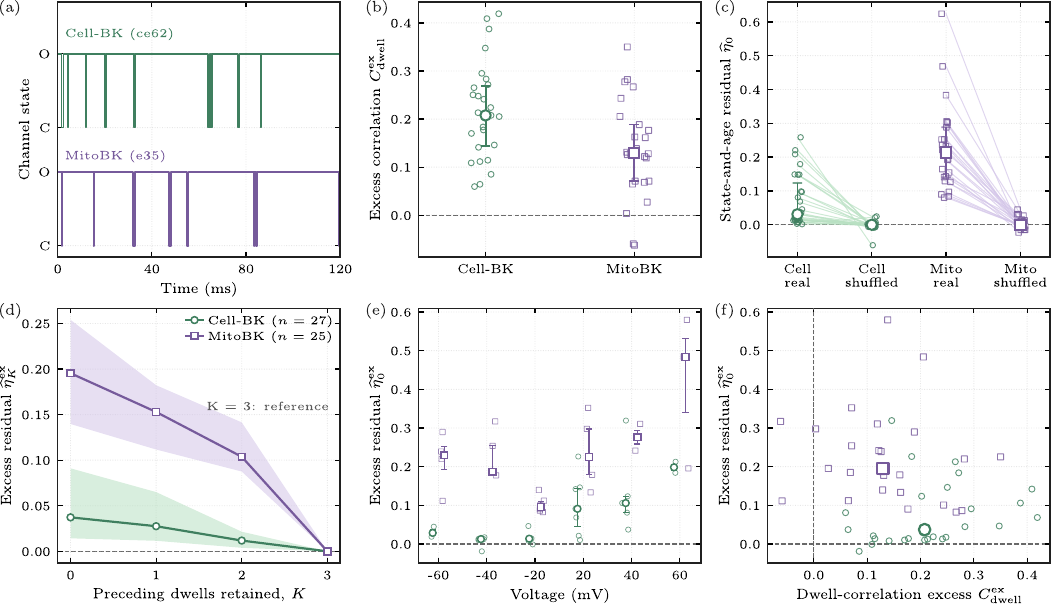}
  \caption{\textbf{Statistical and response-relevant memory in experimental BK-channel trajectories.}
  (a) Illustrative reconstructed cell-BK (ce62) and mitoBK (e35) open-closed trajectory records at $+20$ mV.
  (b) Conventional dwell-correlation excess relative to a state-wise shuffled control.
  (c) Paired real and shuffled state-and-age residuals. Thin lines connect estimates from the same patch.
  (d) Excess residual as preceding dwell durations are added to the candidate state. The three-dwell reference is zero by construction. Shading denotes patch-bootstrap $95\%$ intervals for group medians.
  (e) Voltage-resolved excess residual.
  (f) Dwell-correlation excess versus the state-and-age excess residual. Each point denotes one patch; thick symbols denote patch-level medians.}
  \label{fig:experimental_bk_response_memory}
\end{figure*}

An experimentally resolved open--closed ion-channel trajectory is itself a coarse-grained description of many microscopic conformational states \cite{colquhoun1981stochastic}. Whether such records are Markovian has been debated since early fractal-versus-Markov analyses of single-channel kinetics \cite{liebovitch1987ion,mcmanus1988fractal,korn1988statistical}. The dataset contains 30 cell-BK patches and 26 mitoBK patches recorded at $-60$, $-40$, $-20$, $20$, $40$, and $60\,\mathrm{mV}$ with a sampling frequency of $10\,\mathrm{kHz}$. The reconstructed trajectories contain $433\,523$ cell-BK dwells and $100\,395$ mitoBK dwells. Because each record consists of alternating open and closed dwell durations, it defines a two-state continuous-time trajectory
\begin{equation}
  C \to O \to C \to O \to \cdots.
  \label{eq: experimental_bk_alternating_trajectory}
\end{equation}
The supplied records contain thresholded dwell durations rather than the original current traces. Our analysis therefore concerns memory retained by the experimentally resolved two-state trajectory. Representative reconstructed trajectories are shown in \cref{fig:experimental_bk_response_memory}(a). The panel illustrates the alternating two-state records used in both the dwell-correlation and event--continuation analyses.

\subsubsection{Experimental Construction}
\label{subsubsec:experimental_bk_finite_history_memory}

We translate the response-sufficiency hierarchy into three experimental objects. A candidate memory state specifies which parts of the past are retained. A future open-time observable specifies the consequence of a gating event. An event--continuation comparison estimates how that consequence changes when an opening or closing event occurs.

For each patch, we consider the open time accumulated during a future window of length $\Delta$,
\begin{equation}
  Q_{\Delta}^{\mathrm{open}}(t) \equiv \int_t^{t+\Delta}\mathds{1}_{x_s=O} \,\rmd s.
  \label{eq: experimental_bk_future_open_time}
\end{equation}
Up to the approximately constant single-channel open current, $Q_{\Delta}^{\mathrm{open}}$ is proportional to the ionic charge transported during the future observation window. We use the normalized output $q_{\Delta}(t)\equiv Q_{\Delta}^{\mathrm{open}}(t)/\Delta$. The window length is chosen separately for each patch as four times the sum of its median open and closed dwell durations. The observable therefore covers several subsequent gating cycles while remaining local relative to the full record.

We compare the nested candidate states
\begin{equation}
  Y_{\mathrm{BK},t-}^{(K)} = \left( x_{t-},a_t,\tau_{-1},\ldots,\tau_{-K} \right), \qquad K=0,1,2,3,
  \label{eq: experimental_bk_candidate_memory_states}
\end{equation}
where $a_t$ is the age of the current dwell and $\tau_{-k}$ is the $k$th preceding completed dwell. Thus, $Y_{\mathrm{BK}}^{(0)}$ retains only the present state and residence age. The remaining candidates progressively retain the preceding dwell sequence. We use $Y_{\mathrm{BK}}^{(3)}$ as a finite rich-history reference. It is not assumed to resolve the complete observed history.

At a fixed retained history and fixed future-window length $\Delta$, we compare an opening or closing transition at time $t$ with continuation without a transition at that instant. All subsequent transitions follow the unperturbed dynamics on both branches. Let $e=(j\to i)$ denote either $C\to O$ or $O\to C$. Relative to the finite rich-history state $Y_{\mathrm{BK},t-}^{(3)}$, the observational event-consequence coefficient is
\begin{align}
  K_{e,\Delta}^{(3)}(Y_{\mathrm{BK},t-}^{(3)}) ={}& \mathbb{E}\left[ q_{\Delta}(t) \,\middle|\, Y_{\mathrm{BK},t-}^{(3)},x_{t+}=i \right] \nonumber\\
  &-\mathbb{E}\left[ q_{\Delta}(t) \,\middle|\, Y_{\mathrm{BK},t-}^{(3)},x_{t+}=j \right].
  \label{eq: experimental_bk_observational_event_consequence}
\end{align}
Both expectations are under the unperturbed trajectory distribution. Here $x_{t+}=i$ and $x_{t+}=j$ denote the immediate event and no-jump branches, respectively. The coefficient is therefore a difference between conditional predictions of subsequent open time.

Let $K_{e,\Delta}^{(K)}$ denote the activity-weighted projection of $K_{e,\Delta}^{(3)}$ onto $Y_{\mathrm{BK}}^{(K)}$. Pooling the two edge classes with their empirical event frequencies, we define the finite-history residual fraction
\begin{equation}
  \eta_K^{(3)} \equiv \frac{\left\langle \left[ K_{e,\Delta}^{(3)}-K_{e,\Delta}^{(K)} \right]^2 \right\rangle_{\mathrm{ev}}}{\left\langle \left[ K_{e,\Delta}^{(3)} \right]^2 \right\rangle_{\mathrm{ev}}}.
  \label{eq: experimental_bk_finite_history_residual}
\end{equation}
This quantity is the observational finite-history counterpart of \cref{eq: residual_response_memory_fraction}. By construction, $\eta_3^{(3)}=0$. This equality only defines the three-dwell reference and does not imply that three preceding dwells form an exact response-sufficient state.

To distinguish serial ordering from complex single-dwell statistics, we independently permute the open dwells and the closed dwells within each patch. This state-wise shuffled control preserves the marginal open- and closed-dwell distributions, the open fraction, and the total recording time. It removes their serial order. We report the finite-history excess residual
\begin{equation}
  \widehat{\eta}_{K}^{\mathrm{ex}} \equiv \widehat{\eta}_{K}^{\mathrm{data}}-\widehat{\eta}_{K}^{\mathrm{shuf}}.
  \label{eq: experimental_bk_excess_residual}
\end{equation}
The quantities measured below are finite-history observational counterparts of the intrinsic response-memory hierarchy. The conditional means, activity-weighted projections, split-sample debiasing, quality-control criteria, and statistical tests are detailed in \cref{appsec: experimental_bk_analysis}.

\subsubsection{Opposite Ordering and Memory Depth}
\label{subsubsec:experimental_bk_memory_separation}

After quality control, 52 of the 56 patches remain in the fixed-horizon comparison, comprising 27 cell-BK and 25 mitoBK patches. We first quantify conventional statistical memory using the mean correlation between successive same-state dwell durations,
\begin{align}
  C_{\mathrm{dwell}} = \frac{1}{2}\Big[ &\operatorname{corr}\left( \log\tau_{C,n},\log\tau_{C,n+1} \right) \nonumber\\
  &+\operatorname{corr}\left( \log\tau_{O,n},\log\tau_{O,n+1} \right) \Big],
  \label{eq: experimental_bk_conventional_memory}
\end{align}
and subtract the corresponding state-wise shuffled value. The patch-level median excess correlation is
\begin{equation}
  \operatorname{median} C_{\mathrm{dwell}}^{\mathrm{ex}} =
  \begin{cases}
    0.2077, & \text{cell-BK},\\
    0.1290, & \text{mitoBK}.
  \end{cases}
  \label{eq: experimental_bk_statistical_memory_ordering}
\end{equation}
Thus, the conventional lag-one same-state dwell memory is stronger in cell-BK channels $(p=0.00671$, one-sided Mann--Whitney test$)$.

The response-relevant ordering is the opposite. For the state-and-age candidate $Y_{\mathrm{BK}}^{(0)}=(x_{t-},a_t)$, the median finite-history excess residual is
\begin{equation}
  \operatorname{median}\widehat{\eta}_{0}^{\mathrm{ex}} =
  \begin{cases}
    0.0373, & \text{cell-BK},\\
    0.1954, & \text{mitoBK}.
  \end{cases}
  \label{eq: experimental_bk_opposite_memory_ordering}
\end{equation}
The mitoBK residual is larger than the cell-BK residual with $p=2.79\times10^{-6}$ (one-sided Mann--Whitney test). The original residual also exceeds the state-wise shuffled control in all 25 quality-controlled mitoBK patches $(p=5.96\times10^{-8}$, exact two-sided paired sign test$)$. The strong mitoBK signal therefore cannot be explained by its marginal open- and closed-dwell distributions alone.

The opposite ordering is displayed directly in \cref{fig:experimental_bk_response_memory}(b,c). Panel (b) shows the real-minus-shuffled dwell correlation. This conventional memory signal is larger in cell-BK. Panel (c) instead compares the state-and-age residual in each original trajectory with its state-wise shuffled control. The mitoBK patches exhibit a large and consistent reduction after shuffling, whereas the corresponding cell-BK change is smaller and more heterogeneous. The two observables therefore do not provide different normalizations of one common memory signal; they identify different aspects of the trajectory history.

The hierarchy identifies how much of the finite-history dependence is retained by preceding dwell durations. For mitoBK, the patch-level median excess residual decreases as
\begin{equation}
  \widehat{\eta}_{0}^{\mathrm{ex}}=0.1954, \qquad
  \widehat{\eta}_{1}^{\mathrm{ex}}=0.1529, \qquad
  \widehat{\eta}_{2}^{\mathrm{ex}}=0.1038,
  \label{eq: experimental_bk_memory_hierarchy_values}
\end{equation}
when the first and second preceding dwells are added. The corresponding cell-BK residual remains small. The preceding dwell sequence therefore retains part of the response-relevant history in mitoBK, but the first two preceding dwells do not exhaust the residual relative to the three-dwell reference.

\Cref{fig:experimental_bk_response_memory}(d) displays this candidate-state hierarchy. The mitoBK residual decreases each time a preceding dwell is added, while the cell-BK residual remains small throughout the hierarchy. The final point is the three-dwell reference and is zero by construction. The decrease therefore quantifies how much of the finite-history residual is closed by each added dwell; it does not establish that three preceding dwells form a complete response-sufficient state.

The remaining panels establish robustness. The voltage-resolved medians are shown in \cref{fig:experimental_bk_response_memory}(e). The mitoBK excess residual is positive at all six experimental voltages, with voltage-wise medians between approximately $0.096$ and $0.484$, although no monotone voltage dependence is resolved. \Cref{fig:experimental_bk_response_memory}(f) compares the two memory measures patch by patch. The conventional dwell-memory measure and the finite-history excess residual are not significantly correlated within either channel class. Stronger lag-one dwell correlations therefore do not imply a stronger dependence of event consequences on preceding history.

The fixed-horizon comparison therefore gives the central experimental dissociation. Cell-BK channels show the stronger conventional dwell correlations, but little finite-history residual remains after the present state and age are retained. MitoBK channels show weaker dwell correlations, yet retain a much larger dependence of event consequences on preceding history.

\subsubsection{Timescale of Response-Relevant Memory}
\label{subsubsec:experimental_bk_frequency_memory}

The fixed-horizon analysis establishes the class-level separation but does not reveal when the missing history matters. We therefore resolve the future consequence by frequency. This analysis asks whether the residual left by a candidate state is concentrated in slow propagation, fast propagation, or both.

\begin{figure*}[htbp]
  \centering
  \includegraphics[width=0.95\textwidth]{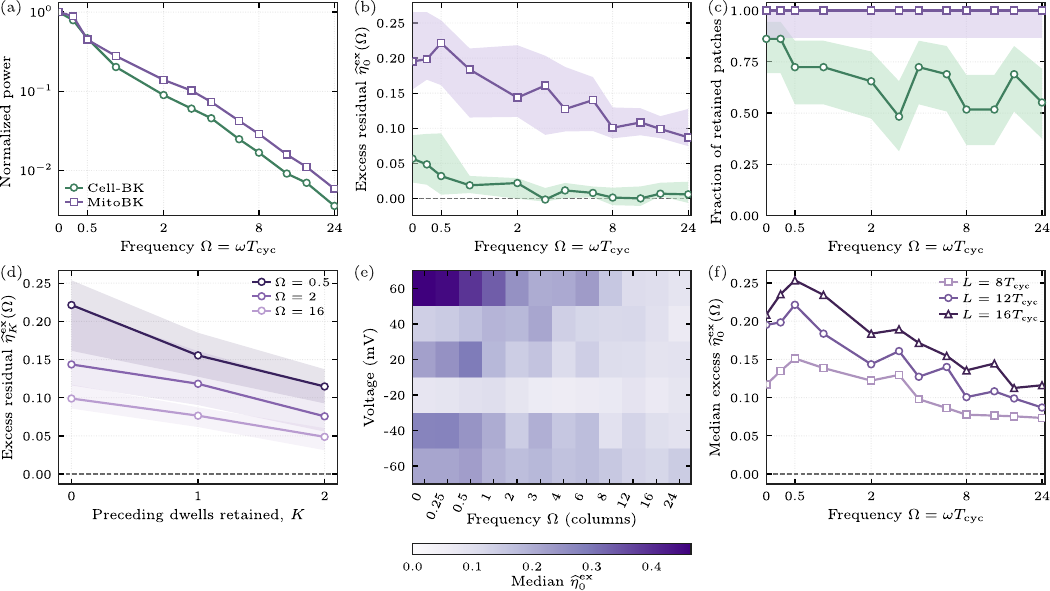}
  \caption{\textbf{Frequency-resolved response-relevant memory in experimental BK-channel trajectories.}
  (a) The group median rich-history event-consequence power normalized by its zero-frequency value.
  (b) Median excess residual spectrum for the state-and-age candidate $Y_{\mathrm{BK}}^{(0)}$.
  (c) Fraction of patches for which the original residual exceeds the mean state-wise shuffled residual. Shaded regions denote Wilson $95\%$ intervals.
  (d) MitoBK residual hierarchy as preceding dwell durations are added.
  (e) Voltage-resolved patch medians for MitoBK.
  (f) Residual spectra obtained with future windows of $8$, $12$, and $16$ characteristic cycles. Lines and symbols denote patch-level medians; shaded regions denote bootstrap $95\%$ intervals where shown.}
  \label{fig:experimental_bk_frequency_memory}
\end{figure*}

For each patch, we introduce the characteristic cycle time and dimensionless angular frequency
\begin{equation}
  T_{\mathrm{cyc}} \equiv \operatorname{median}\tau_O+\operatorname{median}\tau_C = \frac{\Delta}{4}, \qquad \Omega \equiv \omega T_{\mathrm{cyc}}.
  \label{eq: experimental_bk_dimensionless_frequency}
\end{equation}
We replace $q_{\Delta}(t)$ by the tapered Fourier component of the future open-state trajectory,
\begin{equation}
  q_{\Omega}(t) \equiv \frac{1}{T_{\mathrm{cyc}}}\int_0^{12T_{\mathrm{cyc}}} w(s)e^{\mathrm{i}\Omega s/T_{\mathrm{cyc}}}\mathds{1}_{x_{t+s}=O} \,\rmd s,
  \label{eq: experimental_bk_frequency_output}
\end{equation}
where $w(s)$ smoothly suppresses the endpoint of the finite future window. Applying the same event--continuation comparison as in \cref{eq: experimental_bk_observational_event_consequence} gives the frequency-resolved coefficient $\mathcal{K}_{e}^{(3)}(\Omega)$. Its activity-weighted projection onto $Y_{\mathrm{BK}}^{(K)}$ is denoted by $\mathcal{K}_{e}^{(K)}(\Omega)$. We define
\begin{equation}
  \eta_K^{(3)}(\Omega) \equiv \frac{\left\langle \left| \mathcal{K}_{e}^{(3)}(\Omega)-\mathcal{K}_{e}^{(K)}(\Omega) \right|^2 \right\rangle_{\mathrm{ev}}}{\left\langle \left| \mathcal{K}_{e}^{(3)}(\Omega) \right|^2 \right\rangle_{\mathrm{ev}}}
  \label{eq: experimental_bk_frequency_residual}
\end{equation}
and report the corresponding real-minus-shuffled excess $\widehat{\eta}_K^{\mathrm{ex}}(\Omega)$. The finite window, taper, complex-valued debiasing, and frequency-dependent quality controls are detailed in \cref{appsec: experimental_bk_frequency_analysis}. \Cref{fig:experimental_bk_frequency_memory}(a) shows the rich-history event-consequence power over the frequency range used below. Normalizing by the zero-frequency value removes the overall amplitude and displays how the reference event consequence is distributed across timescales.

The residual left by the state-and-age candidate has sharply different frequency dependence in the two channel classes. For mitoBK, the patch-level median excess residual decreases from
\begin{subequations}
\label{eq: experimental_bk_frequency_mito_values}
\begin{align}
  \widehat{\eta}_0^{\mathrm{ex}}(0.5) &= 0.2214, \\
  \widehat{\eta}_0^{\mathrm{ex}}(2) &= 0.1436, \\
  \widehat{\eta}_0^{\mathrm{ex}}(16) &= 0.0989.
\end{align}
\end{subequations}
The missing history is therefore most important at low frequency, but it remains detectable throughout the resolved band. At every sampled frequency, the original residual exceeds the mean shuffled residual in all 25 quality-controlled mitoBK patches at every sampled frequency.

For cell-BK, the median excess residual is $0.0319$ at $\Omega=0.5$, $0.0219$ at $\Omega=2$, and $0.0067$ at $\Omega=16$. It remains close to the shuffled baseline throughout the intermediate- and high-frequency ranges. The weak low-frequency component is not uniform across cell-BK patches, in contrast to the nearly universal mitoBK separation.

\Cref{fig:experimental_bk_frequency_memory}(b,c) separates the magnitude of the frequency-dependent effect from its patchwise consistency. Panel (b) shows the median excess residual spectrum. The mitoBK residual is largest at low frequency and remains positive throughout the resolved band, whereas the cell-BK median rapidly approaches the shuffled baseline. Panel (c) shows that the mitoBK result is not produced by a small subset of patches. Every quality-controlled mitoBK patch has a positive data-minus-shuffle excess at every sampled frequency. The weaker cell-BK low-frequency signal is much less uniform across patches.

Adding preceding dwells reduces the mitoBK residual at both low and high frequencies,
\begin{equation}
  \left( \widehat{\eta}_0^{\mathrm{ex}},\widehat{\eta}_1^{\mathrm{ex}},\widehat{\eta}_2^{\mathrm{ex}} \right)
  =
  \begin{cases}
    (0.2214,0.1554,0.1147), & \Omega=0.5,\\
    (0.0989,0.0765,0.0488), & \Omega=16.
  \end{cases}
  \label{eq: experimental_bk_frequency_hierarchy_values}
\end{equation}
The largest residual and the largest absolute reduction occur in the slow component. Preceding dwell history nevertheless retains information across the full resolved frequency range.

The frequency-resolved hierarchy is shown in \cref{fig:experimental_bk_frequency_memory}(d). Adding preceding dwells lowers the mitoBK residual throughout the resolved band. The largest absolute reduction occurs at low frequency, where the original residual is also largest. The preceding dwell sequence therefore explains part of the slowly propagating event consequence, but it does not merely remove a low-frequency offset: a smaller reduction remains visible at higher frequencies as well.

The remaining panels test the robustness of this frequency structure. The voltage-resolved mitoBK medians are shown in \cref{fig:experimental_bk_frequency_memory}(e). The excess residual remains positive in all six voltage groups, although no monotonic voltage dependence is resolved. \Cref{fig:experimental_bk_frequency_memory}(f) compares future windows of $8$, $12$, and $16$ characteristic cycles. Changing the window modifies the numerical magnitude but preserves the low-frequency enhancement and the separation between the two channel classes.

Within the finite-history resolution of the experiment, the present state and residence age leave little fixed-horizon residual for cell-BK. They also leave little residual at intermediate and high frequencies, although a weak and patch-dependent slow component remains. The same retained state leaves a large residual for mitoBK. The residual is strongest at low frequency, and preceding dwell durations close only part of it.

Response sufficiency is therefore scale-dependent. A retained state can describe the short-time consequence of a gating event while omitting history that controls its slower propagation. Taken together, \cref{fig:experimental_bk_response_memory,fig:experimental_bk_frequency_memory} establish two levels of the experimental result. The fixed-horizon analysis identifies which channel class retains response-relevant memory beyond state and age. The frequency-resolved analysis identifies when that missing history matters. For mitoBK, the missing history acts most strongly on the slow propagation of a gating event, while a weaker component remains at shorter timescales. The analysis is limited by the finite three-dwell reference and by the thresholded dwell detection of the supplied records. The continuation branch and the complex future output are also estimated at finite statistical and temporal resolution.

Within these limits, the experimental conclusion is clear. Cell-BK and mitoBK channels exhibit opposite orderings of conventional dwell memory and response-relevant memory. The mitoBK history that is missed by state and age acts most strongly on the slow consequence of a gating event. Neither the physical relevance nor the characteristic timescale of memory can therefore be inferred from trajectory correlations alone.

\section{Markov limit and dynamical fluctuation-response relations}
\label{sec: markov_dfrr}

Having developed and tested the non-Markovian theory, we finally make explicit how it reduces to the dynamical fluctuation-response relations (DFRRs) of Aslyamov and Esposito in the Markov-local limit \cite{aslyamov2026dynamical}. Consider the Markov limit in which $r_{ij}^{X_{t-}}\rightarrow r_{ij}(t)$, the observable weights depend only on the state, edge, and time, and the perturbation is history-independent. Orient each physical edge $e$ from $s(e)$ to $t(e)$ and denote its two directions by $+e$ and $-e$. Let $q_n^Q(\tau,t)$ be the conditional mean of the remaining contribution to $Q_\tau$ after time $t$, given $x_t=n$. For a state-current observable with jump weights $\pm c_e(t)$, the event-consequence kernels become
\begin{subequations}
\label{eq: markov_dfrr_kernel}
\begin{align}
  K_{+e}^Q(\tau,t) &= c_e(t)+q_{t(e)}^Q(\tau,t) - q_{s(e)}^Q(\tau,t) \equiv \varphi_e^Q(\tau,t), \\
  K_{-e}^Q(\tau,t) &= -\varphi_e^Q(\tau,t).
\end{align}
\end{subequations}

Define the mean forward and reverse fluxes, the net current, and the activity by
\begin{subequations}
\label{eq: markov_dfrr_activity}
\begin{align}
  j_{\pm e}(t) &\equiv \left\langle \gamma_{\pm e}(t) \right\rangle, \\
  j_e(t) &\equiv j_{+e}(t) - j_{-e}(t), \\
  a_e(t) &\equiv j_{+e}(t) + j_{-e}(t).
\end{align}
\end{subequations}
Pairing the two directed contributions in the exact covariance FRR then gives
\begin{align}
  \label{eq: recovered_dfrr_phi}
  \frac{1}{\tau} \operatorname{Cov}\left( Q_\tau, Q'_\tau \right) ={}& \frac{1}{\tau} C_0(Q, Q') \\
  &+ \int_0^\tau \frac{\rmd t}{\tau} \sum_e a_e(t) \varphi_e^Q(\tau,t) \varphi_e^{Q'}(\tau,t). \nonumber
\end{align}
When no initial prehistory is required, $C_0(Q,Q')/\tau$ reduces to the initial-variability term of the DFRR. The activity appears because the kernels of the two directions differ only by sign, while the quadratic variations of their martingale increments add.

For a local edge parameter $\lambda_e(t)$, define $\alpha_{\pm e}(t) \equiv \partial_{\lambda_e}\ln r_{\pm e}(t)$. The corresponding response is
\begin{subequations}
\label{eq: markov_dfrr_response}
\begin{align}
  R_{\lambda_e}^Q(\tau,t) &= \varphi_e^Q(\tau,t) \left[j_{+e}(t) \alpha_{+e}(t) - j_{-e}(t) \alpha_{-e}(t) \right] \\
  &= \varphi_e^Q(\tau, t) \partial_{\lambda_e} j_e(t).
\end{align}
\end{subequations}
Substituting this result into \cref{eq: recovered_dfrr_phi} yields
\begin{align}
  \label{eq: recovered_dfrr_response_form}
  \frac{1}{\tau} \operatorname{Cov}\left(Q_\tau, Q'_\tau\right) ={}& \frac{1}{\tau} C_0(Q,Q') \\
  &+ \int_0^\tau \frac{\rmd t}{\tau} \sum_e \frac{a_e(t) R_{\lambda_e}^Q(\tau,t) R_{\lambda_e}^{Q'}(\tau, t)}{\left[ \partial_{\lambda_e} j_e(t) \right]^2}, \nonumber
\end{align}
which is the DFRR of \cite{aslyamov2026dynamical}. In particular, for $r_{\pm e} \propto \exp[\pm S_e/2]$, one has $\partial_{S_e} j_e = a_e/2$ and $R_{S_e}^Q = a_e \varphi_e^Q / 2$.

The DFRR edge amplitude $\varphi_e^Q$ is therefore the Markov limit of the event-consequence kernel, while $a_e$ is the summed martingale variance of the two directed channels. In this limit, the consequence of a given edge event is independent of the earlier history, and hence $\mathcal{M}_Q=0$. Beyond Markovianity, the same event may carry different consequences on different histories. The response-heterogeneity gap is precisely the additional fluctuation power that prevents the covariance from closing in terms of the ordinary DFRR response kernels alone.

\section{Discussions and outlooks}
\label{sec: discussion}

This work developed fluctuation-response theory at the level where memory actually lives: the observed trajectory. Two structures survive the loss of Markovianity. The stochastic equation of motion closes exactly on the observed history, and the martingale representation theorem makes the compensated event increments a complete orthogonal basis for every trajectory fluctuation generated after the initial condition. The expansion coefficient in this basis is the event-consequence kernel, and its physical identity is the intensity-normalized response of the observable to perturbing one transition on the realized history. Response is therefore the stochastic coordinate of trajectory fluctuations. From this single identification, the rest follows by simple operations. The martingale isometry gives exact finite-time and finite-frequency fluctuation-response relations. Averaging over histories converts them into inequalities whose slack is the response-heterogeneity gap. Projecting onto physical perturbations and bounding their sensitivity gives Fisher-information bounds and response-kinetic uncertainty relations controlled by the dynamical activity.

The question posed in the Introduction now has a precise answer: none of this structure requires Markovianity. Memory enters in exactly two places---the stochastic intensities and the event-consequence kernels---and it introduces no additional source of trajectory randomness. \Cref{sec: markov_dfrr} shows explicitly that the Markov-local limit recovers the dynamical fluctuation-response relations \cite{aslyamov2026dynamical}. Beyond this limit, the response-heterogeneity gap is the exact additional fluctuation power lost when history-resolved event consequences are replaced by ordinary responses. The same gap is the certificate that standard statistics cannot supply. A reduced memory coordinate is response-sufficient precisely when its gap closes. This can be tested from spontaneous trajectories alone, as the BK-channel analysis demonstrates.

This work refreshes our understanding of three familiar ideas. First, a closed theory does not require a Markov state. The record of past events, together with the rate that reads it, is already a closed and exact description. One can do nonequilibrium physics directly on what is observed, without relying on hidden variables. Second, fluctuation and response have been treated since Onsager and Kubo \cite{kubo1957statistical} as two quantities tied together by theorems. Here they are not two quantities. Every fluctuation is built event by event. The weight of each event is its consequence for the observable, and the same quantity sets the response to perturbing that transition. A fluctuating system is continually performing small perturbation experiments on itself. Third, memory is consequence, not correlation, and the two can come apart. A system has memory when its past changes what its present events do. Memory is therefore not a single number attached to a system once and for all. It is the answer to a question, and different questions asked of the same system can receive opposite answers (e.g., which observable, which transition, what timescale).

The experimental program inherits the limitations stated in \cref{sec:response_sufficient_coarse_graining}. The event--continuation comparison is observational rather than interventional, so unresolved variables correlated with both event timing and future output can contribute to the estimated kernels. The rich-history reference retains a finite number of preceding dwells rather than the complete filtration. The supplied trajectories are thresholded dwell sequences rather than raw currents; threshold-based idealization is itself known to distort dwell statistics through missed events \cite{qin1996estimating}. On the theoretical side, the representation is formulated for regular, causal, nonexplosive dynamics on the filtration generated by the observed channels and the initial condition. Additional noise sources require additional basis elements. Singular or explosive dynamics lie outside the present scope.

Several directions follow naturally. The gap separates fluctuations built from a consequence shared by all histories from fluctuations dominated by rare histories with exceptional consequences. This invites a cascade-resolved analysis of self-exciting dynamics, to which we will return in future work. The thermodynamic content of history-dependent intensities---local detailed balance, entropy production, and their uncertainty relations---connects the present kinetic bounds to semi-Markov thermodynamics \cite{ertel2022operationally} and remains to be developed in general. Diffusive and continuous limits should join the present theory to the stochastic calculus of Ref.~\cite{stutzer2026stochastic}. Closed-loop perturbations, in which the control itself is a functional of the realized history, turn the score into a feedback object and connect response theory to control. Markovian fluctuation--response relations have also been used to estimate gradients from sampled trajectories for the numerical optimization of control protocols \cite{lyu2026optimal}. How the present non-Markovian relations can guide the development of analogous optimization methods for systems with memory is a natural question for future work. Finally, the monotone hierarchy of candidate memory states invites data-driven searches for response-sufficient coordinates in systems whose hidden networks cannot be uniquely reconstructed \cite{siegle2010markovian,alexandrovich2016nonparametric,flomenbom2005what}.

Beyond Markovianity, an event is characterized not only by when it occurs but by what it changes. The theory developed here measures the second property directly. Memory is revealed not by how long the past persists, but by how it diversifies the consequences of the present.

\section{Acknowledgements}

This work is supported by the U.S. National Science Foundation under Grant No. DMR-2145256 and Alfred P. Sloan Foundation Matter-to-Life Theory Award under Grant No. G-2025-25194.

\section{Data availability}

No new experiments were performed in this study. The experimental data analyzed in \cref{subsec:experimental_bk_response_memory} were obtained from Refs.~\cite{borys2022new} and \cite{agata_wawrzkiewicz_jalowiecka_2022_6340407}. The present work involved only numerical simulations and data analysis. The simulation data that support the findings of this article are generated by numerical simulation codes that are openly available at \cite{data}.

\appendix

\crefalias{section}{appendix}
\crefalias{subsection}{appendix}
\crefalias{subsubsection}{appendix}

\section{Trajectory-level martingale structure}
\label{app:trajectory_martingale_structure}

This appendix provides the technical basis of the trajectory-level stochastic equation and the martingale representation used in \cref{sec:theory}. We keep the presentation in terms of directly observable counting processes and conditional trajectory averages. The resolved jump process is assumed to be causal, simple, and nonexplosive; its transition intensities are finite and determined by the observed pre-jump history; and all stochasticity entering after the initial condition is carried by the resolved jump channels. If an additional independent stochastic source is present, its martingale increment must be added to the representation.

\subsection{Doob--Meyer Equation for History-Dependent Counts}
\label{app:doob_meyer_equation}

Let $n_{ij}(t)$ count transitions from state $j$ to state $i$, and define the stochastic intensity
\begin{equation}
  \gamma_{ij}(t) = r_{ij}^{X_{t-}}\mathds{1}_{x_{t-}=j}.
  \label{appeq: intensity}
\end{equation}
The defining property of the intensity is
\begin{equation}
  \mathbb{E}\left[\, \rmd n_{ij}(t) \!\mid\! X_{t-}\, \right] = \gamma_{ij}(t)\rmd t.
  \label{appeq: conditional_count}
\end{equation}
Consequently, the accumulated conditional event tendency is
\begin{equation}
  A_{ij}(t) = \int_0^t \gamma_{ij}(s) \,\rmd s.
  \label{appeq: compensator}
\end{equation}
Subtracting this tendency from the realized count defines
\begin{equation}
  \varepsilon_{ij}(t) = n_{ij}(t)-A_{ij}(t).
  \label{appeq: compensated_process}
\end{equation}

For any $t\geq s$,
\begin{align}
  \mathbb{E}\left[\, \varepsilon_{ij}(t) - \varepsilon_{ij}(s) \!\mid\! X_s\, \right] ={}& \mathbb{E}\Big[\, n_{ij}(t) - n_{ij}(s) \\
  &-\int_s^t\gamma_{ij}(u)\rmd u \,\Big|\, X_s\, \Big] \nonumber\\
  ={}& 0.
  \label{appeq: martingale_verification}
\end{align}
Thus, $\varepsilon_{ij}(t)$ is a martingale and
\begin{equation}
  \rmd n_{ij}(t) = \gamma_{ij}(t) \, \rmd t+\rmd\varepsilon_{ij}(t).
  \label{appeq: count_sde}
\end{equation}
No Markov assumption enters this decomposition. The only difference between Markovian and non-Markovian dynamics is whether $\gamma_{ij}(t)$ is determined by the present state and time or by the complete observed history.

\Cref{appeq: count_sde} is closed at the trajectory level. At each time, the intensity is fixed by the history already realized, while the martingale increment contains the remaining random part of the next event. Averaging this equation generally does not produce a closed equation for the instantaneous state probabilities because $\left\langle\gamma_{ij}(t)\right\rangle$ retains history--state correlations.

\subsection{Space--Time Orthogonality of the Martingale Increments}
\label{app:martingale_orthogonality}

For a simple jump process,
\begin{equation}
  \left[\rmd n_{ij}(t)\right]^2 = \rmd n_{ij}(t),
  \label{appeq: count_square}
\end{equation}
and two distinct resolved channels cannot jump simultaneously with probability of order $\rmd t$. Using $\rmd\varepsilon_{ij}(t)=\rmd n_{ij}(t)-\gamma_{ij}(t)\rmd t$, we obtain
\begin{align}
  \mathbb{E}\left[\, \rmd\varepsilon_{ij}(t)\rmd\varepsilon_{kl}(t) \!\mid\! X_{t-}\, \right] &= \mathbb{E}\left[\, \rmd n_{ij}(t)\rmd n_{kl}(t) \!\mid\! X_{t-}\, \right] + o(\rmd t) \nonumber\\
  &= \delta_{ik}\delta_{jl}\gamma_{ij}(t)\rmd t.
  \label{appeq: equal_time_orthogonality}
\end{align}

For $t>s$, the increment $\rmd\varepsilon_{kl}(s)$ is already determined by the history available immediately before time $t$. Therefore,
\begin{align}
  \left\langle \rmd\varepsilon_{ij}(t)\rmd\varepsilon_{kl}(s) \right\rangle &= \left\langle \mathbb{E}\left[\, \rmd\varepsilon_{ij}(t) \!\mid\! X_{t-}\, \right]\rmd\varepsilon_{kl}(s) \right\rangle \nonumber\\
  &=0.
  \label{appeq: different_time_orthogonality}
\end{align}
Combining the equal-time and different-time cases gives
\begin{equation}
  \left\langle \rmd\varepsilon_{ij}(t)\rmd\varepsilon_{kl}(t') \right\rangle = \delta_{ik}\delta_{jl}\delta(t-t')\left\langle\gamma_{ij}(t)\right\rangle \rmd t\rmd t'.
  \label{appeq: full_space_time_orthogonality}
\end{equation}
The martingale increments are therefore orthogonal both in transition space and in time. Their amplitude is history dependent, but no second-moment correlation remains between different edge--time directions.

For two sets of history-dependent coefficients $H_{ij}(t)$ and $G_{ij}(t)$ determined before the increment at time $t$, this orthogonality gives
\begin{align}
  &\mathbb{E}\Bigg[ \left( \sum_{i\neq j}\int_0^\tau H_{ij}(t)\rmd\varepsilon_{ij}(t) \right)\left( \sum_{k\neq l}\int_0^\tau G_{kl}(t)\rmd\varepsilon_{kl}(t) \right) \Bigg] \nonumber\\
  &\qquad = \int_0^\tau \sum_{i\neq j}\left\langle \gamma_{ij}(t)H_{ij}(t)G_{ij}(t) \right\rangle \rmd t.
  \label{appeq: martingale_isometry}
\end{align}
This identity is the continuous-time analogue of expanding two random vectors in the same orthogonal basis and multiplying matching components.

\subsubsection{Equivalent Bracket Notation}
\label{app:bracket_notation}

In standard martingale theory \cite{bremaud1981point,jacod2003limit}, the same local second-moment structure can be summarized by the predictable bracket
\begin{equation}
  \rmd\left\langle \varepsilon_{ij},\varepsilon_{kl} \right\rangle_t = \delta_{ik}\delta_{jl}\gamma_{ij}(t)\rmd t.
  \label{appeq: predictable_bracket}
\end{equation}
We do not use this notation in the main text; it is included only to connect \cref{appeq: equal_time_orthogonality} with the standard point-process formulation.

\subsection{Martingale Property of the History-Conditioned Prediction}
\label{app:doob_martingale_property}

For completeness, we verify that the history-conditioned prediction
\begin{equation}
  M_t^Q = \mathbb{E}\left[\, Q_\tau \!\mid\! X_t\, \right]
  \label{appeq: history_conditioned_prediction}
\end{equation}
is a martingale. As in the main text, conditioning on $X_t$ is shorthand for conditioning on the trajectory information $\mathcal{F}_t^X$ available up to time $t$.

Since $Q_\tau$ is square-integrable, its conditional expectation is also square-integrable. Indeed, the conditional Jensen inequality gives
\begin{equation}
  \mathbb{E}\left[ \left(M_t^Q\right)^2 \right] = \mathbb{E}\left[ \Big( \mathbb{E}\left[\, Q_\tau \!\mid\! X_t\, \right] \Big)^2 \right] \leq \mathbb{E}\left[ Q_\tau^2 \right] < \infty.
  \label{appeq: doob_martingale_square_integrability}
\end{equation}
Moreover, $M_t^Q$ is determined by the trajectory observed up to time $t$ and is therefore adapted to $\mathcal{F}_t^X$.

Consider two times $0\leq s\leq t\leq\tau$. Since the history available at time $s$ is contained in the history available at time $t$, the tower property of conditional expectations gives
\begin{align}
  \mathbb{E}\left[ M_t^Q \Big| X_s \right] &= \mathbb{E}\left[ \mathbb{E}\left[\, Q_\tau \!\mid\! X_t\, \right] \!\mid\! X_s \right] \nonumber\\
  &= \mathbb{E}\left[\, Q_\tau \!\mid\! X_s\, \right]
  \nonumber\\
  &= M_s^Q.
  \label{appeq: doob_martingale_tower_property}
\end{align}
This is precisely the martingale property. Physically, $M_t^Q$ is the current prediction of the fixed terminal observable $Q_\tau$. New events and no-event intervals update this prediction along individual trajectories, but the conditional mean of all possible updated predictions remains equal to the prediction already available at the earlier time.

At the terminal time, the complete trajectory $X_\tau$ determines the observable $Q_\tau$, and hence
\begin{equation}
  M_\tau^Q = \mathbb{E}\left[\, Q_\tau \!\mid\! X_\tau\, \right] = Q_\tau.
  \label{appeq: doob_martingale_terminal_value}
\end{equation}
Therefore, $M_t^Q$ is the square-integrable Doob martingale that interpolates between the initial history-conditioned prediction $\mathbb{E}[\,Q_\tau\!\mid\!X_0\,]$ and the realized terminal observable $Q_\tau$. The proof uses only the nesting of the observed trajectory histories and does not require Markovianity.

\subsection{Martingale Representation of Trajectory Fluctuations}
\label{app:martingale_representation_proof}

Let $Q_\tau$ be a square-integrable observable determined by the observed trajectory up to time $\tau$, and define
\begin{equation}
  M_t^Q = \mathbb{E}\left[\, Q_\tau \!\mid\! X_t\, \right].
  \label{appeq: doob_martingale}
\end{equation}
The quantity $M_t^Q$ is the prediction of the final observable using all trajectory information available at time $t$. It satisfies
\begin{equation}
  \mathbb{E}\left[\, M_t^Q \,\Big|\, X_s\, \right] = M_s^Q, \quad t\geq s,
  \label{appeq: doob_property}
\end{equation}
and is therefore a martingale.

The natural filtration of the observed jump process receives new information through the resolved transition channels: either a transition occurs or the system continues without a transition. Both possibilities are contained in the compensated increments $\rmd\varepsilon_{ij}(t)$. The martingale representation theorem for multichannel counting processes therefore gives
\begin{equation}
  M_t^Q - M_0^Q = \sum_{i\neq j}\int_0^t K_{ij}^Q(\tau,s \!\mid\! X_{s-}) \, \rmd\varepsilon_{ij}(s).
  \label{appeq: finite_time_martingale_representation}
\end{equation}
At the terminal time,
\begin{equation}
  M_\tau^Q = \mathbb{E}\left[\, Q_\tau \!\mid\! X_\tau\, \right] = Q_\tau.
  \label{appeq: terminal_doob_martingale}
\end{equation}
Hence,
\begin{equation}
  Q_\tau-\mathbb{E}\left[\, Q_\tau \!\mid\! X_0\, \right] = \sum_{i\neq j}\int_0^\tau K_{ij}^Q(\tau,t \!\mid\! X_{t-}) \, \rmd\varepsilon_{ij}(t).
  \label{appeq: terminal_martingale_representation}
\end{equation}

This representation separates initial uncertainty from fluctuations generated during the observation interval:
\begin{align}
  Q_\tau-\left\langle Q_\tau\right\rangle ={}& \mathbb{E}\left[\, Q_\tau \!\mid\! X_0\, \right]-\left\langle Q_\tau\right\rangle \nonumber\\
  &+ \sum_{i\neq j}\int_0^\tau K_{ij}^Q(\tau,t \!\mid\! X_{t-}) \, \rmd\varepsilon_{ij}(t).
  \label{appeq: initial_and_dynamic_fluctuation}
\end{align}
The first term is inherited from uncertainty already present at $t=0$. The second contains every fluctuation generated subsequently. For a fixed initial history, the first term vanishes and all observable fluctuations are generated by the martingale increments.

The coefficient $K_{ij}^Q$ is unique on histories for which $\gamma_{ij}(t)>0$. If a transition has zero intensity on a particular history, its coefficient does not affect the representation and may be set to zero. If the physical model contains an additional stochastic source not generated by the observed jump channels, such as an independent continuous noise, the corresponding martingale increment must be included as an additional basis element.

\subsection{Event--Continuation Interpretation of the Kernel}
\label{app:event_consequence_kernel}

We now connect the expansion coefficient in \cref{appeq: terminal_martingale_representation} to the event-consequence interpretation used in the main text. Fix a pre-event history $X_{t-}$ with $x_{t-}=j$. During the infinitesimal interval $[t,t+\rmd t)$, define
\begin{equation}
  q_{kj} = \mathbb{E}\left[ Q_\tau \Big| X_t^{kj} \right], \qquad q_0 = \mathbb{E}\left[ Q_\tau \Big| X_t^0 \right].
  \label{appeq: branch_predictions}
\end{equation}
Here $X_t^{kj}$ denotes the branch on which the transition $j\to k$ occurs, whereas $X_t^0$ denotes the branch on which no transition occurs during the interval.

The conditional future prediction before resolving the interval is
\begin{align}
  \mathbb{E}\left[\, Q_\tau \!\mid\! X_{t-}\, \right] ={}& \left[ 1-\sum_{k\neq j}\gamma_{kj}(t)\rmd t \right]q_0 \nonumber\\
  &+ \sum_{k\neq j}\gamma_{kj}(t)\rmd t\,q_{kj} + o(\rmd t).
  \label{appeq: preevent_prediction}
\end{align}
Moreover,
\begin{equation}
  \mathbb{E}\left[\, Q_\tau\rmd n_{ij}(t) \!\mid\! X_{t-}\, \right] = \gamma_{ij}(t)\rmd t\,q_{ij} + o(\rmd t).
  \label{appeq: event_weighted_prediction}
\end{equation}
Using $\rmd\varepsilon_{ij}(t)=\rmd n_{ij}(t)-\gamma_{ij}(t)\rmd t$, we find
\begin{align}
  \mathbb{E}\left[\, Q_\tau\rmd\varepsilon_{ij}(t) \!\mid\! X_{t-}\, \right] ={}& \gamma_{ij}(t)\left[ q_{ij}-q_0 \right]\rmd t \nonumber\\
  &+ o(\rmd t).
  \label{appeq: branch_martingale_correlation}
\end{align}
Taking the infinitesimal limit gives
\begin{subequations}
\label{appeq: branch_kernel}
\begin{align}
  K_{ij}^Q(\tau,t \!\mid\! X_{t-}) &= q_{ij} - q_0 \\
  &= \mathbb{E}\left[ Q_\tau \Big| X_t^{ij} \right] - \mathbb{E}\left[ Q_\tau \Big| X_t^0 \right].
\end{align}
\end{subequations}
Therefore,
\begin{equation}
  \mathbb{E}\left[\, Q_\tau\rmd\varepsilon_{ij}(t) \!\mid\! X_{t-}\, \right] = \gamma_{ij}(t) \, K_{ij}^Q(\tau,t \!\mid\! X_{t-})\rmd t.
  \label{appeq: conditional_kernel_identity}
\end{equation}

The notation of appending an event at time $t$ is shorthand for this infinitesimal branch comparison. It does not require assigning a finite probability to an event at one predetermined continuous time. Physically, $K_{ij}^Q$ measures how resolving the next infinitesimal interval in favor of the transition $j\to i$, rather than continuation without a transition, changes the predicted future observable.

\subsection{Elementary Observable Correlations}
\label{app:elementary_correlations}

The general identity \cref{appeq: conditional_kernel_identity} generates correlations with occupation and jump observables.

For $t>s$, define the state-influence kernel
\begin{equation}
  I_{ij}^{k}(t,s \!\mid\! X_{s-}) = \mathbb{E}\left[\, \mathds{1}_{x_t=k} \!\mid\! X_s^{ij}\, \right] - \mathbb{E}\left[\, \mathds{1}_{x_t=k} \!\mid\! X_s^0\, \right], \quad t>s.
  \label{appeq: state_kernel}
\end{equation}
Since $\rmd\tau_k(t)=\mathds{1}_{x_t=k}\rmd t$,
\begin{equation}
  \left\langle \rmd\tau_k(t)\rmd\varepsilon_{ij}(s) \right\rangle = \mathds{1}_{t>s}\left\langle \gamma_{ij}(s) \, I_{ij}^{k}(t,s \!\mid\! X_{s-}) \right\rangle \rmd t\rmd s.
  \label{appeq: occupation_martingale}
\end{equation}

Similarly, define the influence of the transition $j\to i$ on a future intensity,
\begin{equation}
  J_{ij}^{kl}(t,s \!\mid\! X_{s-}) = \mathbb{E}\left[ \gamma_{kl}(t) \!\mid\! X_s^{ij} \right] - \mathbb{E}\left[ \gamma_{kl}(t) \!\mid\! X_s^0 \right], \quad t>s.
  \label{appeq: intensity_kernel}
\end{equation}
For $t>s$,
\begin{equation}
  \left\langle \rmd n_{kl}(t)\rmd\varepsilon_{ij}(s) \right\rangle = \left\langle \gamma_{ij}(s) \, J_{ij}^{kl}(t,s \!\mid\! X_{s-}) \right\rangle \rmd t\rmd s.
  \label{appeq: future_jump_martingale}
\end{equation}
At equal times, the jump itself contributes
\begin{align}
  \left\langle \rmd n_{kl}(t)\rmd\varepsilon_{ij}(s) \right\rangle ={}& \delta_{ik}\delta_{jl}\delta(t-s)\left\langle\gamma_{ij}(s)\right\rangle \rmd t\rmd s \nonumber\\
  &+ \mathds{1}_{t>s}\left\langle \gamma_{ij}(s) \, J_{ij}^{kl}(t,s \!\mid\! X_{s-}) \right\rangle \rmd t\rmd s.
  \label{appeq: full_jump_martingale}
\end{align}

For completeness, the ordinary correlations between the elementary occupation and jump increments are given below. The time--time correlation is
\begin{align}
  \left\langle \rmd\tau_k(t)\rmd\tau_l(s) \right\rangle ={}& \mathds{1}_{t>s}\left\langle \mathds{1}_{x_s=l} \, P\left( x_t=k \!\mid\! X_s \right) \right\rangle \rmd t\rmd s \nonumber\\
  &+ \mathds{1}_{t<s}\left\langle \mathds{1}_{x_t=k} \, P\left( x_s=l \!\mid\! X_t \right) \right\rangle \rmd t\rmd s \nonumber\\
  &+ \mathds{1}_{t=s}\delta_{kl}p_k(t) \, \rmd t\rmd s.
  \label{appeq: time_time_correlation}
\end{align}

Using the convention $\rmd\tau_k(t)=\mathds{1}_{x_{t-}=k}\rmd t$, the time--jump correlation is
\begin{align}
  &\left\langle \rmd\tau_k(t)\rmd n_{ij}(s) \right\rangle \nonumber \\
  ={}& \mathds{1}_{t>s}\left\langle \gamma_{ij}(s) \, P\left( x_t=k \!\mid\! x_{s+}=i,X_{s-} \right) \right\rangle \rmd t\rmd s \nonumber\\
  &+ \mathds{1}_{t<s}\left\langle \mathds{1}_{x_t=k} \, \mathbb{E}\left[ \gamma_{ij}(s) \!\mid\! X_t \right] \right\rangle \rmd t\rmd s \nonumber\\
  &+ \mathds{1}_{t=s}\delta_{kj}\left\langle\gamma_{ij}(t)\right\rangle \rmd t\rmd s.
  \label{appeq: time_jump_correlation}
\end{align}

The jump--jump correlation is
\begin{align}
  &\left\langle \rmd n_{kl}(t)\rmd n_{ij}(s) \right\rangle \nonumber \\
  ={}& \mathds{1}_{t>s} \left\langle \gamma_{ij}(s) \, \mathbb{E}\left[ \gamma_{kl}(t) \!\mid\! x_{s+}=i,X_{s-} \right] \right\rangle \rmd t\rmd s \nonumber\\
  &+ \mathds{1}_{t<s} \left\langle \gamma_{kl}(t) \, \mathbb{E}\left[ \gamma_{ij}(s) \!\mid\! x_{t+}=k,X_{t-} \right] \right\rangle \rmd t\rmd s \nonumber\\
  &+ \mathds{1}_{t=s} \, \delta_{ki}\delta_{lj}\delta(t-s)\left\langle\gamma_{ij}(t)\right\rangle \rmd t\rmd s.
  \label{appeq: jump_jump_correlation}
\end{align}
For observables with history-dependent weights, the weights remain inside the corresponding conditional averages. Equivalently, they can be included directly in the future functional entering $K_{ij}^Q$.

\section{Derivations of Trajectory-level Linear Response Theory}
\label{appsec:trajectory_response_theory}

This appendix provides the technical basis of the trajectory-level linear response theory presented in \cref{sec:linear_response}.

\subsection{Trajectory Probability Density}
\label{app:trajectory_density}

Consider a trajectory with transitions at times $0<t_1<t_2<\cdots<t_M<\tau$, where the $m$th transition is $j_m\to i_m$. Between transitions, the probability of remaining on the realized trajectory contributes the survival factor
\begin{equation}
  \exp\left[ -\int_0^\tau \sum_{i\neq j}r_{ij}^{X_{t-}}\mathds{1}_{x_{t-}=j} \,\rmd t \right].
  \label{appeq: trajectory_survival}
\end{equation}
Each realized transition contributes its conditional rate $r_{i_mj_m}^{X_{t_m-}}$. The complete trajectory density is therefore
\begin{equation}
  \mathcal{P}[X_\tau] = p_{\mathrm{ini}}\left[ \prod_{m=1}^{M}r_{i_mj_m}^{X_{t_m-}} \right]\exp\left[ -\int_0^\tau \sum_{i\neq j}r_{ij}^{X_{t-}}\mathds{1}_{x_{t-}=j} \,\rmd t \right].
  \label{appeq: product_trajectory_density}
\end{equation}
The factor $p_{\mathrm{ini}}$ contains the initial state and the prescribed initial-history distribution. Writing the product over jumps as an integral over counting increments gives
\begin{align}
  \mathcal{P}[X_\tau] = p_{\mathrm{ini}}\exp &\Bigg\{ \sum_{i\neq j}\int_0^\tau \ln r_{ij}^{X_{t-}}\rmd n_{ij}(t) \nonumber\\
  &- \int_0^\tau \sum_{i\neq j}r_{ij}^{X_{t-}}\mathds{1}_{x_{t-}=j} \,\rmd t \Bigg\}.
  \label{appeq: exponential_trajectory_density}
\end{align}

For parameter-dependent rates, differentiating this density gives
\begin{align}
  \left.\partial_\lambda\ln\mathcal{P}_{\lambda}[X_\tau]\right|_{\lambda=0} ={}& \sum_{i\neq j}\int_0^\tau \alpha_{ij}^{X_{t-}}\left[ \rmd n_{ij}(t)-\gamma_{ij}(t)\rmd t \right] \nonumber\\
  ={}& \sum_{i\neq j}\int_0^\tau \alpha_{ij}^{X_{t-}}\rmd\varepsilon_{ij}(t),
  \label{appeq: trajectory_score_derivation}
\end{align}
where
\begin{equation}
  \alpha_{ij}^{X_{t-}} = \left.\partial_\lambda\ln r_{ij}^{X_{t-}}(\lambda)\right|_{\lambda=0}.
  \label{appeq: logarithmic_sensitivity}
\end{equation}
Thus, the trajectory score is expressed in the same orthogonal martingale basis as the spontaneous trajectory fluctuations.

The logarithmic sensitivity is defined on the active support of the reference dynamics. We consider perturbations that do not open, at first order, a transition channel that is strictly forbidden in the reference process. Histories with $\gamma_{ij}(t)=0$ do not contribute to the corresponding score or edge response. We also assume in the main text that the initial density $p_{\mathrm{ini}}$ is independent of the perturbation parameter. If it depends on $\lambda$, the trajectory score acquires the additional term $\left.\partial_\lambda\ln p_{\mathrm{ini}}(\lambda)\right|_{\lambda=0}$. The derivative of $r_{ij}^{X_{t-}}(\lambda)$ is always taken at fixed realized history. Changes in the probability of observing that history are already contained in the derivative of the trajectory density.

\subsection{Ordinary Trajectory-Score Identity}
\label{app:trajectory_response_theory}
\label{app:ordinary_score_identity}

For an observable $Q_\tau$ with no explicit parameter dependence, its perturbed mean is
\begin{equation}
  \left\langle Q_\tau \right\rangle_\lambda = \int \mathcal{D}[X_\tau] \, Q_\tau[X_\tau] \mathcal{P}_\lambda[X_\tau].
  \label{appeq: perturbed_observable_mean}
\end{equation}
Differentiating the trajectory density gives
\begin{align}
  \left.\partial_\lambda \left\langle Q_\tau \right\rangle_\lambda \right|_{\lambda=0} &= \int \mathcal{D}[X_\tau] \, Q_\tau[X_\tau] \mathcal{P}[X_\tau]\partial_\lambda \ln\mathcal{P}_\lambda[X_\tau] \Big|_{\lambda=0} \nonumber\\
  &= \left\langle Q_\tau \Lambda_\tau \right\rangle.
  \label{appeq: ordinary_score_identity}
\end{align}
Normalization of the trajectory density implies
\begin{equation}
  \left\langle \Lambda_\tau \right\rangle = \left. \partial_\lambda \int \mathcal{D}[X_\tau] \, \mathcal{P}_\lambda[X_\tau] \right|_{\lambda=0} = 0.
  \label{appeq: zero_mean_score}
\end{equation}
Therefore,
\begin{equation}
  \left. \partial_\lambda \left\langle Q_\tau \right\rangle_\lambda \right|_{\lambda=0} = \operatorname{Cov}(Q_\tau, \Lambda_\tau).
  \label{appeq: ordinary_score_covariance}
\end{equation}
If the observable depends explicitly on $\lambda$, differentiation of both the observable and the trajectory density gives
\begin{align}
  \left. \partial_\lambda \left\langle Q_\tau(\lambda) \right\rangle_\lambda \right|_{\lambda=0} ={}& \left\langle \left. \partial_\lambda Q_\tau(\lambda) \right|_{\lambda=0} \right\rangle \nonumber\\
  &+\operatorname{Cov}(Q_\tau, \Lambda_\tau).
  \label{appeq: score_identity_explicit_observable}
\end{align}

\subsection{Time-Dependent Protocol and Local Score Increment}
\label{app:time_dependent_score}

For the time-dependent perturbation in \cref{eq: time_dependent_rate_perturbation}, differentiating the trajectory density gives
\begin{align}
  \Lambda_{\alpha,h}[X_\tau] &= \sum_{i\neq j} \int_0^\tau h(t) \alpha_{ij}^{X_{t-}} \Big[ \rmd n_{ij}(t) - \gamma_{ij}(t) \, \rmd t \Big] \nonumber\\
  &= \int_0^\tau h(t) \, \rmd\Lambda_\alpha(t),
  \label{appeq: protocol_score_derivation}
\end{align}
where
\begin{equation}
  \rmd\Lambda_\alpha(t) = \sum_{i\neq j} \alpha_{ij}^{X_{t-}} \, \rmd\varepsilon_{ij}(t).
  \label{appeq: local_score_increment}
\end{equation}
The external protocol $h(t)$ selects when the perturbation is applied, while $\alpha_{ij}^{X_{t-}}$ specifies the logarithmic direction in transition space on the history realized at that time.

\subsection{Conditional Trajectory-Score Identity}
\label{app:conditional_score_identity}

We now condition on a fixed pre-perturbation history $X_{t-}$. Consider first a unit pulse supported only during the infinitesimal interval $[t,t+\rmd t)$. The pulse changes only the conditional density of the future continuation from $X_{t-}$; the probability of the already realized past is held fixed. The corresponding conditional score is $\rmd\Lambda_\alpha(t)$. Its conditional mean vanishes,
\begin{equation}
  \mathbb{E}\left[\rmd\Lambda_\alpha(t) \!\mid\! X_{t-}\right] = \sum_{i\neq j} \alpha_{ij}^{X_{t-}} \mathbb{E}\left[ \rmd\varepsilon_{ij}(t) \!\mid\! X_{t-} \right] = 0.
  \label{appeq: zero_mean_conditional_score}
\end{equation}
Differentiating the conditional future mean therefore gives
\begin{equation}
  R_\alpha^Q(\tau,t \!\mid\! X_{t-}) \, \rmd t = \mathbb{E} \left[ Q_\tau \, \rmd\Lambda_\alpha(t) \!\mid\! X_{t-} \right].
  \label{appeq: conditional_score_identity_local}
\end{equation}
Using \cref{appeq: local_score_increment},
\begin{equation}
  R_\alpha^Q(\tau,t \!\mid\! X_{t-}) = \sum_{i\neq j} \alpha_{ij}^{X_{t-}} \mathbb{E} \left[ Q_\tau\frac{\rmd\varepsilon_{ij}(t)}{\rmd t} \!\mid\! X_{t-} \right].
  \label{appeq: conditional_score_identity_expanded}
\end{equation}
A general protocol follows by superposing such infinitesimal pulses over time.

\subsection{Edge-Wise Response and the Event-Consequence Kernel}
\label{app:edge_response_kernel}

For the unit logarithmic perturbation of the transition $j\to i$, the perturbation direction is $\alpha_{kl}^{X_{t-}}=\delta_{ki}\delta_{lj}$. Equation~\eqref{appeq: conditional_score_identity_expanded} then gives
\begin{equation}
  R_{ij}^{Q}(\tau,t \!\mid\! X_{t-})\,\rmd t = \mathbb{E}\left[ Q_\tau\,\rmd\varepsilon_{ij}(t) \!\mid\! X_{t-} \right].
  \label{appeq: conditional_edge_score_identity}
\end{equation}
The event--continuation identity derived in Appendix~A.5 gives
\begin{equation}
  \mathbb{E}\left[Q_\tau\,\rmd\varepsilon_{ij}(t) \!\mid\! X_{t-}\right] = \gamma_{ij}(t) \, K_{ij}^{Q}(\tau,t \!\mid\! X_{t-})\,\rmd t.
  \label{appeq: edge_kernel_correlation}
\end{equation}
Combining the two equations yields
\begin{equation}
  R_{ij}^{Q}(\tau,t \!\mid\! X_{t-}) = \gamma_{ij}(t) \, K_{ij}^{Q}(\tau,t \!\mid\! X_{t-}).
  \label{appeq: edge_response_kernel}
\end{equation}
This is the point at which the fluctuation coefficient introduced in Section~III becomes a response quantity.

\subsection{History Averaging and Causality}
\label{app:history_average_causality}

The perturbation begins at time $t$ and therefore does not change the reference distribution of the histories already realized before $t$. Averaging the conditional derivative over those histories gives
\begin{equation}
  R_\alpha^Q(\tau,t) = \left\langle R_\alpha^Q(\tau,t \!\mid\! X_{t-})\right\rangle.
  \label{appeq: history_average_response}
\end{equation}
For causality, let $Q_t$ be determined by the trajectory up to time $t$. For $s>t$, the martingale property gives
\begin{align}
  \left\langle Q_t\,\rmd\varepsilon_{ij}(s)\right\rangle &= \left\langle Q_t\mathbb{E}\left[\rmd\varepsilon_{ij}(s) \!\mid\! X_t\right]\right\rangle \nonumber\\
  &= 0, \quad s>t.
  \label{appeq: causality_from_martingale}
\end{align}
Consequently, $\left\langle Q_t\,\rmd\Lambda_\alpha(s)\right\rangle=0$ for $s>t$, and a future perturbation cannot change an observable already determined by the past.

\subsection{Responses of Occupation, Jump, and Additive Observables}
\label{app:additive_observable_response}

For the occupation-time increment, the time-resolved score identity gives
\begin{align}
  \frac{\partial}{\partial\epsilon}\!&\left.\frac{\delta\left\langle\rmd\tau_k(s)\right\rangle_{\epsilon h}}{\delta h(t)}\right|_{\epsilon=0} = \sum_{i\neq j}\left\langle\alpha_{ij}^{X_{t-}}\rmd\tau_k(s)\frac{\rmd\varepsilon_{ij}(t)}{\rmd t}\right\rangle \nonumber\\
  &= \mathds{1}_{s>t}\sum_{i\neq j}\left\langle\alpha_{ij}^{X_{t-}}\gamma_{ij}(t)I_{ij}^{k}(s,t \!\mid\! X_{t-})\right\rangle\,\rmd s.
  \label{appeq: occupation_response_derivation}
\end{align}
For the jump increment,
\begin{align}
  \frac{\partial}{\partial\epsilon}\!&\left.\frac{\delta\left\langle\rmd n_{kl}(s)\right\rangle_{\epsilon h}}{\delta h(t)}\right|_{\epsilon=0} = \delta(s-t)\left\langle\alpha_{kl}^{X_{t-}}\gamma_{kl}(t)\right\rangle\,\rmd s \nonumber\\
  &+\mathds{1}_{s>t}\sum_{i\neq j}\left\langle\alpha_{ij}^{X_{t-}}\gamma_{ij}(t)J_{ij}^{kl}(s,t \!\mid\! X_{t-})\right\rangle\,\rmd s.
  \label{appeq: jump_response_derivation}
\end{align}
The equal-time term is present only for a jump observable that directly contains the perturbed event.

For a general additive observable of the form in \cref{eq: general_additive_observable}, with weights that have no explicit dependence on the perturbation parameter,
\begin{equation}
  R_\alpha^Q(\tau,t)\,\rmd t = \left\langle Q_\tau\,\rmd\Lambda_\alpha(t)\right\rangle = \sum_{i\neq j}\left\langle Q_\tau\alpha_{ij}^{X_{t-}}\,\rmd\varepsilon_{ij}(t)\right\rangle.
  \label{appeq: general_additive_response_score}
\end{equation}
Expanding the observable gives
\begin{align}
  R_\alpha^Q(\tau,t) ={}& \sum_{i\neq j}\int_0^\tau\sum_k\left\langle g_k^{X_{s-}}\alpha_{ij}^{X_{t-}}\rmd\tau_k(s)\frac{\rmd\varepsilon_{ij}(t)}{\rmd t}\right\rangle \nonumber\\
  &+\sum_{i\neq j}\int_0^\tau\sum_{k\neq l}\left\langle c_{kl}^{X_{s-}}\alpha_{ij}^{X_{t-}}\rmd n_{kl}(s)\frac{\rmd\varepsilon_{ij}(t)}{\rmd t}\right\rangle.
  \label{appeq: additive_observable_response_expansion}
\end{align}
When $g_k^{X_{s-}}$ or $c_{kl}^{X_{s-}}$ depends on the trajectory history, the weight remains inside the corresponding correlation. If the weights also depend explicitly on the perturbation parameter, the total response contains the additional direct term
\begin{equation}
  R_{\alpha,\mathrm{tot}}^Q(\tau,t) = R_\alpha^Q(\tau,t)+\left\langle\left.\frac{\partial}{\partial\epsilon}\frac{\delta Q_\tau(\epsilon h)}{\delta h(t)}\right|_{\epsilon=0}\right\rangle.
  \label{appeq: direct_observable_response}
\end{equation}

\section{Time-Domain Fluctuation--Response Relations and Kinetic Bounds}
\label{app:time_domain_frr_bounds}

This appendix derives the exact time-domain FRRs, the response-heterogeneity decomposition, the path-space Fisher-information FRI, and the activity-controlled R-KUR stated in Section~\ref{sec:time-domain_frr_bounds}. The only ingredients are the response representation, the space--time orthogonality of the martingale increments, and the assumption from Section~IV that the initial density is not perturbed.

\subsection{Exact Finite-Time Covariance and Variance FRRs}
\label{app:exact_time_domain_frr}

Define the initial and subsequently generated parts of the centered observable by
\begin{subequations}
\label{appeq: initial_dynamic_fluctuation_split}
\begin{align}
  \Delta_0 Q &\equiv \mathbb{E}\left[ Q_\tau \!\mid\! X_0 \right]-\left\langle Q_\tau\right\rangle, \\
  \widetilde{Q}_\tau &\equiv Q_\tau-\mathbb{E}\left[ Q_\tau \!\mid\! X_0 \right].
\end{align}
\end{subequations}
The response representation gives
\begin{equation}
  \widetilde{Q}_\tau = \sum_{i\neq j}\int_0^\tau \frac{R_{ij}^{Q}(\tau,t \!\mid\! X_{t-})}{\gamma_{ij}(t)}\rmd\varepsilon_{ij}(t).
  \label{appeq: response_representation}
\end{equation}
Because the stochastic integral is a martingale beginning at zero,
\begin{align}
  \left\langle \Delta_0 Q\,\widetilde{Q}_\tau'\right\rangle &= \left\langle \Delta_0 Q\,\mathbb{E}\left[\widetilde{Q}_\tau' \!\mid\! X_0\right]\right\rangle \nonumber\\
  &=0.
  \label{appeq: initial_dynamic_orthogonality}
\end{align}
Thus, the covariance separates into an initial contribution and a contribution generated during the observation interval,
\begin{equation}
  \operatorname{Cov}(Q_\tau,Q_\tau') = C_0(Q,Q')+\left\langle\widetilde{Q}_\tau\widetilde{Q}_\tau'\right\rangle.
  \label{appeq: covariance_initial_dynamic_split}
\end{equation}
Using the space--time orthogonality of the martingale increments in \cref{appeq: response_representation},
\begin{align}
  \left\langle\widetilde{Q}_\tau\widetilde{Q}_\tau'\right\rangle = \int_0^\tau\sum_{i\neq j}\left\langle \frac{R_{ij}^{Q}(\tau,t \!\mid\! X_{t-})R_{ij}^{Q'}(\tau,t \!\mid\! X_{t-})}{\gamma_{ij}(t)} \right\rangle\rmd t.
  \label{appeq: exact_dynamic_covariance_frr}
\end{align}
Combining the preceding equations yields \cref{eq: exact_covariance_frr}. Setting $Q_\tau'=Q_\tau$ gives \cref{eq: exact_variance_frr}.

The same result can be written directly in terms of the event-consequence kernels,
\begin{widetext}
\begin{equation}
  \operatorname{Cov}(Q_\tau,Q_\tau') = C_0(Q,Q') + \int_0^\tau \sum_{i\neq j} \left\langle \gamma_{ij}(t)K_{ij}^{Q}(\tau,t \!\mid\! X_{t-}) K_{ij}^{Q'}(\tau,t \!\mid\! X_{t-}) \right\rangle \rmd t.
  \label{appeq: exact_covariance_kernel_form}
\end{equation}
\end{widetext}
The total-variance identity gives
\begin{equation}
  \operatorname{Var}(Q_\tau) = \operatorname{Var}\left(\mathbb{E}\left[ Q_\tau \!\mid\! X_0 \right]\right)+\mathbb{E}\left[\operatorname{Var}(Q_\tau \!\mid\! X_0)\right],
  \label{appeq: total_variance_initial_split}
\end{equation}
so that $C_0(Q,Q)$ is the variance inherited from the initial information and the second term is the variance generated after time zero.

All ratios by $\gamma_{ij}(t)$ are defined only on histories with positive intensity. On histories with $\gamma_{ij}(t)=0$, the corresponding martingale increment has zero conditional variance and that edge makes no contribution; the ratio may therefore be set to zero without changing any identity.

\subsection{Exact History-Projection Decomposition and the Edge-Wise FRI}
\label{app:history_projection_fri}

For a fixed edge and perturbation time with $\left\langle\gamma_{ij}(t)\right\rangle>0$, define
\begin{equation}
  \overline{K}_{ij}^{Q}(\tau,t) = \frac{\left\langle\gamma_{ij}(t)K_{ij}^{Q}(\tau,t \!\mid\! X_{t-})\right\rangle}{\left\langle\gamma_{ij}(t)\right\rangle} = \frac{R_{ij}^{Q}(\tau,t)}{\left\langle\gamma_{ij}(t)\right\rangle}.
  \label{appeq: event_weighted_mean_kernel}
\end{equation}
Expanding the square around this event-weighted mean gives
\begin{widetext}
\begin{equation}
  \left\langle\gamma_{ij}(t)\left[K_{ij}^{Q}(\tau,t \!\mid\! X_{t-})\right]^2\right\rangle = \frac{\left[R_{ij}^{Q}(\tau,t)\right]^2}{\left\langle\gamma_{ij}(t)\right\rangle} + \left\langle\gamma_{ij}(t)\left[K_{ij}^{Q}(\tau,t \!\mid\! X_{t-}) - \overline{K}_{ij}^{Q}(\tau,t)\right]^2\right\rangle.
  \label{appeq: weighted_response_variance_decomposition}
\end{equation}
\end{widetext}
The cross term vanishes because
\begin{equation}
  \left\langle\gamma_{ij}(t)\left[K_{ij}^{Q}(\tau,t \!\mid\! X_{t-})-\overline{K}_{ij}^{Q}(\tau,t)\right]\right\rangle = 0.
  \label{appeq: weighted_kernel_centering}
\end{equation}
Summing over edges and integrating over time yields \cref{eq: frr_fri_decomposition}. Nonnegativity of the second term gives the edge-wise FRI \cref{eq: time_domain_fri_edge} without any additional approximation.

If $\left\langle\gamma_{ij}(t)\right\rangle=0$, nonnegativity of the intensity implies that $\gamma_{ij}(t)=0$ almost surely. The ordinary response and both terms associated with that edge are then zero. The equality condition is
\begin{equation}
  K_{ij}^{Q}(\tau,t \!\mid\! X_{t-}) = \overline{K}_{ij}^{Q}(\tau,t)
  \label{appeq: edge_fri_equality_condition}
\end{equation}
for all histories with positive intensity, up to sets of zero event weight.

The weighting in \cref{appeq: event_weighted_mean_kernel} has a direct event interpretation. Sampling a history in proportion to $\gamma_{ij}(t)$ is equivalent to sampling from the ensemble of potential $j\to i$ events at time $t$. The response-heterogeneity gap is therefore the variance of the event consequence in this event-weighted ensemble. It vanishes whenever the chosen event carries the same consequence on all active histories, even if other aspects of the dynamics remain history dependent.

\subsection{Path-Space Fisher Information and the Parameter-Wise FRI}
\label{app:path_fisher_time_domain_fri}

The trajectory score for the protocol $h(t)$ is
\begin{equation}
  \Lambda_{\alpha,h} = \sum_{i\neq j}\int_0^\tau h(t)\alpha_{ij}^{X_{t-}}\rmd\varepsilon_{ij}(t).
  \label{appeq: trajectory_score_common_basis}
\end{equation}
Using martingale orthogonality,
\begin{equation}
  \mathcal{I}_{\alpha,h} \equiv \left\langle\Lambda_{\alpha,h}^2\right\rangle = \int_0^\tau\sum_{i\neq j}\left\langle\gamma_{ij}(t)\left[h(t)\alpha_{ij}^{X_{t-}}\right]^2\right\rangle\rmd t.
  \label{appeq: fisher_information_derivation}
\end{equation}
Since the score is generated after the initial condition,
\begin{align}
  \left\langle\mathbb{E}\left[ Q_\tau \!\mid\! X_0 \right]\Lambda_{\alpha,h}\right\rangle &= \left\langle\mathbb{E}\left[ Q_\tau \!\mid\! X_0 \right]\mathbb{E}\left[\Lambda_{\alpha,h} \!\mid\! X_0\right]\right\rangle \nonumber\\
  &=0.
  \label{appeq: initial_score_orthogonality}
\end{align}
The score identity may therefore be written as
\begin{equation}
  \left.\partial_\epsilon\left\langle Q_\tau\right\rangle_{\epsilon h}\right|_{\epsilon=0} = \left\langle\left[Q_\tau-\mathbb{E}\left[ Q_\tau \!\mid\! X_0 \right]\right]\Lambda_{\alpha,h}\right\rangle.
  \label{appeq: centered_score_response_identity}
\end{equation}
Cauchy--Schwarz gives
\begin{align}
  \left|\left.\partial_\epsilon\left\langle Q_\tau\right\rangle_{\epsilon h}\right|_{\epsilon=0}\right|^2 &\leq \left\langle\left[Q_\tau-\mathbb{E}\left[ Q_\tau \!\mid\! X_0 \right]\right]^2\right\rangle\left\langle\Lambda_{\alpha,h}^2\right\rangle \nonumber\\
  &= \left[\operatorname{Var}(Q_\tau)-C_0(Q,Q)\right]\mathcal{I}_{\alpha,h},
  \label{appeq: fisher_fri_derivation}
\end{align}
which proves \cref{eq: general_time_domain_fri}. Equality holds if and only if the two centered random variables are proportional,
\begin{equation}
  Q_\tau-\mathbb{E}\left[ Q_\tau \!\mid\! X_0 \right] = c\Lambda_{\alpha,h}
  \label{appeq: fisher_fri_saturation}
\end{equation}
for a constant $c$, apart from degenerate zero-variance cases. This is an alignment condition between the chosen observable and perturbation score; it is independent of whether the edge-wise history-averaging gap vanishes.

For a constant unit logarithmic perturbation of the edge $j\to i$,
\begin{equation}
  h(t)=1, \qquad \alpha_{kl}^{X_{t-}}=\delta_{ki}\delta_{lj},
  \label{appeq: unit_edge_perturbation}
\end{equation}
so that
\begin{equation}
  \mathcal{I}_{ij}(\tau) = \int_0^\tau\left\langle\gamma_{ij}(t)\right\rangle\rmd t = \mathcal{A}_{ij}(\tau).
  \label{appeq: local_edge_fisher_equals_activity}
\end{equation}
Substitution into the Fisher-information FRI gives the local edge bound \cref{eq: local_edge_response_kur}.

\subsection{Activity Bounds and R-KUR Saturation}
\label{app:activity_response_kur}

A more resolved sensitivity envelope may be imposed edge by edge and time by time,
\begin{equation}
  \left|\alpha_{ij}^{X_{t-}}\right| \leq \alpha_{ij}^{\max}(t).
  \label{appeq: edge_dependent_sensitivity_envelope}
\end{equation}
Then
\begin{equation}
  \mathcal{I}_{\alpha,h} \leq \int_0^\tau h(t)^2\sum_{i\neq j}\left[\alpha_{ij}^{\max}(t)\right]^2\left\langle\gamma_{ij}(t)\right\rangle\rmd t.
  \label{appeq: refined_fisher_activity_bound}
\end{equation}
The global bound in the main text follows when $\alpha_{ij}^{\max}(t)\leq\alpha_{\max}$,
\begin{equation}
  \mathcal{I}_{\alpha,h} \leq \alpha_{\max}^2\int_0^\tau h(t)^2\sum_{i\neq j}\left\langle\gamma_{ij}(t)\right\rangle\rmd t = \alpha_{\max}^2\mathcal{A}_{h,\tau}.
  \label{appeq: global_fisher_activity_bound}
\end{equation}
The normalization of the parameter and protocol must be fixed when this inequality is used. Multiplying $h$ by a constant and dividing $\alpha$ by the same constant leaves the physical perturbation unchanged but changes the separate numerical values assigned to the two factors.

The Fisher-to-activity bound is saturated when
\begin{equation}
  \left|\alpha_{ij}^{X_{t-}}\right| = \alpha_{\max}
  \label{appeq: activity_bound_saturation}
\end{equation}
for every edge, history, and time carrying nonzero weight $h(t)^2\gamma_{ij}(t)$. The sign of the sensitivity need not be the same on different edges or histories because only its square enters the Fisher information. Full saturation of the R-KUR additionally requires the observable-score proportionality in \cref{appeq: fisher_fri_saturation}.

For $h(t)=1$,
\begin{equation}
  \mathcal{A}_{h,\tau}=\mathcal{A}_\tau=\int_0^\tau\sum_{i\neq j}\left\langle\gamma_{ij}(t)\right\rangle\rmd t,
  \label{appeq: constant_protocol_total_activity}
\end{equation}
which is the expected total number of resolved transitions during the observation interval.

\subsection{Numerical Evaluation of Fig.~\ref{fig:time_domain_frr_fri}}
\label{app:numerical_time_domain_frr}

The numerical illustration uses the three-state history-dependent model introduced in Fig.~1,
\begin{subequations}
\label{appeq: numerical_history_dependent_model}
\begin{align}
  r_{ij}^{X_{t-}} &= k_{ij}\exp\left[\beta\frac{m_j(t)}{\tau_m}\right], \\
  m_j(t) &= \int_0^t e^{-(t-s)/\tau_m}\mathds{1}_{x(s)=j} \, \rmd s.
\end{align}
\end{subequations}
The observable is $Q_\tau=\int_0^\tau\mathds{1}_{x(s)=1}\rmd s$, and the edge perturbation in panels (b)--(d) acts on $1\to2$. The weak- and strong-memory response clouds use $\beta=0.20$ and $\beta=7.00$, respectively, as indicated in the figure.

Trajectories may be generated event by event by evolving the memory variables deterministically between jumps and sampling the next event from the integrated total hazard. For each sampled pre-event history, $K_{21}^{Q}(\tau,t \!\mid\! X_{t-})$ is estimated from two conditional continuation ensembles: one in which a $1\to2$ event is appended at time $t$, and one in which the trajectory continues without an event during the same infinitesimal interval. The conditioned response is then obtained from $R_{21}^{Q}=\gamma_{21}K_{21}^{Q}$, while the ordinary response is the ensemble average over pre-event histories. The exact FRR, ordinary-response reconstruction, and response-heterogeneity gap are evaluated from \cref{eq: exact_variance_frr,eq: frr_fri_decomposition}. For the unit edge perturbation, the Fisher information is estimated from the mean number of $1\to2$ events, as in \cref{eq: local_edge_fisher_activity}.

\section{Derivations of stationary finite-frequency relations}
\label{appsec: stationary_finite_frequency_derivations}

\subsection{Stationary Assumptions and Spectral Conventions}
\label{appsec: stationary_spectral_assumptions}

We consider a stationary trajectory ensemble with an infinite prehistory. The history-dependent intensities and observable weights are assumed to be invariant under a common shift of the trajectory and time. The observable rates are real generalized processes, so jump contributions may contain delta peaks. Their covariance kernels and causal response kernels are assumed to have finite Fourier transforms, either as ordinary functions or as limits of finite-window transforms.

The notation
\begin{equation}
  \int_{[0,\infty)} f(t)\,\rmd t
  \label{appeq: causal_half_line_convention}
\end{equation}
includes the complete equal-time mass of a jump contribution at $t=0$. Thus, an instantaneous jump response is not divided by two when the causal Fourier transform is taken.

A stationary process may retain a random component fixed by the remote past. Let $\mathcal{F}_{-\infty}$ denote the information common to the entire stationary prehistory and define
\begin{equation}
  Z_Q(t) \equiv \mathbb{E}\!\left[ \dot{Q}(t) \!\mid\! \mathcal{F}_{-\infty} \right]-\left\langle \dot{Q} \right\rangle_{\mathrm{ss}}.
  \label{appeq: remote_past_component}
\end{equation}
The stationary martingale representation then has the general form
\begin{align}
  \dot{Q}(t)-\left\langle \dot{Q} \right\rangle_{\mathrm{ss}} ={}& Z_Q(t) \nonumber\\
  &+\sum_{i\neq j}\int_{-\infty}^{t} K_{ij}^{Q}(t-s \!\mid\! X_{s-})\dot{\varepsilon}_{ij}(s)\,\rmd s.
  \label{appeq: stationary_representation_with_remote_past}
\end{align}
The remote-past component is orthogonal to the martingale-generated contribution because the latter has zero conditional mean given $\mathcal{F}_{-\infty}$. Consequently,
\begin{align}
  \mathcal{S}_{QQ'}(\omega) ={}& \mathcal{S}_{\infty,QQ'}(\omega) \nonumber\\
  &+\sum_{i\neq j}\left\langle \gamma_{ij}(0)\mathcal{K}_{ij}^{Q}(\omega \!\mid\! X_{0-})\mathcal{K}_{ij}^{Q'}(-\omega \!\mid\! X_{0-}) \right\rangle_{\mathrm{ss}},
  \label{appeq: general_stationary_spectral_frr}
\end{align}
where $\mathcal{S}_{\infty,QQ'}$ is the cross spectrum of $Z_Q$ and $Z_{Q'}$. The main text considers stationary regimes with no additional random sector fixed by the remote past, so that $Z_Q(t)=0$ and $\mathcal{S}_{\infty,QQ'}(\omega)=0$. In a stationary mixture with a persistent sector label, this additional term must be retained and may contain singular spectral contributions.

All ratios involving $\gamma_{ij}$ are understood on the active support. If $\left\langle\gamma_{ij}(0)\right\rangle_{\mathrm{ss}}=0$, nonnegativity of the intensity implies that the edge is inactive almost surely and its contribution is defined to be zero.

\subsection{Exact Stationary Finite-frequency FRRs}
\label{appsec: stationary_finite_frequency_frr_derivation}

Under the stationary assumptions of the main text, the observable-rate fluctuation is
\begin{subequations}
\label{appeq: stationary_martingale_response_representation}
\begin{align}
  \dot{Q}(t)-\left\langle \dot{Q} \right\rangle_{\mathrm{ss}} &= \sum_{i\neq j}\int_{-\infty}^{t} K_{ij}^{Q}(t-s \!\mid\! X_{s-})\dot{\varepsilon}_{ij}(s)\,\rmd s \\
  &= \sum_{i\neq j}\int_{-\infty}^{t} \frac{R_{ij}^{Q}(t-s \!\mid\! X_{s-})}{\gamma_{ij}(s)}\dot{\varepsilon}_{ij}(s)\,\rmd s.
\end{align}
\end{subequations}
Using this representation for $\dot{Q}(u)$ and $\dot{Q}'(0)$ and applying martingale orthogonality, we obtain, for any $u\in\mathbb{R}$,
\begin{align}
  \operatorname{Cov}\!\left( \dot{Q}(u),\dot{Q}'(0) \right) ={}& \sum_{i\neq j}\int_{-\infty}^{\min(u,0)} \Bigg\langle \frac{R_{ij}^{Q}(u-s \!\mid\! X_{s-})}{\gamma_{ij}(s)} \nonumber\\
  &\qquad\qquad\times R_{ij}^{Q'}(-s \!\mid\! X_{s-}) \Bigg\rangle_{\mathrm{ss}}\rmd s.
  \label{appeq: stationary_time_correlation_response_form}
\end{align}
The upper limit $\min(u,0)$ ensures that the martingale increment at time $s$ lies in the causal past of both observable rates. Taking the Fourier transform gives
\begin{align}
  \mathcal{S}_{QQ'}(\omega) ={}& \sum_{i\neq j}\int_{-\infty}^{\infty}\rmd u\int_{-\infty}^{\min(u,0)}\rmd s\,e^{\mathrm{i}\omega u} \nonumber\\
  &\times\left\langle \frac{R_{ij}^{Q}(u-s \!\mid\! X_{s-})R_{ij}^{Q'}(-s \!\mid\! X_{s-})}{\gamma_{ij}(s)} \right\rangle_{\mathrm{ss}}.
  \label{appeq: stationary_cross_spectrum_before_change_variables}
\end{align}
We introduce
\begin{equation}
  a=u-s, \qquad b=-s.
  \label{appeq: stationary_frequency_change_variables}
\end{equation}
The conditions $s\leq0$ and $s\leq u$ become $a\geq0$ and $b\geq0$, while $u=a-b$. By stationarity, the shifted history and intensity at time $s$ can be replaced by $X_{0-}$ and $\gamma_{ij}(0)$. Hence,
\begin{subequations}
\label{appeq: stationary_exact_cross_spectrum_derivation}
\begin{align}
  \mathcal{S}_{QQ'}(\omega) &= \sum_{i\neq j}\int_0^\infty\rmd a\int_0^\infty\rmd b\,e^{\mathrm{i}\omega(a-b)} \nonumber\\
  &\qquad\times\left\langle \gamma_{ij}(0)K_{ij}^{Q}(a \!\mid\! X_{0-})K_{ij}^{Q'}(b \!\mid\! X_{0-}) \right\rangle_{\mathrm{ss}} \\
  &= \sum_{i\neq j}\left\langle \gamma_{ij}(0)\mathcal{K}_{ij}^{Q}(\omega \!\mid\! X_{0-})\mathcal{K}_{ij}^{Q'}(-\omega \!\mid\! X_{0-}) \right\rangle_{\mathrm{ss}} \\
  &= \sum_{i\neq j}\left\langle \frac{\mathcal{R}_{ij}^{Q}(\omega \!\mid\! X_{0-})\mathcal{R}_{ij}^{Q'}(-\omega \!\mid\! X_{0-})}{\gamma_{ij}(0)} \right\rangle_{\mathrm{ss}}.
\end{align}
\end{subequations}
This proves \cref{eq: exact_finite_frequency_cross_frr}. Setting $Q'=Q$ gives \cref{eq: exact_finite_frequency_frr}.

For real observables,
\begin{equation}
  \mathcal{S}_{QQ'}(-\omega)=\mathcal{S}_{Q'Q}(\omega)=\mathcal{S}_{QQ'}(\omega)^*.
  \label{appeq: stationary_cross_spectrum_symmetry}
\end{equation}
If $\rmd Q(t)$ contains the jump increment $c_{ij}^{X_{t-}}\rmd n_{ij}(t)$, the response contains an equal-time contribution at the origin. The convention in \cref{appeq: causal_half_line_convention} includes this complete contribution and reproduces the corresponding shot-noise term in the power spectrum.

\subsection{Exact Complex History-Projection Decomposition}
\label{appsec: stationary_finite_frequency_fri_derivation}

For an active edge, define
\begin{align}
  \overline{\mathcal{K}}_{ij}^{Q}(\omega) &= \frac{\left\langle\gamma_{ij}(0)\mathcal{K}_{ij}^{Q}(\omega \!\mid\! X_{0-})\right\rangle_{\mathrm{ss}}}{\left\langle\gamma_{ij}(0)\right\rangle_{\mathrm{ss}}} \nonumber\\
  &= \frac{\mathcal{R}_{ij}^{Q}(\omega)}{\left\langle\gamma_{ij}(0)\right\rangle_{\mathrm{ss}}}.
  \label{appeq: stationary_weighted_mean_complex_kernel}
\end{align}
Expanding around this event-weighted mean gives
\begin{align}
  \Big\langle \gamma_{ij}(0) &\left| \mathcal{K}_{ij}^{Q}(\omega \!\mid\! X_{0-}) \right|^2 \Big\rangle_{\mathrm{ss}} = \frac{\left|\mathcal{R}_{ij}^{Q}(\omega)\right|^2}{\left\langle\gamma_{ij}(0)\right\rangle_{\mathrm{ss}}} \nonumber\\
  &+\Bigg\langle\gamma_{ij}(0)\Bigg|\mathcal{K}_{ij}^{Q}(\omega \!\mid\! X_{0-}) - \overline{\mathcal{K}}_{ij}^{Q}(\omega)\Bigg|^2\Bigg\rangle_{\mathrm{ss}}.
  \label{appeq: stationary_complex_variance_decomposition}
\end{align}
The cross term vanishes because
\begin{equation}
  \left\langle\gamma_{ij}(0)\left[\mathcal{K}_{ij}^{Q}(\omega \!\mid\! X_{0-})-\overline{\mathcal{K}}_{ij}^{Q}(\omega)\right]\right\rangle_{\mathrm{ss}}=0.
  \label{appeq: stationary_weighted_complex_centering}
\end{equation}
Summing over edges and using the exact power-spectrum FRR yields \cref{eq: finite_frequency_frr_fri_decomposition}. Nonnegativity of the second term gives the edge-wise finite-frequency FRI.

Equality holds when
\begin{equation}
  \mathcal{K}_{ij}^{Q}(\omega \!\mid\! X_{0-})=\overline{\mathcal{K}}_{ij}^{Q}(\omega)
  \label{appeq: stationary_edge_fri_saturation}
\end{equation}
for every history carrying nonzero event weight. Because the kernel is complex, this condition requires both its magnitude and phase to be history independent on the active ensemble. It is a condition on the selected observable, edge family, and frequency; it does not imply that the full dynamics is Markovian.

\subsection{Stationary Fisher-Information and Kinetic Bounds}
\label{appsec: stationary_fisher_information_derivation}

For a general stationary perturbation direction,
\begin{equation}
  \mathcal{R}_{\alpha}^{Q}(\omega) = \sum_{i\neq j}\left\langle\gamma_{ij}(0)\alpha_{ij}^{X_{0-}}\mathcal{K}_{ij}^{Q}(\omega \!\mid\! X_{0-})\right\rangle_{\mathrm{ss}}.
  \label{appeq: stationary_general_response_kernel_form}
\end{equation}
Applying complex Cauchy--Schwarz gives
\begin{align}
  \left|\mathcal{R}_{\alpha}^{Q}(\omega)\right|^2 \leq{}& \left[\sum_{i\neq j}\left\langle\gamma_{ij}(0)\left(\alpha_{ij}^{X_{0-}}\right)^2\right\rangle_{\mathrm{ss}}\right] \nonumber\\
  &\times\left[\sum_{i\neq j}\left\langle\gamma_{ij}(0)\left|\mathcal{K}_{ij}^{Q}(\omega \!\mid\! X_{0-})\right|^2\right\rangle_{\mathrm{ss}}\right].
  \label{appeq: stationary_complex_cauchy_schwarz}
\end{align}
The first factor is $\mathcal{I}_{\alpha}$ and the second is $\mathcal{S}_Q(\omega)$ by the exact FRR. Therefore,
\begin{equation}
  \left|\mathcal{R}_{\alpha}^{Q}(\omega)\right|^2 \leq \mathcal{I}_{\alpha}\mathcal{S}_Q(\omega).
  \label{appeq: stationary_fisher_fri_derivation}
\end{equation}
Equality holds when
\begin{equation}
  \mathcal{K}_{ij}^{Q}(\omega \!\mid\! X_{0-})=c(\omega)\alpha_{ij}^{X_{0-}}
  \label{appeq: stationary_fisher_fri_saturation}
\end{equation}
for a complex constant $c(\omega)$ over all active histories and transition channels. The common complex factor fixes the response phase of the observable mode relative to the perturbation direction.

If $|\alpha_{ij}^{X_{0-}}|\leq\alpha_{\max}$, then
\begin{align}
  \mathcal{I}_{\alpha} &= \sum_{i\neq j}\left\langle\gamma_{ij}(0)\left(\alpha_{ij}^{X_{0-}}\right)^2\right\rangle_{\mathrm{ss}} \nonumber\\
  &\leq \alpha_{\max}^2\sum_{i\neq j}\left\langle\gamma_{ij}(0)\right\rangle_{\mathrm{ss}} = \alpha_{\max}^2\mathcal{A}.
  \label{appeq: stationary_activity_bound_derivation}
\end{align}
Combining this result with \cref{appeq: stationary_fisher_fri_derivation} gives the stationary finite-frequency R-KUR. The activity bound is saturated when $|\alpha_{ij}^{X_{0-}}|=\alpha_{\max}$ on every active edge and history. Full saturation additionally requires the alignment condition in \cref{appeq: stationary_fisher_fri_saturation}.

The quantity $\mathcal{I}_{\alpha}$ is a Fisher-information rate per unit protocol power. For the unit-modulus complex harmonic convention $h(t)=e^{-\mathrm{i}\omega t}$, $|h(t)|^2=1$. A physical cosine or sine protocol uses the corresponding real quadrature and its actual squared amplitude in the Fisher information.

\subsection{Zero-Frequency and Long-Time Limits}
\label{appsec: stationary_zero_frequency_limit}

Let
\begin{equation}
  Q_\tau = \int_{(0,\tau]} \rmd Q(t)
  \label{appeq: accumulated_stationary_observable}
\end{equation}
be the accumulated observable. Stationarity gives
\begin{equation}
  \operatorname{Var}(Q_\tau) = \int_{-\tau}^{\tau}(\tau-|u|)\operatorname{Cov}\!\left(\dot{Q}(u),\dot{Q}(0)\right)\,\rmd u.
  \label{appeq: accumulated_variance_stationary_correlation}
\end{equation}
If the covariance kernel is integrable,
\begin{align}
  \lim_{\tau\to\infty}\frac{\operatorname{Var}(Q_\tau)}{\tau} &= \int_{-\infty}^{\infty}\operatorname{Cov}\!\left(\dot{Q}(u),\dot{Q}(0)\right)\,\rmd u \nonumber\\
  &= \mathcal{S}_Q(0).
  \label{appeq: variance_rate_zero_frequency}
\end{align}
Similarly, a constant perturbation applied throughout $[0,\tau]$ produces
\begin{equation}
  \left.\partial_\epsilon\left\langle Q_\tau\right\rangle_\epsilon\right|_{\epsilon=0} = \int_0^\tau\rmd s\int_{[s,\tau]} R_{\alpha}^{Q}(t-s)\,\rmd t.
  \label{appeq: accumulated_stationary_response}
\end{equation}
Changing variables and taking the long-time limit gives
\begin{equation}
  \lim_{\tau\to\infty}\frac{1}{\tau}\left.\partial_\epsilon\left\langle Q_\tau\right\rangle_\epsilon\right|_{\epsilon=0} = \int_{[0,\infty)}R_{\alpha}^{Q}(u)\,\rmd u = \mathcal{R}_{\alpha}^{Q}(0).
  \label{appeq: response_rate_zero_frequency}
\end{equation}
The finite-window spectral power obeys
\begin{equation}
  \frac{1}{\tau}\mathcal{S}_Q^\tau(\omega,\omega) = \int_{-\tau}^{\tau}\left(1-\frac{|u|}{\tau}\right)e^{\mathrm{i}\omega u}\operatorname{Cov}\!\left(\dot{Q}(u),\dot{Q}(0)\right)\,\rmd u,
  \label{appeq: finite_window_stationary_spectrum}
\end{equation}
and therefore converges to $\mathcal{S}_Q(\omega)$ when the stationary spectrum exists.

\subsection{Numerical Evaluation of Fig.~\ref{fig:frequency_domain_response}}
\label{appsec: numerical_stationary_frequency}

The numerical illustration uses the stationary version of the three-state history-dependent model introduced in Fig.~1,
\begin{subequations}
\label{appeq: stationary_numerical_memory_model}
\begin{align}
  r_{ij}^{X_{t-}} &= k_{ij}\exp\!\left[\beta\frac{m_j(t)}{\tau_m}\right], \\
  m_j(t) &= \int_{-\infty}^{t}e^{-(t-s)/\tau_m}\mathds{1}_{x(s)=j}\,\rmd s.
\end{align}
\end{subequations}
In simulation, the infinite prehistory is approximated by a burn-in interval long compared with $\tau_m$. Panels (b)--(d) use the values shown in the figure: weak memory with $\beta=0.20$, strong memory with $\beta=5.00$, and the displayed frequencies $\omega=1.00$ and $\omega=100.00$ in the numerical time units.

For each sampled stationary pre-event history, the causal event-consequence kernel is estimated by two conditional continuation ensembles: one in which the selected event is appended at the perturbation time and one in which the trajectory continues without an event during the same infinitesimal interval. Fourier transformation of the estimated time-domain kernel gives $\mathcal{K}_{21}^{Q}(\omega \!\mid\! X_{0-})$, and the conditioned response is obtained from $\mathcal{R}_{21}^{Q}=\gamma_{21}(0)\mathcal{K}_{21}^{Q}$. The ordinary response is the ensemble average over histories. The exact FRR reconstruction, ordinary-response contribution, and frequency-resolved gap are evaluated from \cref{eq: exact_finite_frequency_frr,eq: finite_frequency_frr_fri_decomposition}.

\section{Derivations of dynamical two-frequency relations}
\label{appsec: dynamical_two_frequency_derivations}

\subsection{Finite-Window Martingale Representation and Initial Spectrum}
\label{appsec: dynamical_two_frequency_frr_derivation}

The time-domain martingale representation of the observable-rate fluctuation is
\begin{align}
  \dot{Q}(t)-\mathbb{E}\!\left[\dot{Q}(t) \!\mid\! X_0\right] ={}& \sum_{i\neq j}\int_0^t \frac{R_{ij}^{Q}(t,s \!\mid\! X_{s-})}{\gamma_{ij}(s)} \nonumber\\
  &\qquad\qquad\times\dot{\varepsilon}_{ij}(s)\,\rmd s.
  \label{appeq: finite_window_rate_representation}
\end{align}
The initial conditional mean is determined at time zero and is orthogonal to the subsequent martingale integral. Hence,
\begin{align}
  \label{appeq: finite_window_time_covariance_response}
  \operatorname{Cov}&\left(\dot{Q}(t),\dot{Q}'(t')\right) = C_{0,QQ'}(t,t') \\
  &+\sum_{i\neq j}\int_0^{\min(t,t')} \Bigg\langle \frac{R_{ij}^{Q}(t,s \!\mid\! X_{s-})}{\gamma_{ij}(s)} R_{ij}^{Q'}(t',s \!\mid\! X_{s-}) \Bigg\rangle \rmd s. \nonumber
\end{align}
Taking the finite-window Fourier transform yields
\begin{widetext}
\begin{equation}
  \mathcal{S}_{QQ'}^{\tau}(\omega,\omega')-\mathcal{S}_{0,QQ'}^{\tau}(\omega,\omega') = \int_0^\tau\rmd t\int_0^\tau\rmd t'\,e^{\mathrm{i}\omega t}e^{-\mathrm{i}\omega't'} \int_0^{\min(t,t')}\sum_{i\neq j}\left\langle \frac{R_{ij}^{Q}(t,s \!\mid\! X_{s-})R_{ij}^{Q'}(t',s \!\mid\! X_{s-})}{\gamma_{ij}(s)} \right\rangle \rmd s.
  \label{appeq: two_frequency_before_exchange}
\end{equation}
\end{widetext}
Exchanging the order of integration, $t$ and $t'$ run independently over $[s,\tau]$ for each perturbation time $s$. Therefore,
\begin{widetext}
\begin{subequations}
\label{appeq: exact_two_frequency_derivation}
\begin{align}
  \mathcal{S}_{QQ'}^{\tau}(\omega,\omega')-\mathcal{S}_{0,QQ'}^{\tau}(\omega,\omega') ={}& \int_0^\tau\sum_{i\neq j} \left\langle \gamma_{ij}(s) \mathcal{K}_{ij}^{Q,\tau}(\omega,s \!\mid\! X_{s-}) \mathcal{K}_{ij}^{Q',\tau}(-\omega',s \!\mid\! X_{s-}) \right\rangle \rmd s \\
  ={}& \int_0^\tau\sum_{i\neq j} \left\langle \frac{\mathcal{R}_{ij}^{Q,\tau}(\omega,s \!\mid\! X_{s-}) \mathcal{R}_{ij}^{Q',\tau}(-\omega',s \!\mid\! X_{s-})}{\gamma_{ij}(s)} \right\rangle \rmd s.
\end{align}
\end{subequations}
\end{widetext}
This proves \cref{eq: dynamical_two_frequency_frr}. For real observables,
\begin{equation}
  \mathcal{S}_{QQ'}^{\tau}(\omega,\omega') = \left[\mathcal{S}_{Q'Q}^{\tau}(\omega',\omega)\right]^*.
  \label{appeq: finite_window_hermitian_symmetry}
\end{equation}
If the complete initial information is fixed, $C_{0,QQ'}(t,t')=0$ and the initial spectrum vanishes.

\subsection{Exact Dynamical Gap and Finite-Window FRI}
\label{appsec: dynamical_two_frequency_fri_derivation}

For an active edge at perturbation time $s$, define
\begin{align}
  \overline{\mathcal{K}}_{ij}^{Q,\tau}(\omega,s) &= \frac{\left\langle\gamma_{ij}(s)\mathcal{K}_{ij}^{Q,\tau}(\omega,s \!\mid\! X_{s-})\right\rangle}{\left\langle\gamma_{ij}(s)\right\rangle} \nonumber\\
  &= \frac{\mathcal{R}_{ij}^{Q,\tau}(\omega,s)}{\left\langle\gamma_{ij}(s)\right\rangle}.
  \label{appeq: dynamical_weighted_mean_kernel}
\end{align}
Completing the square at each $s$ give
\begin{align}
  \label{appeq: dynamical_complex_variance_decomposition}
  \Big\langle \gamma_{ij}(s) &\left| \mathcal{K}_{ij}^{Q,\tau}(\omega,s \!\mid\! X_{s-}) \right|^2 \Big\rangle = \frac{\left| \mathcal{R}_{ij}^{Q,\tau}(\omega,s) \right|^2}{\left\langle \gamma_{ij}(s)\right\rangle} \\
  &+ \left\langle \gamma_{ij}(s) \left| \mathcal{K}_{ij}^{Q,\tau}(\omega,s \!\mid\! X_{s-}) - \overline{\mathcal{K}}_{ij}^{Q,\tau}(\omega,s) \right|^2 \right\rangle. \nonumber
\end{align}
Integrating over $s$ and summing over edges yields the exact decomposition in \cref{eq: dynamical_frr_fri_decomposition}. Nonnegativity gives the finite-window edge-wise FRI. Equality holds when
\begin{equation}
  \mathcal{K}_{ij}^{Q,\tau}(\omega,s \!\mid\! X_{s-})=\overline{\mathcal{K}}_{ij}^{Q,\tau}(\omega,s)
  \label{appeq: dynamical_edge_fri_saturation}
\end{equation}
for all active histories at almost every perturbation time $s$. If $\left\langle\gamma_{ij}(s)\right\rangle=0$, the corresponding edge contribution is zero.

\subsection{General Protocols and Frequency-Conversion Bounds}
\label{appsec: dynamical_frequency_conversion_bounds}

Using $\mathcal{R}_{ij}^{Q,\tau}=\gamma_{ij}\mathcal{K}_{ij}^{Q,\tau}$, the output-frequency response to a general protocol is
\begin{align}
  \mathcal{R}_{\alpha,h}^{Q,\tau}(\omega) ={}& \int_0^\tau\sum_{i\neq j}\Big\langle\gamma_{ij}(s)h(s)\alpha_{ij}^{X_{s-}}(s) \nonumber\\
  &\qquad\qquad\times\mathcal{K}_{ij}^{Q,\tau}(\omega,s \!\mid\! X_{s-})\Big\rangle\rmd s.
  \label{appeq: general_protocol_kernel_response}
\end{align}
Complex Cauchy--Schwarz gives
\begin{align}
  \left| \mathcal{R}_{\alpha,h}^{Q,\tau}(\omega) \right|^2 \leq{}& \mathcal{I}_{\alpha,h}^{\tau} \int_0^\tau \sum_{i\neq j} \Bigg\langle \gamma_{ij}(s) \left| \mathcal{K}_{ij}^{Q,\tau}(\omega,s \!\mid\! X_{s-}) \right|^2 \Bigg\rangle \rmd s \\
  ={}& \mathcal{I}_{\alpha,h}^{\tau} \left[ \mathcal{S}_{Q}^{\tau}(\omega,\omega) - \mathcal{S}_{0,Q}^{\tau}(\omega,\omega) \right].
  \label{appeq: dynamical_complex_cauchy_schwarz}
\end{align}
This proves \cref{eq: dynamical_general_protocol_fri}. Equality holds when
\begin{equation}
  \mathcal{K}_{ij}^{Q,\tau}(\omega,s \!\mid\! X_{s-}) = c(\omega,h)h(s)^*\alpha_{ij}^{X_{s-}}(s)
  \label{appeq: dynamical_fisher_fri_saturation}
\end{equation}
for a complex constant $c(\omega,h)$ over the active edge--history--time support.

For $h(s)=e^{-\mathrm{i}\nu s}$, $|h(s)|^2=1$ and the general expressions reduce to \cref{eq: dynamical_two_frequency_response_alpha,eq: dynamical_fisher_activity_alpha,eq: dynamical_general_fri}. For a real cosine or sine protocol, the ordinary Fisher information is obtained by inserting the squared real protocol rather than replacing it by unity.

If $|\alpha_{ij}^{X_{s-}}(s)|\leq\alpha_{\max}$, then
\begin{equation}
  \mathcal{I}_{\alpha,h}^{\tau} \leq \alpha_{\max}^2\int_0^\tau |h(s)|^2\sum_{i\neq j}\left\langle\gamma_{ij}(s)\right\rangle\rmd s = \alpha_{\max}^2\mathcal{A}_{h,\tau}.
  \label{appeq: dynamical_protocol_activity_bound}
\end{equation}
For a unit-modulus harmonic protocol, $\mathcal{A}_{h,\tau}=\mathcal{A}_{\tau}$, giving the two-frequency R-KUR in the main text. Full saturation requires both \cref{appeq: dynamical_fisher_fri_saturation} and $|\alpha_{ij}^{X_{s-}}(s)|=\alpha_{\max}$ on the protocol-weighted active support.

\subsection{Finite-Window Leakage Versus Genuine Frequency Conversion}
\label{appsec: finite_window_leakage}

A finite observation window can generate off-diagonal output even when the underlying response is stationary. Suppose first that a stationary system is driven harmonically for long enough that boundary transients can be neglected. Its mean response rate is
\begin{equation}
  \delta\left\langle\dot{Q}(t)\right\rangle = \epsilon h_{\nu}\mathcal{R}_{\alpha}^{Q}(\nu)e^{-\mathrm{i}\nu t}.
  \label{appeq: stationary_harmonic_output}
\end{equation}
After multiplication by an observation window $w_{\tau}(t)$, the output-frequency amplitude is
\begin{subequations}
\label{appeq: stationary_window_leakage}
\begin{align}
  \delta\widetilde{Q}_{w}(\omega) &= \epsilon h_{\nu}\mathcal{R}_{\alpha}^{Q}(\nu)\widetilde{w}_{\tau}(\omega-\nu), \\
  \widetilde{w}_{\tau}(\Omega) &= \int w_{\tau}(t)e^{\mathrm{i}\Omega t}\,\rmd t.
\end{align}
\end{subequations}
For a rectangular window on $[0,\tau]$,
\begin{equation}
  \widetilde{w}_{\tau}(\Omega) = e^{\mathrm{i}\Omega\tau/2}\tau\,\operatorname{sinc}\!\left(\frac{\Omega\tau}{2}\right).
  \label{appeq: rectangular_window_transform}
\end{equation}
Thus, a finite window broadens the stationary diagonal $\omega=\nu$ by a known leakage kernel.

For a stationary causal response switched on at the beginning of the window, the exact finite-start benchmark is
\begin{equation}
  \mathcal{R}_{\alpha,\mathrm{stat}}^{Q,\tau}(\omega,\nu) = \int_{[0,\tau]} e^{\mathrm{i}\omega u}R_{\alpha}^{Q}(u)W_{\tau-u}(\omega-\nu)\,\rmd u,
  \label{appeq: stationary_finite_start_frequency_response}
\end{equation}
where
\begin{equation}
  W_T(\Omega) = \int_0^T e^{\mathrm{i}\Omega s}\,\rmd s.
  \label{appeq: finite_interval_window_kernel}
\end{equation}
This expression includes both spectral leakage and the boundary transient associated with starting the drive at $t=0$. In the long-time stationary limit, the response becomes concentrated on the frequency diagonal. Genuine frequency conversion refers to additional input--output mixing that cannot be accounted for by this stationary windowed benchmark. It is this additional structure that the nonstationary two-frequency response resolves.

\section{Monotonicity of the response-sufficiency hierarchy}
\label{appsec: response_sufficiency_monotonicity}

Consider two candidate memory states $Y_{t-}$ and $Y_{t-}'$ satisfying
\begin{equation}
  \sigma(Y_{t-}) \subseteq \sigma(Y_{t-}') \subseteq \mathcal{F}_{t-}.
  \label{appeq: nested_candidate_memory_states}
\end{equation}
For compactness, we suppress the arguments $(\tau,t)$ and write $K_{ij}^{Q}(X_{t-})$, $K_{ij}^{Q}(Y_{t-}')$, and $K_{ij}^{Q}(Y_{t-})$ for the corresponding normalized responses. Their difference can be decomposed as
\begin{align}
  K_{ij}^{Q}(X_{t-})-K_{ij}^{Q}(Y_{t-}) ={}& \left[ K_{ij}^{Q}(X_{t-})-K_{ij}^{Q}(Y_{t-}') \right] \nonumber \\
  &+ \left[ K_{ij}^{Q}(Y_{t-}')-K_{ij}^{Q}(Y_{t-}) \right].
  \label{appeq: nested_response_coefficient_decomposition}
\end{align}
The activity-weighted projection in \cref{eq: reduced_state_activity_weighted_projection} implies
\begin{equation}
  \mathbb{E}\left[ \, \gamma_{ij}(t)\left[ K_{ij}^{Q}(X_{t-})-K_{ij}^{Q}(Y_{t-}') \right] \,\bigg|\, Y_{t-}' \, \right] = 0.
  \label{appeq: weighted_projection_orthogonality}
\end{equation}
Because $K_{ij}^{Q}(Y_{t-}')-K_{ij}^{Q}(Y_{t-})$ is measurable with respect to $Y_{t-}'$, the cross term vanishes,
\begin{align}
  &\Bigg\langle \gamma_{ij}(t) \left[ K_{ij}^{Q}(X_{t-})-K_{ij}^{Q}(Y_{t-}') \right] \\
  &\times \left[ K_{ij}^{Q}(Y_{t-}')-K_{ij}^{Q}(Y_{t-}) \right] \Bigg\rangle \nonumber \\
  ={}& \Bigg\langle \left[ K_{ij}^{Q}(Y_{t-}')-K_{ij}^{Q}(Y_{t-}) \right] \\
  &\times \mathbb{E}\left[ \, \gamma_{ij}(t)\left[ K_{ij}^{Q}(X_{t-})-K_{ij}^{Q}(Y_{t-}') \right] \,\bigg|\, Y_{t-}' \, \right] \Bigg\rangle \nonumber \\
  ={}& 0.
\end{align}
  \label{appeq: nested_response_cross_term}
Squaring \cref{appeq: nested_response_coefficient_decomposition}, multiplying by $\gamma_{ij}(t)$, averaging, summing over edges, and integrating over time therefore gives
\begin{align}
  \mathcal{M}_Q^{(Y)} ={}& \mathcal{M}_Q^{(Y')} + \int_0^\tau \sum_{i \neq j} \Bigg\langle \gamma_{ij}(t) \nonumber \\
  &\times \left[ K_{ij}^{Q}(\tau,t \!\mid\! Y_{t-}')-K_{ij}^{Q}(\tau,t \!\mid\! Y_{t-}) \right]^2 \Bigg\rangle \,\rmd t.
  \label{appeq: nested_memory_gap_decomposition}
\end{align}
The second term is nonnegative, and hence
\begin{equation}
  \mathcal{M}_Q^{(Y')} \leq \mathcal{M}_Q^{(Y)}.
  \label{appeq: nested_memory_gap_monotonicity}
\end{equation}
Thus, refining a candidate memory state cannot increase the residual response-heterogeneity gap. Equality holds when the additional information retained by $Y_{t-}'$ does not change the normalized conditioned response on the active histories.

\section{Experimental Analysis of BK-Channel Gating Trajectories}
\label{appsec: experimental_bk_analysis}

\subsection{Trajectory Reconstruction and Future Open-Time Observable}
\label{appsec: experimental_bk_trajectory_reconstruction}

The public dataset \cite{agata_wawrzkiewicz_jalowiecka_2022_6340407} contains thresholded dwell-time sequences for 30 cell-membrane BK patches and 26 mitochondrial BK patches. The accompanying metadata specify the channel class, membrane voltage, and state of the first dwell. Let $\tau_n>0$ be the $n$th dwell duration and $s_n\in\{C,O\}$ its state. The transition times are
\begin{equation}
  T_n \equiv \sum_{m=1}^{n}\tau_m,
  \label{appeq: experimental_bk_transition_times}
\end{equation}
and the reconstructed trajectory is
\begin{equation}
  x_t=s_n, \qquad T_{n-1}\leq t<T_n.
  \label{appeq: experimental_bk_reconstructed_trajectory}
\end{equation}
The supplied states alternate after every dwell. The sampling interval is $0.1\,\mathrm{ms}$, corresponding to the reported $10\,\mathrm{kHz}$ acquisition frequency. We retain the supplied dwell durations without additional censoring or merging.

For patch $p$, let $\widetilde{\tau}_{O,p}$ and $\widetilde{\tau}_{C,p}$ denote the median open and closed dwell durations. The horizon used in the fixed-time analysis is
\begin{equation}
  \Delta_p \equiv 4\left( \widetilde{\tau}_{O,p}+\widetilde{\tau}_{C,p} \right).
  \label{appeq: experimental_bk_patch_horizon}
\end{equation}
The numerical estimator uses the future open fraction
\begin{equation}
  q_{\Delta_p}(t) \equiv \frac{1}{\Delta_p}\int_t^{t+\Delta_p}\mathds{1}_{x_s=O} \,\rmd s.
  \label{appeq: experimental_bk_future_open_fraction}
\end{equation}
The normalization permits comparisons between patches with different intrinsic time scales and does not change the residual fraction, which is invariant under a constant rescaling of the observable.

At an actual transition time $t=T_n$, the current dwell age is $a_t=\tau_n$. At an interior continuation time $u\in(T_{n-1},T_n)$, it is $a_u=u-T_{n-1}$. The finite-history coordinates used in the regression are
\begin{equation}
  \left( x_{t-},\log a_t,\log\tau_{-1},\ldots,\log\tau_{-3} \right),
  \label{appeq: experimental_bk_history_features}
\end{equation}
The normalized recording time $\frac{t}{T_{\mathrm{rec}}}$ in the rich-history reference and every candidate, where $T_{\mathrm{rec}}$ is the recorded duration. This nuisance coordinate is intended to reduce contamination from slow drift. It is absent from the fixed-horizon implementation. The logarithms are used because the dwell distributions span several orders of magnitude. Within every training fold, continuous features are centered by their median and scaled by their interquartile range.

\subsection{Event and Continuation Ensembles}
\label{appsec: experimental_bk_event_continuation_ensembles}

For each observed transition with three available preceding dwells and a complete future window, we record the pre-event edge, retained history, and future output. These observations form the event ensemble. For the continuation ensemble, we draw a prespecified number of times uniformly from the eligible recording time spent in each pre-event state. Eligibility requires the same available history depth and complete observation of the future window. We then calculate the current age, preceding dwell durations, and future output at each sampled time. There is no fixed number of samples per completed dwell and no preference for times near its endpoint. State-specific sample counts determine only the computational budget. Conditional means are fitted separately for the two edge classes.

For an edge $e$ and rich-history value $y$, define
\begin{subequations}
\label{appeq: experimental_bk_event_continuation_means}
\begin{align}
  m_{e,+}(y) &\equiv \mathbb{E}\left[ \, q_{\Delta}(t) \,\middle|\, Y_{\mathrm{BK},t-}^{(3)}=y,x_{t+}=i \, \right],\\
  m_{e,-}(y) &\equiv \mathbb{E}\left[ \, q_{\Delta}(t) \,\middle|\, Y_{\mathrm{BK},t-}^{(3)}=y,x_{t+}=j \, \right].
\end{align}
\end{subequations}
Both expectations are under the unperturbed trajectories, with the patchwise horizon held fixed. The event/no-jump notation denotes an infinitesimal condition at the comparison time, not a restriction over the full future interval. Observed transition times estimate the event branch; uniformly sampled eligible occupied time estimates the continuation branch before conditioning on the retained history. The finite time resolution of the records and the finite neighborhoods used in regression approximate these ideal conditional means.
The finite-history event consequence is
\begin{equation}
  K_{e,\Delta}^{(3)}(y)=m_{e,+}(y)-m_{e,-}(y).
  \label{appeq: experimental_bk_finite_history_event_consequence}
\end{equation}
The two conditional means are estimated separately for $C\to O$ and $O\to C$ using nearest-neighbor regression in the rich-history coordinates. The fixed-horizon analysis uses five blocked cross-fitting folds along each trajectory, 80 neighbors for the event and continuation conditional means, and 100 neighbors for the projection onto a candidate state. At most $12\,000$ event samples and $24\,000$ continuation samples are retained per patch by stratified subsampling over edge class and temporal fold. Events used to evaluate a prediction are not used to train the corresponding conditional-mean estimator.

For $K<3$, the projected coefficient is
\begin{equation}
  K_{e,\Delta}^{(K)} \equiv \mathbb{E}_{\mathrm{ev}}\left[ \, K_{e,\Delta}^{(3)} \,\middle|\, Y_{\mathrm{BK},t-}^{(K)},e \, \right].
  \label{appeq: experimental_bk_activity_weighted_projection}
\end{equation}
The conditional expectation is taken over actual transition events. Consequently, histories enter with their empirical edge activity, as required by the activity-weighted projection in \cref{eq: reduced_state_activity_weighted_projection}.

\subsection{Split-Sample Debiasing and Fixed-Horizon Residual}
\label{appsec: experimental_bk_debiasing}

A direct square of a nonparametric estimate has a positive estimation-noise bias. To remove its leading contribution, every trajectory is divided into alternating contiguous temporal blocks, which form two disjoint training ensembles $A$ and $B$. The event-continuation regression and candidate-state projection are performed independently in the two ensembles, producing
\begin{equation}
  \widehat{K}_{e,\Delta}^{A,(3)}, \quad \widehat{K}_{e,\Delta}^{B,(3)}, \quad \widehat{K}_{e,\Delta}^{A,(K)}, \quad \widehat{K}_{e,\Delta}^{B,(K)}.
  \label{appeq: experimental_bk_split_estimators}
\end{equation}
If the two estimation errors are conditionally independent and have zero mean, their cross product removes the leading regression-noise contribution. We estimate the rich-history second moment by
\begin{equation}
  \widehat{\mathcal{V}}_{\Delta}^{(3)} \equiv \left\langle \widehat{K}_{e,\Delta}^{A,(3)}\widehat{K}_{e,\Delta}^{B,(3)} \right\rangle_{\mathrm{ev}},
  \label{appeq: experimental_bk_rich_second_moment_estimator}
\end{equation}
and the residual relative to candidate $K$ by
\begin{equation}
  \widehat{\mathcal{M}}_{\Delta}^{(K|3)} \equiv \left\langle \left[ \widehat{K}_{e,\Delta}^{A,(3)} - \widehat{K}_{e,\Delta}^{A,(K)} \right] \left[ \widehat{K}_{e,\Delta}^{B,(3)} - \widehat{K}_{e,\Delta}^{B,(K)} \right] \right\rangle_{\mathrm{ev}}.
  \label{appeq: experimental_bk_residual_cross_product_estimator}
\end{equation}
The reported fixed-horizon residual fraction is
\begin{equation}
  \widehat{\eta}_{K}^{(3)} \equiv \frac{\widehat{\mathcal{M}}_{\Delta}^{(K|3)}}{\widehat{\mathcal{V}}_{\Delta}^{(3)}}.
  \label{appeq: experimental_bk_debiased_residual_fraction}
\end{equation}
Unlike the population residual, the finite-sample cross-product estimator can be slightly negative. Such values are retained rather than truncated at zero.

For the state-wise shuffled control, open dwell durations are permuted among open intervals and closed dwell durations are independently permuted among closed intervals, while the alternating state order is retained. This procedure preserves the complete empirical marginal distributions $p(\tau_O)$ and $p(\tau_C)$, the total open and closed times, the open fraction, and the number of transitions, but removes serial ordering. The excess estimator is
\begin{equation}
  \widehat{\eta}_{K}^{\mathrm{ex}}=\widehat{\eta}_{K}^{\mathrm{data}}-\widehat{\eta}_{K}^{\mathrm{shuf}}.
  \label{appeq: experimental_bk_excess_estimator}
\end{equation}

\subsection{Frequency-Resolved Analysis}
\label{appsec: experimental_bk_frequency_analysis}

For each patch, define the characteristic cycle time
\begin{equation}
  T_{\mathrm{cyc}} \equiv \widetilde{\tau}_{O}+\widetilde{\tau}_{C}
  \label{appeq: experimental_bk_cycle_time}
\end{equation}
and dimensionless angular frequency $\Omega=\omega T_{\mathrm{cyc}}$. The principal analysis uses
\begin{equation}
  \Omega\in\{0,0.25,0.5,1,2,3,4,6,8,12,16,24\}.
  \label{appeq: experimental_bk_frequency_grid}
\end{equation}
For a future window $L=12T_{\mathrm{cyc}}$, we use the half-cosine taper
\begin{equation}
  w_L(s) \equiv \frac{1}{2}\left[ 1+\cos\left( \frac{\pi s}{L} \right) \right], \qquad 0\leq s\leq L,
  \label{appeq: experimental_bk_frequency_taper}
\end{equation}
and define the complex future output
\begin{equation}
  q_{\Omega}(t) \equiv \frac{1}{T_{\mathrm{cyc}}}\int_0^L w_L(s)e^{\mathrm{i}\Omega s/T_{\mathrm{cyc}}}\mathds{1}_{x_{t+s}=O} \,\rmd s.
  \label{appeq: experimental_bk_complex_future_output}
\end{equation}
The factor $T_{\mathrm{cyc}}^{-1}$ makes the output dimensionless. The taper suppresses the artificial endpoint discontinuity of the finite future window. \Cref{appeq: experimental_bk_complex_future_output} is a finite-window regularization rather than the infinite-time Fourier transform of a stationary kernel.

Using the same event and continuation ensembles, we estimate
\begin{align}
  \mathcal{K}_{e}^{(3)}(\Omega;y) ={}& \mathbb{E}\left[ \, q_{\Omega}(t) \,\middle|\, Y_{\mathrm{BK},t-}^{(3)}=y,x_{t+}=i \, \right] \nonumber\\
  &-\mathbb{E}\left[ \, q_{\Omega}(t) \,\middle|\, Y_{\mathrm{BK},t-}^{(3)}=y,x_{t+}=j \, \right].
  \label{appeq: experimental_bk_frequency_kernel}
\end{align}
The frequency analysis uses five blocked folds, $50$ neighbors for the rich-history event and continuation regressions, and $65$ neighbors for the activity-weighted candidate-state projection. At most $2500$ event samples and $5000$ continuation samples are retained per patch and frequency analysis. All frequencies are evaluated on the same sampled event and continuation ensembles.

Two independent temporal training splits give complex estimates $\widehat{\mathcal{K}}_{e}^{A,(3)}(\Omega)$ and $\widehat{\mathcal{K}}_{e}^{B,(3)}(\Omega)$. The debiased rich-history power is
\begin{equation}
  \widehat{\mathcal{V}}^{(3)}(\Omega) \equiv \operatorname{Re}\left\langle \widehat{\mathcal{K}}_{e}^{A,(3)}(\Omega)\widehat{\mathcal{K}}_{e}^{B,(3)}(\Omega)^{*} \right\rangle_{\mathrm{ev}},
  \label{appeq: experimental_bk_frequency_rich_power}
\end{equation}
and the residual is
\begin{align}
  \widehat{\mathcal{M}}^{(K|3)}(\Omega) \equiv \operatorname{Re}\Big\langle &\left[ \widehat{\mathcal{K}}_{e}^{A,(3)}(\Omega)-\widehat{\mathcal{K}}_{e}^{A,(K)}(\Omega) \right] \nonumber\\
  &\times\left[ \widehat{\mathcal{K}}_{e}^{B,(3)}(\Omega)-\widehat{\mathcal{K}}_{e}^{B,(K)}(\Omega) \right]^{*} \Big\rangle_{\mathrm{ev}}.
  \label{appeq: experimental_bk_frequency_residual_estimator}
\end{align}
The frequency-resolved residual fraction is
\begin{equation}
  \widehat{\eta}_{K}^{(3)}(\Omega) \equiv \frac{\widehat{\mathcal{M}}^{(K|3)}(\Omega)}{\widehat{\mathcal{V}}^{(3)}(\Omega)}.
  \label{appeq: experimental_bk_frequency_residual_fraction}
\end{equation}
The real part in \cref{appeq: experimental_bk_frequency_rich_power,appeq: experimental_bk_frequency_residual_estimator} removes the antisymmetric estimation component and targets the real quadratic power. Finite-sample values are not truncated to the population interval $[0,1]$.

For each patch, the principal frequency analysis averages the residual over three independent state-wise shuffled trajectories,
\begin{equation}
  \widehat{\eta}_{K}^{\mathrm{ex}}(\Omega) \equiv \widehat{\eta}_{K}^{\mathrm{data}}(\Omega)-\frac{1}{3}\sum_{b=1}^{3}\widehat{\eta}_{K}^{\mathrm{shuf},b}(\Omega).
  \label{appeq: experimental_bk_frequency_excess_estimator}
\end{equation}
The shuffled trajectories preserve the two marginal dwell-time distributions and all individual dwell values, but remove their original serial order. They therefore provide a patch-specific baseline for regression error and for apparent residuals generated by nonexponential single-dwell statistics alone.

At a given frequency, a patch is retained when it contains at least $2000$ dwells, the debiased rich-history power is positive in both the original and mean shuffled analyses, and the split coherence
\begin{equation}
  \chi_{\mathrm{split}}(\Omega) \equiv \frac{\operatorname{Re}\left\langle \widehat{\mathcal{K}}_{e}^{A,(3)}(\Omega)\widehat{\mathcal{K}}_{e}^{B,(3)}(\Omega)^{*} \right\rangle_{\mathrm{ev}}}{\sqrt{\left\langle \left|\widehat{\mathcal{K}}_{e}^{A,(3)}(\Omega)\right|^2 \right\rangle_{\mathrm{ev}}\left\langle \left|\widehat{\mathcal{K}}_{e}^{B,(3)}(\Omega)\right|^2 \right\rangle_{\mathrm{ev}}}}
  \label{appeq: experimental_bk_frequency_split_coherence}
\end{equation}
is at least $0.10$ for both. The frequency-dependent criterion is applied before the group medians are evaluated. The denominator in \cref{appeq: experimental_bk_frequency_split_coherence} measures the reproducibility of the complex rich-history coefficient between the two disjoint training ensembles.

Patch-level medians and their confidence intervals are evaluated by resampling patches with replacement. The fraction of patches satisfying $\widehat{\eta}_{0}^{\mathrm{data}}(\Omega)>\widehat{\eta}_{0}^{\mathrm{shuf}}(\Omega)$ is accompanied by a Wilson $95\%$ interval. To quantify the frequency trend without treating neighboring frequencies as independent observations, we calculate one Spearman coefficient between $\Omega$ and $\widehat{\eta}_{0}^{\mathrm{ex}}(\Omega)$ for each patch and apply an exact sign test to the patch-level signs. All 25 quality-controlled mitoBK patches have negative frequency trends, with median $\rho=-0.7832$ and exact one-sided sign-test value $p=9.72\times10^{-6}$.

\subsection{Quality Control and Statistical Comparisons}
\label{appsec: experimental_bk_statistics}

For the fixed-horizon analysis, a patch enters the main comparison if it contains at least $2000$ dwells and the two rich-history estimates are reproducible in both the original and shuffled trajectories. Reproducibility is quantified by
\begin{equation}
  \chi_{\mathrm{split}} \equiv \frac{\left\langle \widehat{K}_{e,\Delta}^{A,(3)}\widehat{K}_{e,\Delta}^{B,(3)} \right\rangle_{\mathrm{ev}}}{\sqrt{\left\langle \left[ \widehat{K}_{e,\Delta}^{A,(3)} \right]^2 \right\rangle_{\mathrm{ev}}\left\langle \left[ \widehat{K}_{e,\Delta}^{B,(3)} \right]^2 \right\rangle_{\mathrm{ev}}}},
  \label{appeq: experimental_bk_split_coherence}
\end{equation}
with the threshold $\chi_{\mathrm{split}}\geq0.5$. These criteria retain 27 of 30 cell-BK patches and 25 of 26 mitoBK patches. The excluded records are \texttt{ce58}, \texttt{ce59}, \texttt{ce62}, and \texttt{e41}. Results including all patches do not change the qualitative separation between the two channel classes.

The patch, rather than the individual dwell, is the unit of statistical inference. Conventional dwell memory is evaluated separately in the closed- and open-dwell subsequences as
\begin{equation}
  C_{\mathrm{dwell}}^{\mathrm{ex}} = C_{\mathrm{dwell}}^{\mathrm{data}}-C_{\mathrm{dwell}}^{\mathrm{shuf}},
  \label{appeq: experimental_bk_excess_conventional_memory}
\end{equation}
with $C_{\mathrm{dwell}}$ defined in \cref{eq: experimental_bk_conventional_memory}. We use directional Mann-Whitney tests for the prespecified between-class comparisons and Spearman rank correlations for patch-wise associations between the two memory measures. The original-versus-shuffled mitoBK comparison is reported by an exact paired sign test: all 25 quality-controlled patches have the same positive direction, giving $p=5.96\times10^{-8}$ for the two-sided test.

The principal fixed-horizon summaries are
\begin{table}[t]
  \caption{Patch-level experimental memory measures after quality control. The between-class $p$ values are from the directional Mann-Whitney tests stated in the main text.}
  \label{tab: experimental_bk_patch_summary}
  \begin{ruledtabular}
  \begin{tabular}{lccc}
    Quantity & Cell-BK & MitoBK & $p$ \\
    \midrule
    $C_{\mathrm{dwell}}^{\mathrm{ex}}$ & $0.2077$ & $0.1290$ & $0.00671$ \\
    $\widehat{\eta}_{0}^{\mathrm{ex}}$ & $0.0373$ & $0.1954$ & $2.79\times10^{-6}$ \\
  \end{tabular}
  \end{ruledtabular}
\end{table}
The patch-wise association between $C_{\mathrm{dwell}}^{\mathrm{ex}}$ and $\widehat{\eta}_{0}^{\mathrm{ex}}$ is not significant within either class: for cell-BK, $\rho=0.425$ and $p=0.027$; for mitoBK, $\rho=-0.276$ and $p=0.181$.

For mitoBK, the fixed-horizon voltage-wise medians of $\widehat{\eta}_{0}^{\mathrm{ex}}$ are approximately $0.2295$, $0.1870$, $0.0957$, $0.2253$, $0.2761$, and $0.4837$ at $-60$, $-40$, $-20$, $20$, $40$, and $60\,\mathrm{mV}$, respectively. Different patches were recorded at different voltages, and several voltage groups contain few patches. These values therefore establish only that the effect is not confined to a single voltage condition; they do not support a voltage derivative or a direct voltage-parameter FRI.

\subsection{Window-Length Robustness and Interpretational Scope}
\label{appsec: experimental_bk_scope}

The principal frequency analysis uses the future window $L=12T_{\mathrm{cyc}}$. We repeat the analysis with $L=8T_{\mathrm{cyc}}$ and $L=16T_{\mathrm{cyc}}$. For mitoBK, the median excess residuals at $\Omega=0.5$, $2$, $8$, and $16$ are
\begin{equation}
  \begin{array}{c|cccc}
    L/T_{\mathrm{cyc}} & 0.5 & 2 & 8 & 16 \\
    \hline
    8  & 0.1512 & 0.1222 & 0.0777 & 0.0756 \\
    12 & 0.2214 & 0.1436 & 0.1007 & 0.0989 \\
    16 & 0.2534 & 0.1836 & 0.1357 & 0.1126
  \end{array}
  \label{appeq: experimental_bk_window_robustness_values}
\end{equation}
Longer windows accumulate more delayed event consequences and consequently increase the residual magnitude, but the low-frequency enhancement and broad separation from cell-BK are unchanged. The robustness runs use one shuffled trajectory per patch and are used only for this qualitative window comparison.

The experimental construction differs from the exact response-sufficiency test in several respects. First, $Y_{\mathrm{BK}}^{(3)}$ is a finite reference rather than the complete filtration, so $\widehat{\eta}_{K}^{(3)}$ measures memory resolved within the chosen dwell-history family. Second, the event and continuation branches are obtained by observational matching rather than by observing two futures of the same microscopic realization. Unresolved trajectory variables correlated with both event timing and future output can therefore contribute to the estimated coefficient. Third, the frequency-domain construction is a finite-window tapered transform rather than the infinite-time stationary kernel. Fourth, the supplied trajectories have already been thresholded into open and closed dwells; missed short events and threshold-selection rules can modify serial correlations and the highest resolved frequencies. Finally, frequencies are normalized by each patch-specific $T_{\mathrm{cyc}}$. This permits comparison of intrinsic gating scales, but a group value of $\Omega$ does not represent one common physical frequency across all patches.

The analysis consequently supports a separation between conventional dwell memory and finite-history event-consequence memory, together with a scale dependence of the latter. It does not constitute a model-free experimental measurement of the exact full-history FRR-FRI gap.

\bibliography{manuscript}

\begin{thebibliography}{104}%
\makeatletter
\providecommand \@ifxundefined [1]{%
 \@ifx{#1\undefined}
}%
\providecommand \@ifnum [1]{%
 \ifnum #1\expandafter \@firstoftwo
 \else \expandafter \@secondoftwo
 \fi
}%
\providecommand \@ifx [1]{%
 \ifx #1\expandafter \@firstoftwo
 \else \expandafter \@secondoftwo
 \fi
}%
\providecommand \natexlab [1]{#1}%
\providecommand \enquote  [1]{``#1''}%
\providecommand \bibnamefont  [1]{#1}%
\providecommand \bibfnamefont [1]{#1}%
\providecommand \citenamefont [1]{#1}%
\providecommand \href@noop [0]{\@secondoftwo}%
\providecommand \href [0]{\begingroup \@sanitize@url \@href}%
\providecommand \@href[1]{\@@startlink{#1}\@@href}%
\providecommand \@@href[1]{\endgroup#1\@@endlink}%
\providecommand \@sanitize@url [0]{\catcode `\\12\catcode `\$12\catcode
  `\&12\catcode `\#12\catcode `\^12\catcode `\_12\catcode `\%12\relax}%
\providecommand \@@startlink[1]{}%
\providecommand \@@endlink[0]{}%
\providecommand \url  [0]{\begingroup\@sanitize@url \@url }%
\providecommand \@url [1]{\endgroup\@href {#1}{\urlprefix }}%
\providecommand \urlprefix  [0]{URL }%
\providecommand \Eprint [0]{\href }%
\providecommand \doibase [0]{https://doi.org/}%
\providecommand \selectlanguage [0]{\@gobble}%
\providecommand \bibinfo  [0]{\@secondoftwo}%
\providecommand \bibfield  [0]{\@secondoftwo}%
\providecommand \translation [1]{[#1]}%
\providecommand \BibitemOpen [0]{}%
\providecommand \bibitemStop [0]{}%
\providecommand \bibitemNoStop [0]{.\EOS\space}%
\providecommand \EOS [0]{\spacefactor3000\relax}%
\providecommand \BibitemShut  [1]{\csname bibitem#1\endcsname}%
\let\auto@bib@innerbib\@empty
\bibitem [{\citenamefont {Sekimoto}(2010)}]{sekimoto2010stochastic}%
  \BibitemOpen
  \bibfield  {author} {\bibinfo {author} {\bibfnamefont {K.}~\bibnamefont
  {Sekimoto}},\ }\href {https://books.google.com/books?id=8Fq7BQAAQBAJ} {\emph
  {\bibinfo {title} {Stochastic Energetics}}},\ Lecture Notes in Physics\
  (\bibinfo  {publisher} {Springer Berlin Heidelberg},\ \bibinfo {year}
  {2010})\BibitemShut {NoStop}%
\bibitem [{\citenamefont {Seifert}(2005)}]{seifert2005entropy}%
  \BibitemOpen
  \bibfield  {author} {\bibinfo {author} {\bibfnamefont {U.}~\bibnamefont
  {Seifert}},\ }\bibfield  {title} {\bibinfo {title} {Entropy production along
  a stochastic trajectory and an integral fluctuation theorem},\ }\href@noop {}
  {\bibfield  {journal} {\bibinfo  {journal} {Physical Review Letters}\
  }\textbf {\bibinfo {volume} {95}},\ \bibinfo {pages} {040602} (\bibinfo
  {year} {2005})}\BibitemShut {NoStop}%
\bibitem [{\citenamefont {Seifert}(2012)}]{seifert2012stochastic}%
  \BibitemOpen
  \bibfield  {author} {\bibinfo {author} {\bibfnamefont {U.}~\bibnamefont
  {Seifert}},\ }\bibfield  {title} {\bibinfo {title} {Stochastic
  thermodynamics, fluctuation theorems and molecular machines},\ }\href@noop {}
  {\bibfield  {journal} {\bibinfo  {journal} {Reports on Progress in Physics}\
  }\textbf {\bibinfo {volume} {75}},\ \bibinfo {pages} {126001} (\bibinfo
  {year} {2012})}\BibitemShut {NoStop}%
\bibitem [{\citenamefont {Peliti}\ and\ \citenamefont
  {Pigolotti}(2021)}]{peliti2021stochastic}%
  \BibitemOpen
  \bibfield  {author} {\bibinfo {author} {\bibfnamefont {L.}~\bibnamefont
  {Peliti}}\ and\ \bibinfo {author} {\bibfnamefont {S.}~\bibnamefont
  {Pigolotti}},\ }\href@noop {} {\emph {\bibinfo {title} {Stochastic
  thermodynamics: an introduction}}}\ (\bibinfo  {publisher} {Princeton
  University Press},\ \bibinfo {year} {2021})\BibitemShut {NoStop}%
\bibitem [{\citenamefont {Jarzynski}(1997)}]{jarzynski1997nonequilibrium}%
  \BibitemOpen
  \bibfield  {author} {\bibinfo {author} {\bibfnamefont {C.}~\bibnamefont
  {Jarzynski}},\ }\bibfield  {title} {\bibinfo {title} {Nonequilibrium equality
  for free energy differences},\ }\href@noop {} {\bibfield  {journal} {\bibinfo
   {journal} {Physical review letters}\ }\textbf {\bibinfo {volume} {78}},\
  \bibinfo {pages} {2690} (\bibinfo {year} {1997})}\BibitemShut {NoStop}%
\bibitem [{\citenamefont {Crooks}(1999)}]{crooks1999entropy}%
  \BibitemOpen
  \bibfield  {author} {\bibinfo {author} {\bibfnamefont {G.~E.}\ \bibnamefont
  {Crooks}},\ }\bibfield  {title} {\bibinfo {title} {Entropy production
  fluctuation theorem and the nonequilibrium work relation for free energy
  differences},\ }\href@noop {} {\bibfield  {journal} {\bibinfo  {journal}
  {Physical Review E}\ }\textbf {\bibinfo {volume} {60}},\ \bibinfo {pages}
  {2721} (\bibinfo {year} {1999})}\BibitemShut {NoStop}%
\bibitem [{\citenamefont {Lebowitz}\ and\ \citenamefont
  {Spohn}(1999)}]{lebowitz1999gallavotti}%
  \BibitemOpen
  \bibfield  {author} {\bibinfo {author} {\bibfnamefont {J.~L.}\ \bibnamefont
  {Lebowitz}}\ and\ \bibinfo {author} {\bibfnamefont {H.}~\bibnamefont
  {Spohn}},\ }\bibfield  {title} {\bibinfo {title} {A gallavotti--cohen-type
  symmetry in the large deviation functional for stochastic dynamics},\
  }\href@noop {} {\bibfield  {journal} {\bibinfo  {journal} {Journal of
  Statistical Physics}\ }\textbf {\bibinfo {volume} {95}},\ \bibinfo {pages}
  {333} (\bibinfo {year} {1999})}\BibitemShut {NoStop}%
\bibitem [{\citenamefont {Evans}\ and\ \citenamefont
  {Searles}(2002)}]{evans2002fluctuation}%
  \BibitemOpen
  \bibfield  {author} {\bibinfo {author} {\bibfnamefont {D.~J.}\ \bibnamefont
  {Evans}}\ and\ \bibinfo {author} {\bibfnamefont {D.~J.}\ \bibnamefont
  {Searles}},\ }\bibfield  {title} {\bibinfo {title} {The fluctuation
  theorem},\ }\href@noop {} {\bibfield  {journal} {\bibinfo  {journal}
  {Advances in Physics}\ }\textbf {\bibinfo {volume} {51}},\ \bibinfo {pages}
  {1529} (\bibinfo {year} {2002})}\BibitemShut {NoStop}%
\bibitem [{\citenamefont {Sagawa}\ and\ \citenamefont
  {Ueda}(2010)}]{sagawa2010generalized}%
  \BibitemOpen
  \bibfield  {author} {\bibinfo {author} {\bibfnamefont {T.}~\bibnamefont
  {Sagawa}}\ and\ \bibinfo {author} {\bibfnamefont {M.}~\bibnamefont {Ueda}},\
  }\bibfield  {title} {\bibinfo {title} {Generalized jarzynski equality under
  nonequilibrium feedback control},\ }\href@noop {} {\bibfield  {journal}
  {\bibinfo  {journal} {Physical review letters}\ }\textbf {\bibinfo {volume}
  {104}},\ \bibinfo {pages} {090602} (\bibinfo {year} {2010})}\BibitemShut
  {NoStop}%
\bibitem [{\citenamefont {Parrondo}\ \emph {et~al.}(2015)\citenamefont
  {Parrondo}, \citenamefont {Horowitz},\ and\ \citenamefont
  {Sagawa}}]{parrondo2015thermodynamics}%
  \BibitemOpen
  \bibfield  {author} {\bibinfo {author} {\bibfnamefont {J.~M.}\ \bibnamefont
  {Parrondo}}, \bibinfo {author} {\bibfnamefont {J.~M.}\ \bibnamefont
  {Horowitz}},\ and\ \bibinfo {author} {\bibfnamefont {T.}~\bibnamefont
  {Sagawa}},\ }\bibfield  {title} {\bibinfo {title} {Thermodynamics of
  information},\ }\href@noop {} {\bibfield  {journal} {\bibinfo  {journal}
  {Nature physics}\ }\textbf {\bibinfo {volume} {11}},\ \bibinfo {pages} {131}
  (\bibinfo {year} {2015})}\BibitemShut {NoStop}%
\bibitem [{\citenamefont {Horowitz}\ and\ \citenamefont
  {Esposito}(2014)}]{horowitz2014thermodynamics}%
  \BibitemOpen
  \bibfield  {author} {\bibinfo {author} {\bibfnamefont {J.~M.}\ \bibnamefont
  {Horowitz}}\ and\ \bibinfo {author} {\bibfnamefont {M.}~\bibnamefont
  {Esposito}},\ }\bibfield  {title} {\bibinfo {title} {Thermodynamics with
  continuous information flow},\ }\href@noop {} {\bibfield  {journal} {\bibinfo
   {journal} {Physical Review X}\ }\textbf {\bibinfo {volume} {4}},\ \bibinfo
  {pages} {031015} (\bibinfo {year} {2014})}\BibitemShut {NoStop}%
\bibitem [{\citenamefont {Mandal}\ and\ \citenamefont
  {Jarzynski}(2012)}]{mandal2012work}%
  \BibitemOpen
  \bibfield  {author} {\bibinfo {author} {\bibfnamefont {D.}~\bibnamefont
  {Mandal}}\ and\ \bibinfo {author} {\bibfnamefont {C.}~\bibnamefont
  {Jarzynski}},\ }\bibfield  {title} {\bibinfo {title} {Work and information
  processing in a solvable model of maxwell’s demon},\ }\href@noop {}
  {\bibfield  {journal} {\bibinfo  {journal} {Proceedings of the National
  Academy of Sciences}\ }\textbf {\bibinfo {volume} {109}},\ \bibinfo {pages}
  {11641} (\bibinfo {year} {2012})}\BibitemShut {NoStop}%
\bibitem [{\citenamefont {Gingrich}\ \emph {et~al.}(2016)\citenamefont
  {Gingrich}, \citenamefont {Horowitz}, \citenamefont {Perunov},\ and\
  \citenamefont {England}}]{gingrich2016dissipation}%
  \BibitemOpen
  \bibfield  {author} {\bibinfo {author} {\bibfnamefont {T.~R.}\ \bibnamefont
  {Gingrich}}, \bibinfo {author} {\bibfnamefont {J.~M.}\ \bibnamefont
  {Horowitz}}, \bibinfo {author} {\bibfnamefont {N.}~\bibnamefont {Perunov}},\
  and\ \bibinfo {author} {\bibfnamefont {J.~L.}\ \bibnamefont {England}},\
  }\bibfield  {title} {\bibinfo {title} {Dissipation bounds all steady-state
  current fluctuations},\ }\href@noop {} {\bibfield  {journal} {\bibinfo
  {journal} {Physical Review Letters}\ }\textbf {\bibinfo {volume} {116}},\
  \bibinfo {pages} {120601} (\bibinfo {year} {2016})}\BibitemShut {NoStop}%
\bibitem [{\citenamefont {Horowitz}\ and\ \citenamefont
  {Gingrich}(2020)}]{horowitz2020thermodynamic}%
  \BibitemOpen
  \bibfield  {author} {\bibinfo {author} {\bibfnamefont {J.~M.}\ \bibnamefont
  {Horowitz}}\ and\ \bibinfo {author} {\bibfnamefont {T.~R.}\ \bibnamefont
  {Gingrich}},\ }\bibfield  {title} {\bibinfo {title} {Thermodynamic
  uncertainty relations constrain non-equilibrium fluctuations},\ }\href@noop
  {} {\bibfield  {journal} {\bibinfo  {journal} {Nature Physics}\ }\textbf
  {\bibinfo {volume} {16}},\ \bibinfo {pages} {15} (\bibinfo {year}
  {2020})}\BibitemShut {NoStop}%
\bibitem [{\citenamefont {Hasegawa}\ and\ \citenamefont
  {Van~Vu}(2019)}]{hasegawa2019fluctuation}%
  \BibitemOpen
  \bibfield  {author} {\bibinfo {author} {\bibfnamefont {Y.}~\bibnamefont
  {Hasegawa}}\ and\ \bibinfo {author} {\bibfnamefont {T.}~\bibnamefont
  {Van~Vu}},\ }\bibfield  {title} {\bibinfo {title} {Fluctuation theorem
  uncertainty relation},\ }\href@noop {} {\bibfield  {journal} {\bibinfo
  {journal} {Physical review letters}\ }\textbf {\bibinfo {volume} {123}},\
  \bibinfo {pages} {110602} (\bibinfo {year} {2019})}\BibitemShut {NoStop}%
\bibitem [{\citenamefont {Dieball}\ and\ \citenamefont
  {Godec}(2023)}]{dieball2023direct}%
  \BibitemOpen
  \bibfield  {author} {\bibinfo {author} {\bibfnamefont {C.}~\bibnamefont
  {Dieball}}\ and\ \bibinfo {author} {\bibfnamefont {A.}~\bibnamefont
  {Godec}},\ }\bibfield  {title} {\bibinfo {title} {Direct route to
  thermodynamic uncertainty relations and their saturation},\ }\href@noop {}
  {\bibfield  {journal} {\bibinfo  {journal} {Physical Review Letters}\
  }\textbf {\bibinfo {volume} {130}},\ \bibinfo {pages} {087101} (\bibinfo
  {year} {2023})}\BibitemShut {NoStop}%
\bibitem [{\citenamefont {Shiraishi}\ \emph {et~al.}(2018)\citenamefont
  {Shiraishi}, \citenamefont {Funo},\ and\ \citenamefont
  {Saito}}]{shiraishi2018speed}%
  \BibitemOpen
  \bibfield  {author} {\bibinfo {author} {\bibfnamefont {N.}~\bibnamefont
  {Shiraishi}}, \bibinfo {author} {\bibfnamefont {K.}~\bibnamefont {Funo}},\
  and\ \bibinfo {author} {\bibfnamefont {K.}~\bibnamefont {Saito}},\ }\bibfield
   {title} {\bibinfo {title} {Speed limit for classical stochastic processes},\
  }\href@noop {} {\bibfield  {journal} {\bibinfo  {journal} {arXiv preprint
  arXiv:1802.06554}\ } (\bibinfo {year} {2018})}\BibitemShut {NoStop}%
\bibitem [{\citenamefont {Ito}\ and\ \citenamefont
  {Dechant}(2020)}]{ito2020stochastic}%
  \BibitemOpen
  \bibfield  {author} {\bibinfo {author} {\bibfnamefont {S.}~\bibnamefont
  {Ito}}\ and\ \bibinfo {author} {\bibfnamefont {A.}~\bibnamefont {Dechant}},\
  }\bibfield  {title} {\bibinfo {title} {Stochastic time evolution, information
  geometry, and the cram{\'e}r-rao bound},\ }\href@noop {} {\bibfield
  {journal} {\bibinfo  {journal} {Physical Review X}\ }\textbf {\bibinfo
  {volume} {10}},\ \bibinfo {pages} {021056} (\bibinfo {year}
  {2020})}\BibitemShut {NoStop}%
\bibitem [{\citenamefont {Neher}\ and\ \citenamefont
  {Sakmann}(1976)}]{neher1976single}%
  \BibitemOpen
  \bibfield  {author} {\bibinfo {author} {\bibfnamefont {E.}~\bibnamefont
  {Neher}}\ and\ \bibinfo {author} {\bibfnamefont {B.}~\bibnamefont
  {Sakmann}},\ }\bibfield  {title} {\bibinfo {title} {Single-channel currents
  recorded from membrane of denervated frog muscle fibres},\ }\href@noop {}
  {\bibfield  {journal} {\bibinfo  {journal} {Nature}\ }\textbf {\bibinfo
  {volume} {260}},\ \bibinfo {pages} {799} (\bibinfo {year}
  {1976})}\BibitemShut {NoStop}%
\bibitem [{\citenamefont {Mickus}\ \emph {et~al.}(1999)\citenamefont {Mickus},
  \citenamefont {Jung},\ and\ \citenamefont {Spruston}}]{mickus1999properties}%
  \BibitemOpen
  \bibfield  {author} {\bibinfo {author} {\bibfnamefont {T.}~\bibnamefont
  {Mickus}}, \bibinfo {author} {\bibfnamefont {H.-y.}\ \bibnamefont {Jung}},\
  and\ \bibinfo {author} {\bibfnamefont {N.}~\bibnamefont {Spruston}},\
  }\bibfield  {title} {\bibinfo {title} {Properties of slow, cumulative sodium
  channel inactivation in rat hippocampal ca1 pyramidal neurons},\ }\href@noop
  {} {\bibfield  {journal} {\bibinfo  {journal} {Biophysical journal}\ }\textbf
  {\bibinfo {volume} {76}},\ \bibinfo {pages} {846} (\bibinfo {year}
  {1999})}\BibitemShut {NoStop}%
\bibitem [{\citenamefont {Lu}\ \emph {et~al.}(1998)\citenamefont {Lu},
  \citenamefont {Xun},\ and\ \citenamefont {Xie}}]{lu1998single}%
  \BibitemOpen
  \bibfield  {author} {\bibinfo {author} {\bibfnamefont {H.~P.}\ \bibnamefont
  {Lu}}, \bibinfo {author} {\bibfnamefont {L.}~\bibnamefont {Xun}},\ and\
  \bibinfo {author} {\bibfnamefont {X.~S.}\ \bibnamefont {Xie}},\ }\bibfield
  {title} {\bibinfo {title} {Single-molecule enzymatic dynamics},\ }\href@noop
  {} {\bibfield  {journal} {\bibinfo  {journal} {Science}\ }\textbf {\bibinfo
  {volume} {282}},\ \bibinfo {pages} {1877} (\bibinfo {year}
  {1998})}\BibitemShut {NoStop}%
\bibitem [{\citenamefont {English}\ \emph {et~al.}(2006)\citenamefont
  {English}, \citenamefont {Min}, \citenamefont {Van~Oijen}, \citenamefont
  {Lee}, \citenamefont {Luo}, \citenamefont {Sun}, \citenamefont {Cherayil},
  \citenamefont {Kou},\ and\ \citenamefont {Xie}}]{english2006ever}%
  \BibitemOpen
  \bibfield  {author} {\bibinfo {author} {\bibfnamefont {B.~P.}\ \bibnamefont
  {English}}, \bibinfo {author} {\bibfnamefont {W.}~\bibnamefont {Min}},
  \bibinfo {author} {\bibfnamefont {A.~M.}\ \bibnamefont {Van~Oijen}}, \bibinfo
  {author} {\bibfnamefont {K.~T.}\ \bibnamefont {Lee}}, \bibinfo {author}
  {\bibfnamefont {G.}~\bibnamefont {Luo}}, \bibinfo {author} {\bibfnamefont
  {H.}~\bibnamefont {Sun}}, \bibinfo {author} {\bibfnamefont {B.~J.}\
  \bibnamefont {Cherayil}}, \bibinfo {author} {\bibfnamefont {S.}~\bibnamefont
  {Kou}},\ and\ \bibinfo {author} {\bibfnamefont {X.~S.}\ \bibnamefont {Xie}},\
  }\bibfield  {title} {\bibinfo {title} {Ever-fluctuating single enzyme
  molecules: Michaelis-menten equation revisited},\ }\href@noop {} {\bibfield
  {journal} {\bibinfo  {journal} {Nature chemical biology}\ }\textbf {\bibinfo
  {volume} {2}},\ \bibinfo {pages} {87} (\bibinfo {year} {2006})}\BibitemShut
  {NoStop}%
\bibitem [{\citenamefont {Farkhooi}\ \emph {et~al.}(2009)\citenamefont
  {Farkhooi}, \citenamefont {Strube-Bloss},\ and\ \citenamefont
  {Nawrot}}]{farkhooi2009serial}%
  \BibitemOpen
  \bibfield  {author} {\bibinfo {author} {\bibfnamefont {F.}~\bibnamefont
  {Farkhooi}}, \bibinfo {author} {\bibfnamefont {M.~F.}\ \bibnamefont
  {Strube-Bloss}},\ and\ \bibinfo {author} {\bibfnamefont {M.~P.}\ \bibnamefont
  {Nawrot}},\ }\bibfield  {title} {\bibinfo {title} {Serial correlation in
  neural spike trains: Experimental evidence, stochastic modeling, and single
  neuron variability},\ }\href@noop {} {\bibfield  {journal} {\bibinfo
  {journal} {Physical Review E—Statistical, Nonlinear, and Soft Matter
  Physics}\ }\textbf {\bibinfo {volume} {79}},\ \bibinfo {pages} {021905}
  (\bibinfo {year} {2009})}\BibitemShut {NoStop}%
\bibitem [{\citenamefont {Markram}\ and\ \citenamefont
  {Tsodyks}(1996)}]{markram1996redistribution}%
  \BibitemOpen
  \bibfield  {author} {\bibinfo {author} {\bibfnamefont {H.}~\bibnamefont
  {Markram}}\ and\ \bibinfo {author} {\bibfnamefont {M.}~\bibnamefont
  {Tsodyks}},\ }\bibfield  {title} {\bibinfo {title} {Redistribution of
  synaptic efficacy between neocortical pyramidal neurons},\ }\href@noop {}
  {\bibfield  {journal} {\bibinfo  {journal} {Nature}\ }\textbf {\bibinfo
  {volume} {382}},\ \bibinfo {pages} {807} (\bibinfo {year}
  {1996})}\BibitemShut {NoStop}%
\bibitem [{\citenamefont {Suter}\ \emph {et~al.}(2011)\citenamefont {Suter},
  \citenamefont {Molina}, \citenamefont {Gatfield}, \citenamefont {Schneider},
  \citenamefont {Schibler},\ and\ \citenamefont {Naef}}]{suter2011mammalian}%
  \BibitemOpen
  \bibfield  {author} {\bibinfo {author} {\bibfnamefont {D.~M.}\ \bibnamefont
  {Suter}}, \bibinfo {author} {\bibfnamefont {N.}~\bibnamefont {Molina}},
  \bibinfo {author} {\bibfnamefont {D.}~\bibnamefont {Gatfield}}, \bibinfo
  {author} {\bibfnamefont {K.}~\bibnamefont {Schneider}}, \bibinfo {author}
  {\bibfnamefont {U.}~\bibnamefont {Schibler}},\ and\ \bibinfo {author}
  {\bibfnamefont {F.}~\bibnamefont {Naef}},\ }\bibfield  {title} {\bibinfo
  {title} {Mammalian genes are transcribed with widely different bursting
  kinetics},\ }\href@noop {} {\bibfield  {journal} {\bibinfo  {journal}
  {science}\ }\textbf {\bibinfo {volume} {332}},\ \bibinfo {pages} {472}
  (\bibinfo {year} {2011})}\BibitemShut {NoStop}%
\bibitem [{\citenamefont {Molina}\ \emph {et~al.}(2013)\citenamefont {Molina},
  \citenamefont {Suter}, \citenamefont {Cannavo}, \citenamefont {Zoller},
  \citenamefont {Gotic},\ and\ \citenamefont {Naef}}]{molina2013stimulus}%
  \BibitemOpen
  \bibfield  {author} {\bibinfo {author} {\bibfnamefont {N.}~\bibnamefont
  {Molina}}, \bibinfo {author} {\bibfnamefont {D.~M.}\ \bibnamefont {Suter}},
  \bibinfo {author} {\bibfnamefont {R.}~\bibnamefont {Cannavo}}, \bibinfo
  {author} {\bibfnamefont {B.}~\bibnamefont {Zoller}}, \bibinfo {author}
  {\bibfnamefont {I.}~\bibnamefont {Gotic}},\ and\ \bibinfo {author}
  {\bibfnamefont {F.}~\bibnamefont {Naef}},\ }\bibfield  {title} {\bibinfo
  {title} {Stimulus-induced modulation of transcriptional bursting in a single
  mammalian gene},\ }\href@noop {} {\bibfield  {journal} {\bibinfo  {journal}
  {Proceedings of the National Academy of Sciences}\ }\textbf {\bibinfo
  {volume} {110}},\ \bibinfo {pages} {20563} (\bibinfo {year}
  {2013})}\BibitemShut {NoStop}%
\bibitem [{\citenamefont {Brokmann}\ \emph {et~al.}(2003)\citenamefont
  {Brokmann}, \citenamefont {Hermier}, \citenamefont {Messin}, \citenamefont
  {Desbiolles}, \citenamefont {Bouchaud},\ and\ \citenamefont
  {Dahan}}]{brokmann2003statistical}%
  \BibitemOpen
  \bibfield  {author} {\bibinfo {author} {\bibfnamefont {X.}~\bibnamefont
  {Brokmann}}, \bibinfo {author} {\bibfnamefont {J.-P.}\ \bibnamefont
  {Hermier}}, \bibinfo {author} {\bibfnamefont {G.}~\bibnamefont {Messin}},
  \bibinfo {author} {\bibfnamefont {P.}~\bibnamefont {Desbiolles}}, \bibinfo
  {author} {\bibfnamefont {J.-P.}\ \bibnamefont {Bouchaud}},\ and\ \bibinfo
  {author} {\bibfnamefont {M.}~\bibnamefont {Dahan}},\ }\bibfield  {title}
  {\bibinfo {title} {Statistical aging and nonergodicity in the fluorescence of
  single nanocrystals},\ }\href@noop {} {\bibfield  {journal} {\bibinfo
  {journal} {Physical review letters}\ }\textbf {\bibinfo {volume} {90}},\
  \bibinfo {pages} {120601} (\bibinfo {year} {2003})}\BibitemShut {NoStop}%
\bibitem [{\citenamefont {Shi}\ \emph {et~al.}(2021)\citenamefont {Shi},
  \citenamefont {Sun}, \citenamefont {Utzat}, \citenamefont {Farahvash},
  \citenamefont {Gao}, \citenamefont {Zhang}, \citenamefont {Barotov},
  \citenamefont {Willard}, \citenamefont {Nelson},\ and\ \citenamefont
  {Bawendi}}]{shi2021all}%
  \BibitemOpen
  \bibfield  {author} {\bibinfo {author} {\bibfnamefont {J.}~\bibnamefont
  {Shi}}, \bibinfo {author} {\bibfnamefont {W.}~\bibnamefont {Sun}}, \bibinfo
  {author} {\bibfnamefont {H.}~\bibnamefont {Utzat}}, \bibinfo {author}
  {\bibfnamefont {A.}~\bibnamefont {Farahvash}}, \bibinfo {author}
  {\bibfnamefont {F.~Y.}\ \bibnamefont {Gao}}, \bibinfo {author} {\bibfnamefont
  {Z.}~\bibnamefont {Zhang}}, \bibinfo {author} {\bibfnamefont
  {U.}~\bibnamefont {Barotov}}, \bibinfo {author} {\bibfnamefont {A.~P.}\
  \bibnamefont {Willard}}, \bibinfo {author} {\bibfnamefont {K.~A.}\
  \bibnamefont {Nelson}},\ and\ \bibinfo {author} {\bibfnamefont {M.~G.}\
  \bibnamefont {Bawendi}},\ }\bibfield  {title} {\bibinfo {title} {All-optical
  fluorescence blinking control in quantum dots with ultrafast mid-infrared
  pulses},\ }\href@noop {} {\bibfield  {journal} {\bibinfo  {journal} {Nature
  Nanotechnology}\ }\textbf {\bibinfo {volume} {16}},\ \bibinfo {pages} {1355}
  (\bibinfo {year} {2021})}\BibitemShut {NoStop}%
\bibitem [{\citenamefont {Lapolla}\ and\ \citenamefont
  {Godec}(2019)}]{lapolla2019manifestations}%
  \BibitemOpen
  \bibfield  {author} {\bibinfo {author} {\bibfnamefont {A.}~\bibnamefont
  {Lapolla}}\ and\ \bibinfo {author} {\bibfnamefont {A.}~\bibnamefont
  {Godec}},\ }\bibfield  {title} {\bibinfo {title} {Manifestations of
  projection-induced memory: General theory and the tilted single file},\
  }\href@noop {} {\bibfield  {journal} {\bibinfo  {journal} {Frontiers in
  Physics}\ }\textbf {\bibinfo {volume} {7}},\ \bibinfo {pages} {182} (\bibinfo
  {year} {2019})}\BibitemShut {NoStop}%
\bibitem [{\citenamefont {Zwanzig}(1961)}]{zwanzig1961memory}%
  \BibitemOpen
  \bibfield  {author} {\bibinfo {author} {\bibfnamefont {R.}~\bibnamefont
  {Zwanzig}},\ }\bibfield  {title} {\bibinfo {title} {Memory effects in
  irreversible thermodynamics},\ }\href@noop {} {\bibfield  {journal} {\bibinfo
   {journal} {Physical Review}\ }\textbf {\bibinfo {volume} {124}},\ \bibinfo
  {pages} {983} (\bibinfo {year} {1961})}\BibitemShut {NoStop}%
\bibitem [{\citenamefont {Mori}(1965)}]{mori1965transport}%
  \BibitemOpen
  \bibfield  {author} {\bibinfo {author} {\bibfnamefont {H.}~\bibnamefont
  {Mori}},\ }\bibfield  {title} {\bibinfo {title} {Transport, collective
  motion, and brownian motion},\ }\href@noop {} {\bibfield  {journal} {\bibinfo
   {journal} {Progress of theoretical physics}\ }\textbf {\bibinfo {volume}
  {33}},\ \bibinfo {pages} {423} (\bibinfo {year} {1965})}\BibitemShut
  {NoStop}%
\bibitem [{\citenamefont {Zwanzig}(1992)}]{zwanzig1992dynamical}%
  \BibitemOpen
  \bibfield  {author} {\bibinfo {author} {\bibfnamefont {R.}~\bibnamefont
  {Zwanzig}},\ }\bibfield  {title} {\bibinfo {title} {Dynamical disorder:
  Passage through a fluctuating bottleneck},\ }\href@noop {} {\bibfield
  {journal} {\bibinfo  {journal} {The Journal of chemical physics}\ }\textbf
  {\bibinfo {volume} {97}},\ \bibinfo {pages} {3587} (\bibinfo {year}
  {1992})}\BibitemShut {NoStop}%
\bibitem [{\citenamefont {Benda}\ and\ \citenamefont
  {Herz}(2003)}]{benda2003universal}%
  \BibitemOpen
  \bibfield  {author} {\bibinfo {author} {\bibfnamefont {J.}~\bibnamefont
  {Benda}}\ and\ \bibinfo {author} {\bibfnamefont {A.~V.}\ \bibnamefont
  {Herz}},\ }\bibfield  {title} {\bibinfo {title} {A universal model for
  spike-frequency adaptation},\ }\href@noop {} {\bibfield  {journal} {\bibinfo
  {journal} {Neural computation}\ }\textbf {\bibinfo {volume} {15}},\ \bibinfo
  {pages} {2523} (\bibinfo {year} {2003})}\BibitemShut {NoStop}%
\bibitem [{\citenamefont {Chacron}\ \emph {et~al.}(2001)\citenamefont
  {Chacron}, \citenamefont {Longtin},\ and\ \citenamefont
  {Maler}}]{chacron2001negative}%
  \BibitemOpen
  \bibfield  {author} {\bibinfo {author} {\bibfnamefont {M.~J.}\ \bibnamefont
  {Chacron}}, \bibinfo {author} {\bibfnamefont {A.}~\bibnamefont {Longtin}},\
  and\ \bibinfo {author} {\bibfnamefont {L.}~\bibnamefont {Maler}},\ }\bibfield
   {title} {\bibinfo {title} {Negative interspike interval correlations
  increase the neuronal capacity for encoding time-dependent stimuli},\
  }\href@noop {} {\bibfield  {journal} {\bibinfo  {journal} {The Journal of
  Neuroscience}\ }\textbf {\bibinfo {volume} {21}},\ \bibinfo {pages} {5328}
  (\bibinfo {year} {2001})}\BibitemShut {NoStop}%
\bibitem [{\citenamefont {Dobrunz}\ and\ \citenamefont
  {Stevens}(1997)}]{dobrunz1997heterogeneity}%
  \BibitemOpen
  \bibfield  {author} {\bibinfo {author} {\bibfnamefont {L.~E.}\ \bibnamefont
  {Dobrunz}}\ and\ \bibinfo {author} {\bibfnamefont {C.~F.}\ \bibnamefont
  {Stevens}},\ }\bibfield  {title} {\bibinfo {title} {Heterogeneity of release
  probability, facilitation, and depletion at central synapses},\ }\href@noop
  {} {\bibfield  {journal} {\bibinfo  {journal} {Neuron}\ }\textbf {\bibinfo
  {volume} {18}},\ \bibinfo {pages} {995} (\bibinfo {year} {1997})}\BibitemShut
  {NoStop}%
\bibitem [{\citenamefont {Tsodyks}\ and\ \citenamefont
  {Markram}(1997)}]{tsodyks1997neural}%
  \BibitemOpen
  \bibfield  {author} {\bibinfo {author} {\bibfnamefont {M.~V.}\ \bibnamefont
  {Tsodyks}}\ and\ \bibinfo {author} {\bibfnamefont {H.}~\bibnamefont
  {Markram}},\ }\bibfield  {title} {\bibinfo {title} {The neural code between
  neocortical pyramidal neurons depends on neurotransmitter release
  probability},\ }\href@noop {} {\bibfield  {journal} {\bibinfo  {journal}
  {Proceedings of the national academy of sciences}\ }\textbf {\bibinfo
  {volume} {94}},\ \bibinfo {pages} {719} (\bibinfo {year} {1997})}\BibitemShut
  {NoStop}%
\bibitem [{\citenamefont {Abbott}\ \emph {et~al.}(1997)\citenamefont {Abbott},
  \citenamefont {Varela}, \citenamefont {Sen},\ and\ \citenamefont
  {Nelson}}]{abbott1997synaptic}%
  \BibitemOpen
  \bibfield  {author} {\bibinfo {author} {\bibfnamefont {L.~F.}\ \bibnamefont
  {Abbott}}, \bibinfo {author} {\bibfnamefont {J.~A.}\ \bibnamefont {Varela}},
  \bibinfo {author} {\bibfnamefont {K.}~\bibnamefont {Sen}},\ and\ \bibinfo
  {author} {\bibfnamefont {S.~B.}\ \bibnamefont {Nelson}},\ }\bibfield  {title}
  {\bibinfo {title} {Synaptic depression and cortical gain control},\
  }\href@noop {} {\bibfield  {journal} {\bibinfo  {journal} {Science}\ }\textbf
  {\bibinfo {volume} {275}},\ \bibinfo {pages} {221} (\bibinfo {year}
  {1997})}\BibitemShut {NoStop}%
\bibitem [{\citenamefont {Berry}\ and\ \citenamefont
  {Meister}(1997)}]{berry1997refractoriness}%
  \BibitemOpen
  \bibfield  {author} {\bibinfo {author} {\bibfnamefont {M.}~\bibnamefont
  {Berry}}\ and\ \bibinfo {author} {\bibfnamefont {M.}~\bibnamefont
  {Meister}},\ }\bibfield  {title} {\bibinfo {title} {Refractoriness and neural
  precision},\ }\href@noop {} {\bibfield  {journal} {\bibinfo  {journal}
  {Advances in neural information processing systems}\ }\textbf {\bibinfo
  {volume} {10}} (\bibinfo {year} {1997})}\BibitemShut {NoStop}%
\bibitem [{\citenamefont {Montroll}\ and\ \citenamefont
  {Weiss}(1965)}]{montroll1965random}%
  \BibitemOpen
  \bibfield  {author} {\bibinfo {author} {\bibfnamefont {E.~W.}\ \bibnamefont
  {Montroll}}\ and\ \bibinfo {author} {\bibfnamefont {G.~H.}\ \bibnamefont
  {Weiss}},\ }\bibfield  {title} {\bibinfo {title} {Random walks on lattices.
  ii},\ }\href@noop {} {\bibfield  {journal} {\bibinfo  {journal} {Journal of
  Mathematical Physics}\ }\textbf {\bibinfo {volume} {6}},\ \bibinfo {pages}
  {167} (\bibinfo {year} {1965})}\BibitemShut {NoStop}%
\bibitem [{\citenamefont {Kubo}(1957)}]{kubo1957statistical}%
  \BibitemOpen
  \bibfield  {author} {\bibinfo {author} {\bibfnamefont {R.}~\bibnamefont
  {Kubo}},\ }\bibfield  {title} {\bibinfo {title} {Statistical-mechanical
  theory of irreversible processes. i. general theory and simple applications
  to magnetic and conduction problems},\ }\href@noop {} {\bibfield  {journal}
  {\bibinfo  {journal} {Journal of the Physical Society of Japan}\ }\textbf
  {\bibinfo {volume} {12}},\ \bibinfo {pages} {570} (\bibinfo {year}
  {1957})}\BibitemShut {NoStop}%
\bibitem [{\citenamefont {Marconi}\ \emph {et~al.}(2008)\citenamefont
  {Marconi}, \citenamefont {Puglisi}, \citenamefont {Rondoni},\ and\
  \citenamefont {Vulpiani}}]{marconi2008fluctuation}%
  \BibitemOpen
  \bibfield  {author} {\bibinfo {author} {\bibfnamefont {U.~M.~B.}\
  \bibnamefont {Marconi}}, \bibinfo {author} {\bibfnamefont {A.}~\bibnamefont
  {Puglisi}}, \bibinfo {author} {\bibfnamefont {L.}~\bibnamefont {Rondoni}},\
  and\ \bibinfo {author} {\bibfnamefont {A.}~\bibnamefont {Vulpiani}},\
  }\bibfield  {title} {\bibinfo {title} {Fluctuation-dissipation: Response
  theory in statistical physics},\ }\href@noop {} {\bibfield  {journal}
  {\bibinfo  {journal} {Physics Reports}\ }\textbf {\bibinfo {volume} {461}},\
  \bibinfo {pages} {111} (\bibinfo {year} {2008})}\BibitemShut {NoStop}%
\bibitem [{\citenamefont {Maes}(2020)}]{maes2020response}%
  \BibitemOpen
  \bibfield  {author} {\bibinfo {author} {\bibfnamefont {C.}~\bibnamefont
  {Maes}},\ }\bibfield  {title} {\bibinfo {title} {Response theory: a
  trajectory-based approach},\ }\href@noop {} {\bibfield  {journal} {\bibinfo
  {journal} {Frontiers in Physics}\ }\textbf {\bibinfo {volume} {8}},\ \bibinfo
  {pages} {229} (\bibinfo {year} {2020})}\BibitemShut {NoStop}%
\bibitem [{\citenamefont {Baiesi}\ \emph {et~al.}(2009)\citenamefont {Baiesi},
  \citenamefont {Maes},\ and\ \citenamefont
  {Wynants}}]{baiesi2009fluctuations}%
  \BibitemOpen
  \bibfield  {author} {\bibinfo {author} {\bibfnamefont {M.}~\bibnamefont
  {Baiesi}}, \bibinfo {author} {\bibfnamefont {C.}~\bibnamefont {Maes}},\ and\
  \bibinfo {author} {\bibfnamefont {B.}~\bibnamefont {Wynants}},\ }\bibfield
  {title} {\bibinfo {title} {Fluctuations and response of nonequilibrium
  states},\ }\href@noop {} {\bibfield  {journal} {\bibinfo  {journal} {Physical
  Review Letters}\ }\textbf {\bibinfo {volume} {103}},\ \bibinfo {pages}
  {010602} (\bibinfo {year} {2009})}\BibitemShut {NoStop}%
\bibitem [{\citenamefont {Seifert}\ and\ \citenamefont
  {Speck}(2010)}]{seifert2010fluctuation}%
  \BibitemOpen
  \bibfield  {author} {\bibinfo {author} {\bibfnamefont {U.}~\bibnamefont
  {Seifert}}\ and\ \bibinfo {author} {\bibfnamefont {T.}~\bibnamefont
  {Speck}},\ }\bibfield  {title} {\bibinfo {title} {Fluctuation-dissipation
  theorem in nonequilibrium steady states},\ }\href@noop {} {\bibfield
  {journal} {\bibinfo  {journal} {EPL (Europhysics Letters)}\ }\textbf
  {\bibinfo {volume} {89}},\ \bibinfo {pages} {10007} (\bibinfo {year}
  {2010})}\BibitemShut {NoStop}%
\bibitem [{\citenamefont {Speck}\ and\ \citenamefont
  {Seifert}(2006)}]{speck2006restoring}%
  \BibitemOpen
  \bibfield  {author} {\bibinfo {author} {\bibfnamefont {T.}~\bibnamefont
  {Speck}}\ and\ \bibinfo {author} {\bibfnamefont {U.}~\bibnamefont
  {Seifert}},\ }\bibfield  {title} {\bibinfo {title} {Restoring a
  fluctuation-dissipation theorem in a nonequilibrium steady state},\
  }\href@noop {} {\bibfield  {journal} {\bibinfo  {journal} {EPL (Europhysics
  Letters)}\ }\textbf {\bibinfo {volume} {74}},\ \bibinfo {pages} {391}
  (\bibinfo {year} {2006})}\BibitemShut {NoStop}%
\bibitem [{\citenamefont {Harada}\ and\ \citenamefont
  {Sasa}(2005)}]{harada2005equality}%
  \BibitemOpen
  \bibfield  {author} {\bibinfo {author} {\bibfnamefont {T.}~\bibnamefont
  {Harada}}\ and\ \bibinfo {author} {\bibfnamefont {S.-i.}\ \bibnamefont
  {Sasa}},\ }\bibfield  {title} {\bibinfo {title} {Equality connecting energy
  dissipation with a violation of the fluctuation-response relation},\
  }\href@noop {} {\bibfield  {journal} {\bibinfo  {journal} {Physical Review
  Letters}\ }\textbf {\bibinfo {volume} {95}},\ \bibinfo {pages} {130602}
  (\bibinfo {year} {2005})}\BibitemShut {NoStop}%
\bibitem [{\citenamefont {Prost}\ \emph {et~al.}(2009)\citenamefont {Prost},
  \citenamefont {Joanny},\ and\ \citenamefont
  {Parrondo}}]{prost2009generalized}%
  \BibitemOpen
  \bibfield  {author} {\bibinfo {author} {\bibfnamefont {J.}~\bibnamefont
  {Prost}}, \bibinfo {author} {\bibfnamefont {J.-F.}\ \bibnamefont {Joanny}},\
  and\ \bibinfo {author} {\bibfnamefont {J.~M.~R.}\ \bibnamefont {Parrondo}},\
  }\bibfield  {title} {\bibinfo {title} {Generalized fluctuation-dissipation
  theorem for steady-state systems},\ }\href@noop {} {\bibfield  {journal}
  {\bibinfo  {journal} {Physical Review Letters}\ }\textbf {\bibinfo {volume}
  {103}},\ \bibinfo {pages} {090601} (\bibinfo {year} {2009})}\BibitemShut
  {NoStop}%
\bibitem [{\citenamefont {Aslyamov}\ \emph {et~al.}(2025)\citenamefont
  {Aslyamov}, \citenamefont {Ptaszy{\'n}ski},\ and\ \citenamefont
  {Esposito}}]{aslyamov2025nonequilibrium}%
  \BibitemOpen
  \bibfield  {author} {\bibinfo {author} {\bibfnamefont {T.}~\bibnamefont
  {Aslyamov}}, \bibinfo {author} {\bibfnamefont {K.}~\bibnamefont
  {Ptaszy{\'n}ski}},\ and\ \bibinfo {author} {\bibfnamefont {M.}~\bibnamefont
  {Esposito}},\ }\bibfield  {title} {\bibinfo {title} {Nonequilibrium
  fluctuation-response relations: From identities to bounds},\ }\href@noop {}
  {\bibfield  {journal} {\bibinfo  {journal} {Physical Review Letters}\
  }\textbf {\bibinfo {volume} {134}},\ \bibinfo {pages} {157101} (\bibinfo
  {year} {2025})}\BibitemShut {NoStop}%
\bibitem [{\citenamefont {Ptaszy{\'n}ski}\ \emph {et~al.}(2026)\citenamefont
  {Ptaszy{\'n}ski}, \citenamefont {Aslyamov},\ and\ \citenamefont
  {Esposito}}]{ptaszynski2026nonequilibrium}%
  \BibitemOpen
  \bibfield  {author} {\bibinfo {author} {\bibfnamefont {K.}~\bibnamefont
  {Ptaszy{\'n}ski}}, \bibinfo {author} {\bibfnamefont {T.}~\bibnamefont
  {Aslyamov}},\ and\ \bibinfo {author} {\bibfnamefont {M.}~\bibnamefont
  {Esposito}},\ }\bibfield  {title} {\bibinfo {title} {Nonequilibrium
  fluctuation-response relations for state-current correlations},\ }\href@noop
  {} {\bibfield  {journal} {\bibinfo  {journal} {Physical Review E}\ }\textbf
  {\bibinfo {volume} {113}},\ \bibinfo {pages} {024131} (\bibinfo {year}
  {2026})}\BibitemShut {NoStop}%
\bibitem [{\citenamefont {Aslyamov}\ and\ \citenamefont
  {Esposito}(2026)}]{aslyamov2026dynamical}%
  \BibitemOpen
  \bibfield  {author} {\bibinfo {author} {\bibfnamefont {T.}~\bibnamefont
  {Aslyamov}}\ and\ \bibinfo {author} {\bibfnamefont {M.}~\bibnamefont
  {Esposito}},\ }\bibfield  {title} {\bibinfo {title} {Dynamical
  fluctuation-response relations},\ }\href@noop {} {\bibfield  {journal}
  {\bibinfo  {journal} {arXiv preprint arXiv:2604.24626}\ } (\bibinfo {year}
  {2026})}\BibitemShut {NoStop}%
\bibitem [{\citenamefont {Bao}\ and\ \citenamefont
  {Liang}(2024)}]{bao2024nonlinear}%
  \BibitemOpen
  \bibfield  {author} {\bibinfo {author} {\bibfnamefont {R.}~\bibnamefont
  {Bao}}\ and\ \bibinfo {author} {\bibfnamefont {S.}~\bibnamefont {Liang}},\
  }\bibfield  {title} {\bibinfo {title} {Nonlinear response identities and
  bounds for nonequilibrium steady states},\ }\href@noop {} {\bibfield
  {journal} {\bibinfo  {journal} {arXiv preprint arXiv:2412.19602}\ } (\bibinfo
  {year} {2024})}\BibitemShut {NoStop}%
\bibitem [{\citenamefont {Dechant}\ and\ \citenamefont
  {Sasa}(2020)}]{dechant2020fluctuation}%
  \BibitemOpen
  \bibfield  {author} {\bibinfo {author} {\bibfnamefont {A.}~\bibnamefont
  {Dechant}}\ and\ \bibinfo {author} {\bibfnamefont {S.-i.}\ \bibnamefont
  {Sasa}},\ }\bibfield  {title} {\bibinfo {title} {Fluctuation--response
  inequality out of equilibrium},\ }\href@noop {} {\bibfield  {journal}
  {\bibinfo  {journal} {Proceedings of the National Academy of Sciences}\
  }\textbf {\bibinfo {volume} {117}},\ \bibinfo {pages} {6430} (\bibinfo {year}
  {2020})}\BibitemShut {NoStop}%
\bibitem [{\citenamefont {Zheng}\ and\ \citenamefont
  {Lu}(2025{\natexlab{a}})}]{zheng2025universal}%
  \BibitemOpen
  \bibfield  {author} {\bibinfo {author} {\bibfnamefont {J.}~\bibnamefont
  {Zheng}}\ and\ \bibinfo {author} {\bibfnamefont {Z.}~\bibnamefont {Lu}},\
  }\bibfield  {title} {\bibinfo {title} {Universal response inequalities beyond
  steady states via trajectory information geometry},\ }\href@noop {}
  {\bibfield  {journal} {\bibinfo  {journal} {Physical Review E}\ }\textbf
  {\bibinfo {volume} {112}},\ \bibinfo {pages} {L012103} (\bibinfo {year}
  {2025}{\natexlab{a}})}\BibitemShut {NoStop}%
\bibitem [{\citenamefont {Zheng}\ and\ \citenamefont
  {Lu}(2025{\natexlab{b}})}]{zheng2025unified}%
  \BibitemOpen
  \bibfield  {author} {\bibinfo {author} {\bibfnamefont {J.}~\bibnamefont
  {Zheng}}\ and\ \bibinfo {author} {\bibfnamefont {Z.}~\bibnamefont {Lu}},\
  }\bibfield  {title} {\bibinfo {title} {Unified linear fluctuation-response
  theory arbitrarily far from equilibrium},\ }\href@noop {} {\bibfield
  {journal} {\bibinfo  {journal} {Physical Review E}\ }\textbf {\bibinfo
  {volume} {112}},\ \bibinfo {pages} {064103} (\bibinfo {year}
  {2025}{\natexlab{b}})}\BibitemShut {NoStop}%
\bibitem [{\citenamefont {Zheng}\ and\ \citenamefont
  {Lu}(2026{\natexlab{a}})}]{zheng2026nonlinear}%
  \BibitemOpen
  \bibfield  {author} {\bibinfo {author} {\bibfnamefont {J.}~\bibnamefont
  {Zheng}}\ and\ \bibinfo {author} {\bibfnamefont {Z.}~\bibnamefont {Lu}},\
  }\bibfield  {title} {\bibinfo {title} {Nonlinear response relations and
  fluctuation-response inequalities for nonequilibrium stochastic systems},\
  }\href@noop {} {\bibfield  {journal} {\bibinfo  {journal} {The Journal of
  Chemical Physics}\ }\textbf {\bibinfo {volume} {164}} (\bibinfo {year}
  {2026}{\natexlab{a}})}\BibitemShut {NoStop}%
\bibitem [{\citenamefont {Dechant}(2026)}]{dechant2026finite}%
  \BibitemOpen
  \bibfield  {author} {\bibinfo {author} {\bibfnamefont {A.}~\bibnamefont
  {Dechant}},\ }\bibfield  {title} {\bibinfo {title} {Finite-frequency
  fluctuation-response inequality},\ }\href@noop {} {\bibfield  {journal}
  {\bibinfo  {journal} {Physical Review Letters}\ }\textbf {\bibinfo {volume}
  {136}},\ \bibinfo {pages} {207101} (\bibinfo {year} {2026})}\BibitemShut
  {NoStop}%
\bibitem [{\citenamefont {Zheng}\ and\ \citenamefont
  {Lu}(2026{\natexlab{b}})}]{zheng2026thermodynamic}%
  \BibitemOpen
  \bibfield  {author} {\bibinfo {author} {\bibfnamefont {J.}~\bibnamefont
  {Zheng}}\ and\ \bibinfo {author} {\bibfnamefont {Z.}~\bibnamefont {Lu}},\
  }\bibfield  {title} {\bibinfo {title} {Thermodynamic and kinetic bounds for
  finite-frequency fluctuation-response},\ }\href@noop {} {\bibfield  {journal}
  {\bibinfo  {journal} {arXiv preprint arXiv:2602.18631}\ } (\bibinfo {year}
  {2026}{\natexlab{b}})}\BibitemShut {NoStop}%
\bibitem [{\citenamefont {Owen}\ \emph {et~al.}(2020)\citenamefont {Owen},
  \citenamefont {Gingrich},\ and\ \citenamefont
  {Horowitz}}]{owen2020universal}%
  \BibitemOpen
  \bibfield  {author} {\bibinfo {author} {\bibfnamefont {J.~A.}\ \bibnamefont
  {Owen}}, \bibinfo {author} {\bibfnamefont {T.~R.}\ \bibnamefont {Gingrich}},\
  and\ \bibinfo {author} {\bibfnamefont {J.~M.}\ \bibnamefont {Horowitz}},\
  }\bibfield  {title} {\bibinfo {title} {Universal thermodynamic bounds on
  nonequilibrium response with biochemical applications},\ }\href@noop {}
  {\bibfield  {journal} {\bibinfo  {journal} {Physical Review X}\ }\textbf
  {\bibinfo {volume} {10}},\ \bibinfo {pages} {011066} (\bibinfo {year}
  {2020})}\BibitemShut {NoStop}%
\bibitem [{\citenamefont {Fernandes~Martins}\ and\ \citenamefont
  {Horowitz}(2023)}]{fernandes2023topologically}%
  \BibitemOpen
  \bibfield  {author} {\bibinfo {author} {\bibfnamefont {G.}~\bibnamefont
  {Fernandes~Martins}}\ and\ \bibinfo {author} {\bibfnamefont {J.~M.}\
  \bibnamefont {Horowitz}},\ }\bibfield  {title} {\bibinfo {title}
  {Topologically constrained fluctuations and thermodynamics regulate
  nonequilibrium response},\ }\href@noop {} {\bibfield  {journal} {\bibinfo
  {journal} {Physical Review E}\ }\textbf {\bibinfo {volume} {108}},\ \bibinfo
  {pages} {044113} (\bibinfo {year} {2023})}\BibitemShut {NoStop}%
\bibitem [{\citenamefont {Owen}\ and\ \citenamefont
  {Horowitz}(2023)}]{owen2023size}%
  \BibitemOpen
  \bibfield  {author} {\bibinfo {author} {\bibfnamefont {J.~A.}\ \bibnamefont
  {Owen}}\ and\ \bibinfo {author} {\bibfnamefont {J.~M.}\ \bibnamefont
  {Horowitz}},\ }\bibfield  {title} {\bibinfo {title} {Size limits the
  sensitivity of kinetic schemes},\ }\href@noop {} {\bibfield  {journal}
  {\bibinfo  {journal} {Nature Communications}\ }\textbf {\bibinfo {volume}
  {14}},\ \bibinfo {pages} {1280} (\bibinfo {year} {2023})}\BibitemShut
  {NoStop}%
\bibitem [{\citenamefont {Floyd}\ \emph {et~al.}(2025)\citenamefont {Floyd},
  \citenamefont {Dinner},\ and\ \citenamefont
  {Vaikuntanathan}}]{floyd2025local}%
  \BibitemOpen
  \bibfield  {author} {\bibinfo {author} {\bibfnamefont {C.}~\bibnamefont
  {Floyd}}, \bibinfo {author} {\bibfnamefont {A.~R.}\ \bibnamefont {Dinner}},\
  and\ \bibinfo {author} {\bibfnamefont {S.}~\bibnamefont {Vaikuntanathan}},\
  }\bibfield  {title} {\bibinfo {title} {Local imperfect feedback control in
  non-equilibrium biophysical systems enabled by thermodynamic constraints},\
  }\href@noop {} {\bibfield  {journal} {\bibinfo  {journal} {arXiv preprint
  arXiv:2507.07295}\ } (\bibinfo {year} {2025})}\BibitemShut {NoStop}%
\bibitem [{\citenamefont {Harunari}\ \emph {et~al.}(2024)\citenamefont
  {Harunari}, \citenamefont {Dal~Cengio}, \citenamefont {Lecomte},\ and\
  \citenamefont {Polettini}}]{harunari2024mutual}%
  \BibitemOpen
  \bibfield  {author} {\bibinfo {author} {\bibfnamefont {P.~E.}\ \bibnamefont
  {Harunari}}, \bibinfo {author} {\bibfnamefont {S.}~\bibnamefont
  {Dal~Cengio}}, \bibinfo {author} {\bibfnamefont {V.}~\bibnamefont
  {Lecomte}},\ and\ \bibinfo {author} {\bibfnamefont {M.}~\bibnamefont
  {Polettini}},\ }\bibfield  {title} {\bibinfo {title} {Mutual linearity of
  nonequilibrium network currents},\ }\href@noop {} {\bibfield  {journal}
  {\bibinfo  {journal} {Physical Review Letters}\ }\textbf {\bibinfo {volume}
  {133}},\ \bibinfo {pages} {047401} (\bibinfo {year} {2024})}\BibitemShut
  {NoStop}%
\bibitem [{\citenamefont {Bebon}\ and\ \citenamefont
  {Speck}(2026)}]{bebon2026mutual}%
  \BibitemOpen
  \bibfield  {author} {\bibinfo {author} {\bibfnamefont {R.}~\bibnamefont
  {Bebon}}\ and\ \bibinfo {author} {\bibfnamefont {T.}~\bibnamefont {Speck}},\
  }\bibfield  {title} {\bibinfo {title} {Mutual linearity is a generic property
  of steady-state markov networks},\ }\href@noop {} {\bibfield  {journal}
  {\bibinfo  {journal} {Physical Review Letters}\ }\textbf {\bibinfo {volume}
  {136}},\ \bibinfo {pages} {137401} (\bibinfo {year} {2026})}\BibitemShut
  {NoStop}%
\bibitem [{\citenamefont {Zheng}\ and\ \citenamefont
  {Lu}(2026{\natexlab{c}})}]{zheng2026mutualpre}%
  \BibitemOpen
  \bibfield  {author} {\bibinfo {author} {\bibfnamefont {J.}~\bibnamefont
  {Zheng}}\ and\ \bibinfo {author} {\bibfnamefont {Z.}~\bibnamefont {Lu}},\
  }\bibfield  {title} {\bibinfo {title} {Mutual linearity in and out of
  stationarity for markov jump processes: A trajectory-based approach},\
  }\href@noop {} {\bibfield  {journal} {\bibinfo  {journal} {Physical Review
  E}\ }\textbf {\bibinfo {volume} {114}},\ \bibinfo {pages} {014126} (\bibinfo
  {year} {2026}{\natexlab{c}})}\BibitemShut {NoStop}%
\bibitem [{\citenamefont {Zheng}\ and\ \citenamefont
  {Lu}(2026{\natexlab{d}})}]{zheng2026mutual}%
  \BibitemOpen
  \bibfield  {author} {\bibinfo {author} {\bibfnamefont {J.}~\bibnamefont
  {Zheng}}\ and\ \bibinfo {author} {\bibfnamefont {Z.}~\bibnamefont {Lu}},\
  }\bibfield  {title} {\bibinfo {title} {Mutual linearity in and out of
  stationarity for markov jump processes: A trajectory-based approach},\
  }\href@noop {} {\bibfield  {journal} {\bibinfo  {journal} {Physical Review
  E}\ }\textbf {\bibinfo {volume} {114}},\ \bibinfo {pages} {014126} (\bibinfo
  {year} {2026}{\natexlab{d}})}\BibitemShut {NoStop}%
\bibitem [{\citenamefont {Stutzer}\ \emph {et~al.}(2026)\citenamefont
  {Stutzer}, \citenamefont {Dieball},\ and\ \citenamefont
  {Godec}}]{stutzer2026stochastic}%
  \BibitemOpen
  \bibfield  {author} {\bibinfo {author} {\bibfnamefont {L.~T.}\ \bibnamefont
  {Stutzer}}, \bibinfo {author} {\bibfnamefont {C.}~\bibnamefont {Dieball}},\
  and\ \bibinfo {author} {\bibfnamefont {A.}~\bibnamefont {Godec}},\ }\bibfield
   {title} {\bibinfo {title} {Stochastic calculus for pathwise observables of
  markov-jump processes: Unification of diffusion and jump dynamics},\
  }\href@noop {} {\bibfield  {journal} {\bibinfo  {journal} {Physical Review
  X}\ }\textbf {\bibinfo {volume} {16}},\ \bibinfo {pages} {021038} (\bibinfo
  {year} {2026})}\BibitemShut {NoStop}%
\bibitem [{\citenamefont {H{\"a}nggi}\ and\ \citenamefont
  {Thomas}(1982)}]{hanggi1982stochastic}%
  \BibitemOpen
  \bibfield  {author} {\bibinfo {author} {\bibfnamefont {P.}~\bibnamefont
  {H{\"a}nggi}}\ and\ \bibinfo {author} {\bibfnamefont {H.}~\bibnamefont
  {Thomas}},\ }\bibfield  {title} {\bibinfo {title} {Stochastic processes: Time
  evolution, symmetries and linear response},\ }\href@noop {} {\bibfield
  {journal} {\bibinfo  {journal} {Physics Reports}\ }\textbf {\bibinfo {volume}
  {88}},\ \bibinfo {pages} {207} (\bibinfo {year} {1982})}\BibitemShut
  {NoStop}%
\bibitem [{\citenamefont {Barbi}\ \emph {et~al.}(2005)\citenamefont {Barbi},
  \citenamefont {Bologna},\ and\ \citenamefont {Grigolini}}]{barbi2005linear}%
  \BibitemOpen
  \bibfield  {author} {\bibinfo {author} {\bibfnamefont {F.}~\bibnamefont
  {Barbi}}, \bibinfo {author} {\bibfnamefont {M.}~\bibnamefont {Bologna}},\
  and\ \bibinfo {author} {\bibfnamefont {P.}~\bibnamefont {Grigolini}},\
  }\bibfield  {title} {\bibinfo {title} {Linear response to perturbation of
  nonexponential renewal processes},\ }\href@noop {} {\bibfield  {journal}
  {\bibinfo  {journal} {Physical Review Letters}\ }\textbf {\bibinfo {volume}
  {95}},\ \bibinfo {pages} {220601} (\bibinfo {year} {2005})}\BibitemShut
  {NoStop}%
\bibitem [{\citenamefont {Andrieux}\ and\ \citenamefont
  {Gaspard}(2008)}]{andrieux2008fluctuation}%
  \BibitemOpen
  \bibfield  {author} {\bibinfo {author} {\bibfnamefont {D.}~\bibnamefont
  {Andrieux}}\ and\ \bibinfo {author} {\bibfnamefont {P.}~\bibnamefont
  {Gaspard}},\ }\bibfield  {title} {\bibinfo {title} {The fluctuation theorem
  for currents in semi-{M}arkov processes},\ }\href@noop {} {\bibfield
  {journal} {\bibinfo  {journal} {Journal of Statistical Mechanics: Theory and
  Experiment}\ ,\ \bibinfo {pages} {P11007}} (\bibinfo {year}
  {2008})}\BibitemShut {NoStop}%
\bibitem [{\citenamefont {Maes}\ \emph {et~al.}(2009)\citenamefont {Maes},
  \citenamefont {Neto{\v c}n{\'y}},\ and\ \citenamefont
  {Wynants}}]{maes2009dynamical}%
  \BibitemOpen
  \bibfield  {author} {\bibinfo {author} {\bibfnamefont {C.}~\bibnamefont
  {Maes}}, \bibinfo {author} {\bibfnamefont {K.}~\bibnamefont {Neto{\v
  c}n{\'y}}},\ and\ \bibinfo {author} {\bibfnamefont {B.}~\bibnamefont
  {Wynants}},\ }\bibfield  {title} {\bibinfo {title} {Dynamical fluctuations
  for semi-{M}arkov processes},\ }\href@noop {} {\bibfield  {journal} {\bibinfo
   {journal} {Journal of Physics A: Mathematical and Theoretical}\ }\textbf
  {\bibinfo {volume} {42}},\ \bibinfo {pages} {365002} (\bibinfo {year}
  {2009})}\BibitemShut {NoStop}%
\bibitem [{\citenamefont {Esposito}\ and\ \citenamefont
  {Lindenberg}(2008)}]{esposito2008continuous}%
  \BibitemOpen
  \bibfield  {author} {\bibinfo {author} {\bibfnamefont {M.}~\bibnamefont
  {Esposito}}\ and\ \bibinfo {author} {\bibfnamefont {K.}~\bibnamefont
  {Lindenberg}},\ }\bibfield  {title} {\bibinfo {title} {Continuous-time random
  walk for open systems: Fluctuation theorems and counting statistics},\
  }\href@noop {} {\bibfield  {journal} {\bibinfo  {journal} {Physical Review
  E}\ }\textbf {\bibinfo {volume} {77}},\ \bibinfo {pages} {051119} (\bibinfo
  {year} {2008})}\BibitemShut {NoStop}%
\bibitem [{\citenamefont {Ertel}\ \emph {et~al.}(2022)\citenamefont {Ertel},
  \citenamefont {van~der Meer},\ and\ \citenamefont
  {Seifert}}]{ertel2022operationally}%
  \BibitemOpen
  \bibfield  {author} {\bibinfo {author} {\bibfnamefont {B.}~\bibnamefont
  {Ertel}}, \bibinfo {author} {\bibfnamefont {J.}~\bibnamefont {van~der
  Meer}},\ and\ \bibinfo {author} {\bibfnamefont {U.}~\bibnamefont {Seifert}},\
  }\bibfield  {title} {\bibinfo {title} {Operationally accessible uncertainty
  relations for thermodynamically consistent semi-markov processes},\
  }\href@noop {} {\bibfield  {journal} {\bibinfo  {journal} {Physical Review
  E}\ }\textbf {\bibinfo {volume} {105}},\ \bibinfo {pages} {044113} (\bibinfo
  {year} {2022})}\BibitemShut {NoStop}%
\bibitem [{\citenamefont {Coghi}\ and\ \citenamefont
  {Garrahan}(2026)}]{coghi2026level}%
  \BibitemOpen
  \bibfield  {author} {\bibinfo {author} {\bibfnamefont {F.}~\bibnamefont
  {Coghi}}\ and\ \bibinfo {author} {\bibfnamefont {J.~P.}\ \bibnamefont
  {Garrahan}},\ }\bibfield  {title} {\bibinfo {title} {Level 2.5 large
  deviations and uncertainty relations for self-interacting jump processes:
  Tilting constructions and the emergence of time-scale separation},\
  }\href@noop {} {\bibfield  {journal} {\bibinfo  {journal} {arXiv preprint
  arXiv:2603.19411}\ } (\bibinfo {year} {2026})}\BibitemShut {NoStop}%
\bibitem [{\citenamefont {Crutchfield}\ and\ \citenamefont
  {Young}(1989)}]{crutchfield1989inferring}%
  \BibitemOpen
  \bibfield  {author} {\bibinfo {author} {\bibfnamefont {J.~P.}\ \bibnamefont
  {Crutchfield}}\ and\ \bibinfo {author} {\bibfnamefont {K.}~\bibnamefont
  {Young}},\ }\bibfield  {title} {\bibinfo {title} {Inferring statistical
  complexity},\ }\href@noop {} {\bibfield  {journal} {\bibinfo  {journal}
  {Physical review letters}\ }\textbf {\bibinfo {volume} {63}},\ \bibinfo
  {pages} {105} (\bibinfo {year} {1989})}\BibitemShut {NoStop}%
\bibitem [{\citenamefont {Shalizi}\ and\ \citenamefont
  {Crutchfield}(2001)}]{shalizi2001computational}%
  \BibitemOpen
  \bibfield  {author} {\bibinfo {author} {\bibfnamefont {C.~R.}\ \bibnamefont
  {Shalizi}}\ and\ \bibinfo {author} {\bibfnamefont {J.~P.}\ \bibnamefont
  {Crutchfield}},\ }\bibfield  {title} {\bibinfo {title} {Computational
  mechanics: Pattern and prediction, structure and simplicity},\ }\href@noop {}
  {\bibfield  {journal} {\bibinfo  {journal} {Journal of statistical physics}\
  }\textbf {\bibinfo {volume} {104}},\ \bibinfo {pages} {817} (\bibinfo {year}
  {2001})}\BibitemShut {NoStop}%
\bibitem [{\citenamefont {Kanazawa}\ and\ \citenamefont
  {Sornette}(2024)}]{kanazawa2024standard}%
  \BibitemOpen
  \bibfield  {author} {\bibinfo {author} {\bibfnamefont {K.}~\bibnamefont
  {Kanazawa}}\ and\ \bibinfo {author} {\bibfnamefont {D.}~\bibnamefont
  {Sornette}},\ }\bibfield  {title} {\bibinfo {title} {Standard form of master
  equations for general non-markovian jump processes: The laplace-space
  embedding framework and asymptotic solution},\ }\href@noop {} {\bibfield
  {journal} {\bibinfo  {journal} {Physical Review Research}\ }\textbf {\bibinfo
  {volume} {6}},\ \bibinfo {pages} {023270} (\bibinfo {year}
  {2024})}\BibitemShut {NoStop}%
\bibitem [{\citenamefont {Goerlich}\ \emph {et~al.}(2026)\citenamefont
  {Goerlich}, \citenamefont {Tartar}, \citenamefont {Roichman},\ and\
  \citenamefont {Sokolov}}]{goerlich2026fluctuation}%
  \BibitemOpen
  \bibfield  {author} {\bibinfo {author} {\bibfnamefont {R.}~\bibnamefont
  {Goerlich}}, \bibinfo {author} {\bibfnamefont {A.}~\bibnamefont {Tartar}},
  \bibinfo {author} {\bibfnamefont {Y.}~\bibnamefont {Roichman}},\ and\
  \bibinfo {author} {\bibfnamefont {I.~M.}\ \bibnamefont {Sokolov}},\
  }\bibfield  {title} {\bibinfo {title} {Fluctuation-response relation for a
  nonequilibrium system with resolved {M}arkovian embedding},\ }\href@noop {}
  {\bibfield  {journal} {\bibinfo  {journal} {Physical Review E}\ }\textbf
  {\bibinfo {volume} {114}},\ \bibinfo {pages} {014138} (\bibinfo {year}
  {2026})}\BibitemShut {NoStop}%
\bibitem [{\citenamefont {Siegle}\ \emph {et~al.}(2010)\citenamefont {Siegle},
  \citenamefont {Goychuk}, \citenamefont {Talkner},\ and\ \citenamefont
  {H{\"a}nggi}}]{siegle2010markovian}%
  \BibitemOpen
  \bibfield  {author} {\bibinfo {author} {\bibfnamefont {P.}~\bibnamefont
  {Siegle}}, \bibinfo {author} {\bibfnamefont {I.}~\bibnamefont {Goychuk}},
  \bibinfo {author} {\bibfnamefont {P.}~\bibnamefont {Talkner}},\ and\ \bibinfo
  {author} {\bibfnamefont {P.}~\bibnamefont {H{\"a}nggi}},\ }\bibfield  {title}
  {\bibinfo {title} {Markovian embedding of non-markovian superdiffusion},\
  }\href@noop {} {\bibfield  {journal} {\bibinfo  {journal} {Physical Review
  E}\ }\textbf {\bibinfo {volume} {81}},\ \bibinfo {pages} {011136} (\bibinfo
  {year} {2010})}\BibitemShut {NoStop}%
\bibitem [{\citenamefont {Alexandrovich}\ \emph {et~al.}(2016)\citenamefont
  {Alexandrovich}, \citenamefont {Holzmann},\ and\ \citenamefont
  {Leister}}]{alexandrovich2016nonparametric}%
  \BibitemOpen
  \bibfield  {author} {\bibinfo {author} {\bibfnamefont {G.}~\bibnamefont
  {Alexandrovich}}, \bibinfo {author} {\bibfnamefont {H.}~\bibnamefont
  {Holzmann}},\ and\ \bibinfo {author} {\bibfnamefont {A.}~\bibnamefont
  {Leister}},\ }\bibfield  {title} {\bibinfo {title} {Nonparametric
  identification and maximum likelihood estimation for hidden markov models},\
  }\href@noop {} {\bibfield  {journal} {\bibinfo  {journal} {Biometrika}\
  }\textbf {\bibinfo {volume} {103}},\ \bibinfo {pages} {423} (\bibinfo {year}
  {2016})}\BibitemShut {NoStop}%
\bibitem [{\citenamefont {Flomenbom}\ \emph {et~al.}(2005)\citenamefont
  {Flomenbom}, \citenamefont {Klafter},\ and\ \citenamefont
  {Szabo}}]{flomenbom2005what}%
  \BibitemOpen
  \bibfield  {author} {\bibinfo {author} {\bibfnamefont {O.}~\bibnamefont
  {Flomenbom}}, \bibinfo {author} {\bibfnamefont {J.}~\bibnamefont {Klafter}},\
  and\ \bibinfo {author} {\bibfnamefont {A.}~\bibnamefont {Szabo}},\ }\bibfield
   {title} {\bibinfo {title} {What can one learn from two-state single-molecule
  trajectories?},\ }\href@noop {} {\bibfield  {journal} {\bibinfo  {journal}
  {Biophysical Journal}\ }\textbf {\bibinfo {volume} {88}},\ \bibinfo {pages}
  {3780} (\bibinfo {year} {2005})}\BibitemShut {NoStop}%
\bibitem [{\citenamefont {Br{\'e}maud}(1981)}]{bremaud1981point}%
  \BibitemOpen
  \bibfield  {author} {\bibinfo {author} {\bibfnamefont {P.}~\bibnamefont
  {Br{\'e}maud}},\ }\href@noop {} {\emph {\bibinfo {title} {Point processes and
  queues: Martingale dynamics}}}\ (\bibinfo  {publisher} {Springer},\ \bibinfo
  {year} {1981})\BibitemShut {NoStop}%
\bibitem [{\citenamefont {Jacod}\ and\ \citenamefont
  {Shiryaev}(2003)}]{jacod2003limit}%
  \BibitemOpen
  \bibfield  {author} {\bibinfo {author} {\bibfnamefont {J.}~\bibnamefont
  {Jacod}}\ and\ \bibinfo {author} {\bibfnamefont {A.~N.}\ \bibnamefont
  {Shiryaev}},\ }\href@noop {} {\emph {\bibinfo {title} {Limit theorems for
  stochastic processes}}},\ Vol.\ \bibinfo {volume} {288}\ (\bibinfo
  {publisher} {Springer Science \& Business Media},\ \bibinfo {year}
  {2003})\BibitemShut {NoStop}%
\bibitem [{\citenamefont {Daley}\ and\ \citenamefont
  {Vere-Jones}(2008)}]{daley2008introduction}%
  \BibitemOpen
  \bibfield  {author} {\bibinfo {author} {\bibfnamefont {D.~J.}\ \bibnamefont
  {Daley}}\ and\ \bibinfo {author} {\bibfnamefont {D.}~\bibnamefont
  {Vere-Jones}},\ }\href@noop {} {\emph {\bibinfo {title} {An introduction to
  the theory of point processes: Volume II: General theory and structure}}}\
  (\bibinfo  {publisher} {Springer Science \& Business Media},\ \bibinfo {year}
  {2008})\BibitemShut {NoStop}%
\bibitem [{\citenamefont {Neri}\ \emph {et~al.}(2017)\citenamefont {Neri},
  \citenamefont {Rold{\'a}n},\ and\ \citenamefont
  {J{\"u}licher}}]{neri2017statistics}%
  \BibitemOpen
  \bibfield  {author} {\bibinfo {author} {\bibfnamefont {I.}~\bibnamefont
  {Neri}}, \bibinfo {author} {\bibfnamefont {{\'E}.}~\bibnamefont
  {Rold{\'a}n}},\ and\ \bibinfo {author} {\bibfnamefont {F.}~\bibnamefont
  {J{\"u}licher}},\ }\bibfield  {title} {\bibinfo {title} {Statistics of infima
  and stopping times of entropy production and applications to active molecular
  processes},\ }\href@noop {} {\bibfield  {journal} {\bibinfo  {journal}
  {Physical Review X}\ }\textbf {\bibinfo {volume} {7}},\ \bibinfo {pages}
  {011019} (\bibinfo {year} {2017})}\BibitemShut {NoStop}%
\bibitem [{\citenamefont {Ch{\'e}trite}\ and\ \citenamefont
  {Gupta}(2011)}]{chetrite2011two}%
  \BibitemOpen
  \bibfield  {author} {\bibinfo {author} {\bibfnamefont {R.}~\bibnamefont
  {Ch{\'e}trite}}\ and\ \bibinfo {author} {\bibfnamefont {S.}~\bibnamefont
  {Gupta}},\ }\bibfield  {title} {\bibinfo {title} {Two refreshing views of
  fluctuation theorems through kinematics elements and exponential
  martingale},\ }\href@noop {} {\bibfield  {journal} {\bibinfo  {journal}
  {Journal of Statistical Physics}\ }\textbf {\bibinfo {volume} {143}},\
  \bibinfo {pages} {543} (\bibinfo {year} {2011})}\BibitemShut {NoStop}%
\bibitem [{\citenamefont {Rold{\'a}n}\ \emph {et~al.}(2023)\citenamefont
  {Rold{\'a}n}, \citenamefont {Neri}, \citenamefont {Chetrite}, \citenamefont
  {Gupta}, \citenamefont {Pigolotti}, \citenamefont {J{\"u}licher},\ and\
  \citenamefont {Sekimoto}}]{roldan2023martingales}%
  \BibitemOpen
  \bibfield  {author} {\bibinfo {author} {\bibfnamefont {{\'E}.}~\bibnamefont
  {Rold{\'a}n}}, \bibinfo {author} {\bibfnamefont {I.}~\bibnamefont {Neri}},
  \bibinfo {author} {\bibfnamefont {R.}~\bibnamefont {Chetrite}}, \bibinfo
  {author} {\bibfnamefont {S.}~\bibnamefont {Gupta}}, \bibinfo {author}
  {\bibfnamefont {S.}~\bibnamefont {Pigolotti}}, \bibinfo {author}
  {\bibfnamefont {F.}~\bibnamefont {J{\"u}licher}},\ and\ \bibinfo {author}
  {\bibfnamefont {K.}~\bibnamefont {Sekimoto}},\ }\bibfield  {title} {\bibinfo
  {title} {Martingales for physicists: a treatise on stochastic thermodynamics
  and beyond},\ }\href@noop {} {\bibfield  {journal} {\bibinfo  {journal}
  {Advances in Physics}\ }\textbf {\bibinfo {volume} {72}},\ \bibinfo {pages}
  {1} (\bibinfo {year} {2023})}\BibitemShut {NoStop}%
\bibitem [{\citenamefont {Esposito}(2012)}]{esposito2012stochastic}%
  \BibitemOpen
  \bibfield  {author} {\bibinfo {author} {\bibfnamefont {M.}~\bibnamefont
  {Esposito}},\ }\bibfield  {title} {\bibinfo {title} {Stochastic
  thermodynamics under coarse graining},\ }\href@noop {} {\bibfield  {journal}
  {\bibinfo  {journal} {Physical Review E}\ }\textbf {\bibinfo {volume} {85}},\
  \bibinfo {pages} {041125} (\bibinfo {year} {2012})}\BibitemShut {NoStop}%
\bibitem [{\citenamefont {Hartich}\ and\ \citenamefont
  {Godec}(2021)}]{hartich2021emergent}%
  \BibitemOpen
  \bibfield  {author} {\bibinfo {author} {\bibfnamefont {D.}~\bibnamefont
  {Hartich}}\ and\ \bibinfo {author} {\bibfnamefont {A.}~\bibnamefont
  {Godec}},\ }\bibfield  {title} {\bibinfo {title} {Emergent memory and kinetic
  hysteresis in strongly driven networks},\ }\href@noop {} {\bibfield
  {journal} {\bibinfo  {journal} {Physical Review X}\ }\textbf {\bibinfo
  {volume} {11}},\ \bibinfo {pages} {041047} (\bibinfo {year}
  {2021})}\BibitemShut {NoStop}%
\bibitem [{\citenamefont {Colquhoun}\ and\ \citenamefont
  {Hawkes}(1987)}]{colquhoun1987note}%
  \BibitemOpen
  \bibfield  {author} {\bibinfo {author} {\bibfnamefont {D.}~\bibnamefont
  {Colquhoun}}\ and\ \bibinfo {author} {\bibfnamefont {A.~G.}\ \bibnamefont
  {Hawkes}},\ }\bibfield  {title} {\bibinfo {title} {A note on correlations in
  single ion channel records},\ }\href@noop {} {\bibfield  {journal} {\bibinfo
  {journal} {Proceedings of the Royal Society of London. Series B}\ }\textbf
  {\bibinfo {volume} {230}},\ \bibinfo {pages} {15} (\bibinfo {year}
  {1987})}\BibitemShut {NoStop}%
\bibitem [{\citenamefont {Magleby}\ and\ \citenamefont
  {Song}(1992)}]{magleby1992dependency}%
  \BibitemOpen
  \bibfield  {author} {\bibinfo {author} {\bibfnamefont {K.~L.}\ \bibnamefont
  {Magleby}}\ and\ \bibinfo {author} {\bibfnamefont {L.}~\bibnamefont {Song}},\
  }\bibfield  {title} {\bibinfo {title} {Dependency plots suggest the kinetic
  structure of ion channels},\ }\href@noop {} {\bibfield  {journal} {\bibinfo
  {journal} {Proceedings of the Royal Society of London. Series B}\ }\textbf
  {\bibinfo {volume} {249}},\ \bibinfo {pages} {133} (\bibinfo {year}
  {1992})}\BibitemShut {NoStop}%
\bibitem [{\citenamefont {Davis}(1976)}]{davis1976representation}%
  \BibitemOpen
  \bibfield  {author} {\bibinfo {author} {\bibfnamefont {M.~H.~A.}\
  \bibnamefont {Davis}},\ }\bibfield  {title} {\bibinfo {title} {The
  representation of martingales of jump processes},\ }\href@noop {} {\bibfield
  {journal} {\bibinfo  {journal} {SIAM Journal on Control and Optimization}\
  }\textbf {\bibinfo {volume} {14}},\ \bibinfo {pages} {623} (\bibinfo {year}
  {1976})}\BibitemShut {NoStop}%
\bibitem [{Note1()}]{Note1}%
  \BibitemOpen
  \bibinfo {note} {Here ``additive'' means that the observable is accumulated
  along the trajectory as a time integral of state-dependent contributions and
  a sum of jump-dependent contributions.}\BibitemShut {Stop}%
\bibitem [{\citenamefont {Plyasunov}\ and\ \citenamefont
  {Arkin}(2007)}]{plyasunov2007efficient}%
  \BibitemOpen
  \bibfield  {author} {\bibinfo {author} {\bibfnamefont {S.}~\bibnamefont
  {Plyasunov}}\ and\ \bibinfo {author} {\bibfnamefont {A.~P.}\ \bibnamefont
  {Arkin}},\ }\bibfield  {title} {\bibinfo {title} {Efficient stochastic
  sensitivity analysis of discrete event systems},\ }\href@noop {} {\bibfield
  {journal} {\bibinfo  {journal} {Journal of Computational Physics}\ }\textbf
  {\bibinfo {volume} {221}},\ \bibinfo {pages} {724} (\bibinfo {year}
  {2007})}\BibitemShut {NoStop}%
\bibitem [{\citenamefont {Warren}\ and\ \citenamefont
  {Allen}(2012)}]{warren2012steady}%
  \BibitemOpen
  \bibfield  {author} {\bibinfo {author} {\bibfnamefont {P.~B.}\ \bibnamefont
  {Warren}}\ and\ \bibinfo {author} {\bibfnamefont {R.~J.}\ \bibnamefont
  {Allen}},\ }\bibfield  {title} {\bibinfo {title} {Steady-state parameter
  sensitivity in stochastic modeling via trajectory reweighting},\ }\href@noop
  {} {\bibfield  {journal} {\bibinfo  {journal} {The Journal of Chemical
  Physics}\ }\textbf {\bibinfo {volume} {136}},\ \bibinfo {pages} {104106}
  (\bibinfo {year} {2012})}\BibitemShut {NoStop}%
\bibitem [{\citenamefont {Marzen}\ and\ \citenamefont
  {Crutchfield}(2017)}]{marzen2017structure}%
  \BibitemOpen
  \bibfield  {author} {\bibinfo {author} {\bibfnamefont {S.~E.}\ \bibnamefont
  {Marzen}}\ and\ \bibinfo {author} {\bibfnamefont {J.~P.}\ \bibnamefont
  {Crutchfield}},\ }\bibfield  {title} {\bibinfo {title} {Structure and
  randomness of continuous-time, discrete-event processes},\ }\href@noop {}
  {\bibfield  {journal} {\bibinfo  {journal} {Journal of Statistical Physics}\
  }\textbf {\bibinfo {volume} {169}},\ \bibinfo {pages} {303} (\bibinfo {year}
  {2017})}\BibitemShut {NoStop}%
\bibitem [{\citenamefont {Borys}\ \emph {et~al.}(2022)\citenamefont {Borys},
  \citenamefont {Trybek}, \citenamefont {Dworakowska}, \citenamefont
  {Bednarczyk},\ and\ \citenamefont
  {Wawrzkiewicz-Ja{\l}owiecka}}]{borys2022new}%
  \BibitemOpen
  \bibfield  {author} {\bibinfo {author} {\bibfnamefont {P.}~\bibnamefont
  {Borys}}, \bibinfo {author} {\bibfnamefont {P.}~\bibnamefont {Trybek}},
  \bibinfo {author} {\bibfnamefont {B.}~\bibnamefont {Dworakowska}}, \bibinfo
  {author} {\bibfnamefont {P.}~\bibnamefont {Bednarczyk}},\ and\ \bibinfo
  {author} {\bibfnamefont {A.}~\bibnamefont {Wawrzkiewicz-Ja{\l}owiecka}},\
  }\bibfield  {title} {\bibinfo {title} {New diagnostic tool for ion channel
  activity hidden behind the dwell-time correlations},\ }\href@noop {}
  {\bibfield  {journal} {\bibinfo  {journal} {The Journal of Physical Chemistry
  B}\ }\textbf {\bibinfo {volume} {126}},\ \bibinfo {pages} {4236} (\bibinfo
  {year} {2022})}\BibitemShut {NoStop}%
\bibitem [{\citenamefont {Wawrzkiewicz-Ja{\l}owiecka}\ \emph
  {et~al.}(2022)\citenamefont {Wawrzkiewicz-Ja{\l}owiecka}, \citenamefont
  {Borys}, \citenamefont {Trybek}, \citenamefont {Dworakowska},\ and\
  \citenamefont {Bednarczyk}}]{agata_wawrzkiewicz_jalowiecka_2022_6340407}%
  \BibitemOpen
  \bibfield  {author} {\bibinfo {author} {\bibfnamefont {A.}~\bibnamefont
  {Wawrzkiewicz-Ja{\l}owiecka}}, \bibinfo {author} {\bibfnamefont
  {P.}~\bibnamefont {Borys}}, \bibinfo {author} {\bibfnamefont
  {P.}~\bibnamefont {Trybek}}, \bibinfo {author} {\bibfnamefont
  {B.}~\bibnamefont {Dworakowska}},\ and\ \bibinfo {author} {\bibfnamefont
  {P.}~\bibnamefont {Bednarczyk}},\ }\href
  {https://doi.org/10.5281/zenodo.6340407} {\bibinfo {title} {A new powerful
  diagnostic tool for ion channels activity hidden behind the dwell-times
  correlations; software and metadata}} (\bibinfo {year} {2022})\BibitemShut
  {NoStop}%
\bibitem [{\citenamefont {Colquhoun}\ and\ \citenamefont
  {Hawkes}(1981)}]{colquhoun1981stochastic}%
  \BibitemOpen
  \bibfield  {author} {\bibinfo {author} {\bibfnamefont {D.}~\bibnamefont
  {Colquhoun}}\ and\ \bibinfo {author} {\bibfnamefont {A.~G.}\ \bibnamefont
  {Hawkes}},\ }\bibfield  {title} {\bibinfo {title} {On the stochastic
  properties of single ion channels},\ }\href@noop {} {\bibfield  {journal}
  {\bibinfo  {journal} {Proceedings of the Royal Society of London. Series B}\
  }\textbf {\bibinfo {volume} {211}},\ \bibinfo {pages} {205} (\bibinfo {year}
  {1981})}\BibitemShut {NoStop}%
\bibitem [{\citenamefont {Liebovitch}\ \emph {et~al.}(1987)\citenamefont
  {Liebovitch}, \citenamefont {Fischbarg},\ and\ \citenamefont
  {Koniarek}}]{liebovitch1987ion}%
  \BibitemOpen
  \bibfield  {author} {\bibinfo {author} {\bibfnamefont {L.~S.}\ \bibnamefont
  {Liebovitch}}, \bibinfo {author} {\bibfnamefont {J.}~\bibnamefont
  {Fischbarg}},\ and\ \bibinfo {author} {\bibfnamefont {J.~P.}\ \bibnamefont
  {Koniarek}},\ }\bibfield  {title} {\bibinfo {title} {Ion channel kinetics: A
  model based on fractal scaling rather than multistate {M}arkov processes},\
  }\href@noop {} {\bibfield  {journal} {\bibinfo  {journal} {Mathematical
  Biosciences}\ }\textbf {\bibinfo {volume} {84}},\ \bibinfo {pages} {37}
  (\bibinfo {year} {1987})}\BibitemShut {NoStop}%
\bibitem [{\citenamefont {McManus}\ \emph {et~al.}(1988)\citenamefont
  {McManus}, \citenamefont {Weiss}, \citenamefont {Spivak}, \citenamefont
  {Blatz},\ and\ \citenamefont {Magleby}}]{mcmanus1988fractal}%
  \BibitemOpen
  \bibfield  {author} {\bibinfo {author} {\bibfnamefont {O.~B.}\ \bibnamefont
  {McManus}}, \bibinfo {author} {\bibfnamefont {D.~S.}\ \bibnamefont {Weiss}},
  \bibinfo {author} {\bibfnamefont {C.~E.}\ \bibnamefont {Spivak}}, \bibinfo
  {author} {\bibfnamefont {A.~L.}\ \bibnamefont {Blatz}},\ and\ \bibinfo
  {author} {\bibfnamefont {K.~L.}\ \bibnamefont {Magleby}},\ }\bibfield
  {title} {\bibinfo {title} {Fractal models are inadequate for the kinetics of
  four different ion channels},\ }\href@noop {} {\bibfield  {journal} {\bibinfo
   {journal} {Biophysical Journal}\ }\textbf {\bibinfo {volume} {54}},\
  \bibinfo {pages} {859} (\bibinfo {year} {1988})}\BibitemShut {NoStop}%
\bibitem [{\citenamefont {Korn}\ and\ \citenamefont
  {Horn}(1988)}]{korn1988statistical}%
  \BibitemOpen
  \bibfield  {author} {\bibinfo {author} {\bibfnamefont {S.~J.}\ \bibnamefont
  {Korn}}\ and\ \bibinfo {author} {\bibfnamefont {R.}~\bibnamefont {Horn}},\
  }\bibfield  {title} {\bibinfo {title} {Statistical discrimination of fractal
  and {M}arkov models of single-channel gating},\ }\href@noop {} {\bibfield
  {journal} {\bibinfo  {journal} {Biophysical Journal}\ }\textbf {\bibinfo
  {volume} {54}},\ \bibinfo {pages} {871} (\bibinfo {year} {1988})}\BibitemShut
  {NoStop}%
\bibitem [{\citenamefont {Qin}\ \emph {et~al.}(1996)\citenamefont {Qin},
  \citenamefont {Auerbach},\ and\ \citenamefont {Sachs}}]{qin1996estimating}%
  \BibitemOpen
  \bibfield  {author} {\bibinfo {author} {\bibfnamefont {F.}~\bibnamefont
  {Qin}}, \bibinfo {author} {\bibfnamefont {A.}~\bibnamefont {Auerbach}},\ and\
  \bibinfo {author} {\bibfnamefont {F.}~\bibnamefont {Sachs}},\ }\bibfield
  {title} {\bibinfo {title} {Estimating single-channel kinetic parameters from
  idealized patch-clamp data containing missed events},\ }\href@noop {}
  {\bibfield  {journal} {\bibinfo  {journal} {Biophysical Journal}\ }\textbf
  {\bibinfo {volume} {70}},\ \bibinfo {pages} {264} (\bibinfo {year}
  {1996})}\BibitemShut {NoStop}%
\bibitem [{\citenamefont {Lyu}\ \emph {et~al.}(2026)\citenamefont {Lyu},
  \citenamefont {Ray},\ and\ \citenamefont {Crutchfield}}]{lyu2026optimal}%
  \BibitemOpen
  \bibfield  {author} {\bibinfo {author} {\bibfnamefont {J.}~\bibnamefont
  {Lyu}}, \bibinfo {author} {\bibfnamefont {K.~J.}\ \bibnamefont {Ray}},\ and\
  \bibinfo {author} {\bibfnamefont {J.~P.}\ \bibnamefont {Crutchfield}},\
  }\bibfield  {title} {\bibinfo {title} {Optimal computation from fluctuation
  responses},\ }\href@noop {} {\bibfield  {journal} {\bibinfo  {journal}
  {Physical Review Research}\ }\textbf {\bibinfo {volume} {8}},\ \bibinfo
  {pages} {023253} (\bibinfo {year} {2026})}\BibitemShut {NoStop}%
\bibitem [{\citenamefont {Zheng}(2026)}]{data}%
  \BibitemOpen
  \bibfield  {author} {\bibinfo {author} {\bibfnamefont {J.}~\bibnamefont
  {Zheng}},\ }\href@noop {} {} (\bibinfo {year} {2026}),\ \bibinfo {note}
  {https://github.com/Axeho2/The-Memory-Hidden-in-Response-Fluctuations.}\BibitemShut
  {Stop}%
\end{thebibliography}%

\end{document}